\documentclass[preprints,article,accept,moreauthors,pdftex]{mdpi}

\usepackage{url, color}
\usepackage{graphicx}
\usepackage{blindtext,fontenc, inputenc, calc, indentfirst, fancyhdr}
\usepackage{bookmark}
\usepackage{hyperref}

\hypersetup{linkcolor=red,citecolor=blue,filecolor=cyan,urlcolor=magenta}

\firstpage{1}
\pubvolume{xx}
\issuenum{1}
\articlenumber{5}
\pubyear{2020}
\copyrightyear{2020}

\history{Received: date; Accepted: date; Published: date}

\Title{Constraining
 LQG   Graph    with  Light Surfaces: Properties of  BH Thermodynamics  for  Mini-Super-Space, Semi-Classical Polymeric  BH}

\Author{{Daniela Pugliese} $^{1,}$* and Giovanni Montani $^{2,3}$}

\address{%
$^{1}$ \quad Research Centre of Theoretical Physics and Astrophysics, Institute of Physics, Silesian University in Opava, Bezru\v{c}ovo n\'{a}m\v{e}st\'{i} 13, CZ-74601 Opava, Czech Republic\\
$^{2}$ \quad ENEA- R.C. Frascati, UTFUS-MAG, Via Enrico Fermi 45, Frascati, Roma 00044, Italy; giovanni.montani@enea.it\\
$^{3}$ \quad {Physics Department,} "Sapienza" University of Rome, P.le Aldo Moro 5, Roma 00185, Italy}

\corres{Correspondence: daniela.pugliese@fpf.slu.cz}

\abstract{{This work  participates  in the research for   potential  areas of  observational  evidence of quantum effects on geometry  in a black hole astrophysical context.}
We consider  properties of a  family of loop quantum  corrected regular black hole {(BHs)} solutions  and their horizons,  focusing  on the geometry symmetries.
{We study here  a recently developed  model, where the
geometry is determined by a   metric quantum modification outside the horizon.}  This is a regular static spherical solution  of mini-super-space BH metric  with {Loop Quantum Gravity (LQG)}  corrections.
The solutions are characterized delineating certain   polymeric  functions on the basis of the properties  of the horizons and the emergence of a singularity in the limiting  case of the  Schwarzschild  geometry.
  {We discuss  particular metric  solutions on the  base of  the parameters of the polymeric model related to similar properties of structures,  the  metric Killing bundles (or metric bundles {MBs}), related to  the   BH  horizons' properties}.
 A comparison with the Reissner--Norstr\"om geometry  and the Kerr geometry with which analogies exist from the point of their respective  {MBs} properties is done.
The analysis provides a way to recognize these geometries  and detect their main  distinctive {phenomenological evidence of} {LQG} origin  on the basis of the detection  of stationary/static observers and  the properties of light-like  orbits within the analysis of   the (conformal invariant) {MBs} related to the   (local) causal structure.
 This approach could  be applied  in other  quantum corrected BH solutions,   constraining  the characteristics of the underlining {LQG}-{graph}, as the minimal loop area, through the analysis of the null-like  orbits and photons detection.  {The study of light surfaces associated with a diversified and wide range of  BH phenomenology  and  grounding {MBs} definition provides  a  channel to search for possible  astrophysical evidence.} The main   {BHs} thermodynamic characteristics are studied as luminosity, surface gravity, and  temperature.  Ultimately, the application of  this method to this spherically symmetric  approximate  solution  provides us with a way  to clarify some formal aspects of {MBs}, in the presence of static, spherical symmetric spacetimes.}

\keyword{quantum gravity; loop quantum gravity (LQG); graphs; polymeric black hole regular black hole; Killing horizons; event horizons; stationary observers}

\begin{document}
\newcommand{\ti}[1]{\mbox{\tiny{#1}}}
\newcommand{\im}{\mathop{\mathrm{Im}}}
\def\be{\begin{equation}}
\def\ee{\end{equation}}
\def\bew{\begin{widetext}}
\def\eew{\end{widetext}}
\def\Rem{\textbf{Remark}}
\def\bea{\begin{eqnarray}}
\def\eea{\end{eqnarray}}
\newcommand{\oft}{\mathcal{o}_T}
\newcommand{\ofx}{\mathcal{o}_X}

\newcommand{\ttb}[1]{\textbf{#1}}
\newcommand{\ctimes}{\overset{\mathbf{\cdot}}{\times}}
\newcommand{\tb}[1]{\textbf{{#1}}}
\newcommand{\rtb}[1]{\textcolor[rgb]{1.00,0.00,0.00}{\tb{#1}}}
\newcommand{\gtb}[1]{\textcolor[rgb]{0.17,0.72,0.40}{\tb{#1}}}
\newcommand{\ptb}[1]{\textcolor[rgb]{0.77,0.04,0.95}{\tb{#1}}}

\newcommand{\otb}[1]{\textcolor[rgb]{1.00,0.50,0.25}{\tb{#1}}}
\newcommand{\non}[1]{{\LARGE{\not}}{#1}}
\newcommand{\nnon}[1]{{\Huge{\not}}{#1}}
\newcommand{\parr}[1]{\breve{\partial}}
\newcommand{\body}[1]{\|\mathbf{#1}\|}
\newcommand{\ff}[1]{\left\lfloor #1\right\rfloor}
\newcommand{\Hl}{{\large{ \mathrm{H}}}}
\newcommand{\srt}[1]{{\scriptsize{#1}}}

\newcommand{\bp}[1]{$\eth$}
\newcommand{\Ga}{\mathrm{G}}
\newcommand{\hh}[1]{\left\lceil #1\right\rceil}
\newcommand{\kets}[1]{\ket{#1}}
\newcommand{\kett}[1]{\kets_{#1}}
\newcommand{\ssp}{{\footnotesize{\textsf{\textbf{{S}}}}}}
\newcommand{\cc}{\mathrm{C}}
\newcommand{\il}{~}
\newcommand{\ddp}[1]{\vec{\dda}^{\scriptscriptstyle{#1}}}
\newcommand{\rc}{\rho_{\ti{C}}}
\newcommand{\ex}{\exists}
\newcommand{\vex}{\vec{\exists}}
\newcommand{\hex}{\hat{\exists}}
\newcommand{\oftt}{\mathcal{o}_{\widetilde{T}}}
\newcommand{\ofs}{\mathcal{o_S}}
\newcommand{\la}{\mathcal{A}}
\newcommand{\scr}{${\scriptsize{$\circ$}}$}

\newcommand{\Sa}{\mathcal{\mathbf{S}}}
\newcommand{\Ca}{\mathcal{\mathbf{C}}}
\newcommand{\Ha}{\mathcal{H}}
\newcommand{\dda}{\mathcal{D}}
\newcommand{\Qa}{\mathcal{Q}}
\newcommand{\1}{\mathds{1}}

\newcommand{\loops}{\mathbf{\mathrlap{\circlearrowleft}{\circlearrowright}}}

  \newcommand{\downmapsto}{\rotatebox[origin=c]{-90}{${\longmapsto}$}\mkern2mu}
   \newcommand{\vpoints}{\rotatebox[origin=c]{-90}{$...$}\mkern2mu}
  \newcommand{\upmapsto}{\rotatebox[origin=c]{90}{${\longmapsto}$}\mkern2mu}

\newcommand{\Tem}{T^{\rm{em}}}
\newcommand{\laa}{\mathcal{L}}


\section{Introduction}
{Providing astrophysical evidence of any quantum effects on the  large scales of  the general relativistic (GR) geometry  is a pressing and long-sought channel of analysis of the current  research efforts  in the development  of the theory   as well as  in observations and data analysis. Observational feedbacks,
 confirming  or, vice versa, diverging expectations,  could  provide indications, directions, or confirmations to the theory.   The astrophysical phenomenology  of the  black hole (BH) horizons  is an increasingly attractive and promising channel that has revealed itself to be surprisingly rich of applications, both on the phenomenological and theoretical level.
There is  great expectation concerning the two new phenomenological windows on the High Energy  Universe,
represented by the Event Horizon Telescope {(https://eventhorizontelescope.org/)}  and  the Gravitational Waves detection  combined with data from  their electromagnetic  counterparts.  There is hope that new data
 will  shed light on unexpected aspects of BH theory and particularly in  the  near  horizon geometry.
 This work fits into this investigation by discussing  a special  structure, the metric Killing bundles. We present this   novel frame  applied to a Loop Quantum Gravity ({LQG}) approximate (semiclassical)  BH solution.
We consider   properties of the light-surfaces defined from the metrics symmetries  and the associated relativistical photon  orbital frequencies. The objects   brought into play   here, as the base of the newly introduced structures, Killing   metric bundles,   ground in fact many constraints of the   BH  (and  Killing horizons) astrophysics  ranging from physics of accretion to the jet emission, from BH magnetosphere to some  models of   Quasi-Periodic Oscillation  ({QPOs}) emission. As a consequence of this, this frame and the new approach could reveal ample  applications. The introduction of metric bundles   potentially   highlights
 small but finite discrepancies, expectably detectible from the observations of the photon orbital frequencies, and provides an alternative and new  framework  for the investigation of the BH geometries.
 One goal of this analysis is  to constrain the graph   bridging  the  quantum (discrete) and (semi-)classical (continuum) geometry  with special light-surfaces. The graph, in a general  {LQG} application, is in fact geometrical construction associated with loop quantum gravity states which can in fact be interpreted as the basis of states of a quantum  (discrete) geometry, grounding from  original  Penrose's spin networks. In the Penrose original combinatorial  graphs, spins labelled the graph, and graphs turned to be  just the mathematical   (geometrical) objects used to describe the quanta of space  for a  spin
network theory.   Such original   graph has been modified later in different loop inspired models.
Graphs clearly have a direct  application   in  the  relational approaches, where the adjacency (links, lines) describes  relations in
quanta of space.
Within the grounding idea of gravity geometrization,
  quanta of space are accordingly  also  gravity quanta and the texture of  the spacetime structures  (hence framing geometry into a relational approach where relations are  embedded and translated into the graph structure). In many approaches, graphs are not dynamical but rather
"generalized  lattices",
polyhedron (tetrahedron) modeling 3D spaces, providing eventually an area parameter.
 As in this analysis, we link the area parameter of the model  to  the BH area  through the frequencies defined with  the bundles.
The new frame is  used to  relate   geometries, defined  by different values of {LQG} derived graph   parameters, by the light--surfaces  frequencies. Such frequencies have replicas in different geometries of the  metric family defined by different values. These replicas are present also in  different points of the one geometry, connecting  therefore regions  close to and far away from the black hole: they  connect two regions of the same spacetime,  possibly revealing essential discrepancies with respect to the onset provided by the classic  GR solution  of reference. An observer can detect certain aspects of  regions close to the BH  through replicas  in the distant regions.  These replicas  appear also  related to structures emerged  in different  analyses and called  "horizons  memory" and "horizons  remnants" also in   naked singularity extension of the  geometries.
}

Loop Quantum Gravity ({LQG}) is a non-perturbative and background-independent quantum theory of gravity. In its  standard formulation, "space" is described by a spin network.
The  formulation of  the {LQG}  spin network is represented  with a graph that is generally   closed and colored (with values attached to edged or vertices of its faces according to the different realizations and {LQG} models).
The  basis states of
{LQG} are the graphs, with  (valued) graph edges and nodes (vertices),  associated with
 irreducible SU(2) representations the first and interwiners the second. More precisely,  the graph  vertices represent 3-volume quanta. As the  graph is fully connected, the edges  are in fact the  "quanta" of area
$A(j)= 8 \pi \gamma _{BI} \sqrt{j(j+1)}$, being the  half-integer $j$ the  \emph{edge} value, and
$\gamma_{BI}$ is  the Berbero--Immirzi parameter.
The model is defined by a  smallest possible quanta that corresponds to the minimal (quantized) area
$A_{min}$,
depending on the  product of the $\gamma_{BI}$   and a polymeric {parameter} ($\delta$).

In this work, we focus  on special loop quantum corrected (polymeric) regular  black hole solutions  ({LBHs}), a special family of spherical symmetric  regular loop BH solutions, whose  general relativistic limit for some values of the parameters is the Schwarzschild solution.
The {model developed in}\cite{Aleshi2012zz,12,13,14,15} is a metric modification outside the horizon, with  the minimal area derived from  {LQG},  and  the (free) parameter $\delta$,   found   from a  mini-super-space   {LQG} approximation.
The metric approximation is based on quantum modifications outside the horizon assuming a regular lattice with edges of lengths $d_b $and $d_c$, reduced then to one independent $\delta$ parameter by considering the minimal area to be  the {LQG}
minimal area.
In this mini-super-space model, the {LQG} corrections regularize (solve) the central black hole singularity problem.

 {One task of this analysis is the investigation of}  the constraints for the graph properties by constraining the values of the main graph parameters  emerging from evaluations of quantities related to the {LBHs}, as a minimal area parameter  $a_o$ or eventually $\epsilon=\gamma_{BI}\delta$, or the loop mass $m$ which is a function here of the polymeric parameter $P$ and the ADM mass $M$.
For this purpose, we use \emph{metric Killing bundles} (or metric bundles {(MBs)}) which are a collection of metric solutions of the parameterized family of {LBHs}, characterized by certain identical properties of the light-like particles orbital frequency that can be measured by observers at infinity, informing on some properties of the geometries close to the horizons, and connecting the different metrics of the bundles. These structures define also some properties of the local causal structure and thermodynamical properties of {BHs} as the surface gravity, temperature, and luminosity.
The {LBHs} geometry shares similarities with the
 Reissner--Nordstr\"om ({RN}) line tensor. We consider here this analogy in the analysis of {MBs} applied to {LBHs}.

 These  regular quantum corrected
 {BHs} constitute an important spacetime environment to test {LQG}, and  to  search quantum-gravity effects  for the MB applications.
We constrain the loop graph  characteristics  (minimal area,  polymeric parameter) on the horizons property within characteristics  locally measurable from an observer in the region $r>r_+$, where $r_+$ is the outer horizon of the {LBH} solution, as  the orbital light-like particle frequencies, and, depending on the mass parameters, here  the ADM mass $M$  and the  polymeric mass  $m(M,P)$, which we consider separately.
The orbital frequency can be measured related to the  {local} causal structure by an observer locally, and in a point of the \emph{extended plane} which is a plane $\mathcal{P}-r$, where the {MBs} are defined as curves, $\mathcal{P}$ is the metrics family parameter, and $r$ is a radial distance. In this application, as in \cite{renmants,observers,ergon,Pugliese:2019rfz,
Pugliese:2019efv}, the \emph{bundles} are defined as the  sets of all geometries having equal limiting light-like orbital frequency, which is also an asymptotic limiting value for time-like stationary observers (as measured at infinity).
{MBs} are conformal invariant and can be easily read in terms of the light surfaces ({LS}), related to the analysis of many aspects of {BHs} physics, as "BH" images   and several  processes constraining energy extraction as the  BH jet emission and jet  collimation.
The role of {MBs} is clear in the geometries with Killing horizons as the Kerr geometries, and more generally in the axially symmetric spacetimes as the Kerr--Newman ({KN}) geometry and Kerr--{de- Sitter} geometry--\cite{renmants,ergon,observers,Pugliese:2019rfz,Pugliese:2019efv}.
The {MBs} definition to the spherical symmetric cases considered here for the regular BH is not immediate.
Spherically symmetric {BHs} solutions have generally a direct astrophysical interest  as limiting conditions for the spinning {BHs}. In \cite{renmants,ergon,observers}, the Schwarzschild geometry in {MBs} analysis has been considered as limiting solution for the axially symmetric   Kerr geometry or the  Kerr--Newman geometry, or also the Reissner--Nordstr\"om family. Schwarzschild geometry is  represented as a point on the horizon in  the extended plane for all these solutions \cite{Kerr-Newman,Kerr}.
In this respect, the {LBH}  metric  is interesting from the MB point of view  as in fact these solutions are spherically symmetric and regular {BHs} that are   asymptotically related (in the {MBs} sense) to the Schwarzschild solution, allowing for clarifying bundles' characteristics.

Below we discuss more precisely some main notions on metric bundles.
Introduced in\cite{renmants} to explain some properties of black holes
and naked singularities ({NSs}),  {MBs}  establish  a relation between these. Their definition was first   based   on the  investigation  of the limiting frequencies of stationary observers,  and define  the  Killing horizons for the Kerr   black holes and then extended the  to other  exact solutions as the cosmological Kerr--de-Sitter,  the Kerr Newman spacetimes, and the limiting case of Reissner--Norstr\"om solutions \cite{renmants,ergon,observers,Pugliese:2019rfz,Pugliese:2019efv}.
The
\emph{metric bundles}, characterized by a particular relation between the metric parameters, are sets of geometries defined by one characteristic light-like (circular) orbital frequency $\omega_{\pm}$, which is  the bundle characteristic frequency, and, in the spinning geometries, also coincides with the horizons frequencies/angular velocities.
A metric bundle is represented by a curve on the so-called extended plane \cite{renmants,Pugliese:2019rfz,
Pugliese:2019efv}. The  extended plane  contains the entire collection of the parameterized family of metric solutions. We can define  here the  \emph{extended plane} as a plane $\mathcal{P}-r$ where  $r$ is the radial distance (in polar spherical coordinate or conveniently chosen adapted to infinity Boyer--Lindquist coordinates in the Kerr, Kerr-Newman or Kerr--de --Sitter), and $\mathcal{P}$ is a metric family parameter (or, eventually, a set of parameters).
In the extended plane, the horizons of  all {BHs} solutions of the family  can be represented as one curve or a set of curves.
In the axially symmetric spacetimes with Killing horizons, the {MBs} are   all
tangent to the horizon curve  on the extended plane. Then, the horizon curve
emerges as the envelope surface of the set of metric bundles.
In the spherically symmetric spacetimes, {MBs} approach asymptotically (for some special values) the horizons.
The tangency condition  of {MBs} with the horizons' curves characteristic of the axially symmetric Killing horizons  spacetimes reduces to an approximation condition for the spherically symmetric {LBH} we consider here as well as the {RN} solutions, considered in \cite{renmants} as limiting static solution of the {KN}. On the other hand, a special adaptation of the   main idea underling the MB definitions adapted  to more general  horizons concepts is certainly possible.

Investigating  metric bundles,  we explore in
an alternative way some aspects of the geometries defining  the bundle as measured by an observer at infinity.
The metric bundle concept can be
significant  in the study of BH physics, in the interpretation of {NSs} solutions
and BH thermodynamics.
In this work, we present the definition of metric bundles and discuss their properties
also in the context  of loop BH thermodynamics.

These structures essentially explicate some properties of the Killing horizons in the axially symmetric spacetimes and events horizons in the spherically symmetric case.  The horizons in this last case  are  limiting surfaces of the {MBs} in the extended plane. Each geometry of the set  has, at  a certain radius $r$,  equal   {characteristic bundle frequency}.

{MBs} are characterized by several special properties: the horizons remnants, structures of the bundles typical of  certain {NSs}, the horizons replicas,  and  the idea of an "horizon  memory"  firstly introduced in \cite{renmants} provide through the {MBs}  a different perspective to explore these spacetimes, conferring a global vision  including the possibility to study the transition of  one geometry evolving in a different  solution of the family. In this sense, the  extended plane, where bundles are defined as curves, is endowed with a  "certain plasticity" (whose typical expression is for example in the {NSs} remnants).
Each spacetime in one point has  some properties that are replicated in different  geometries (BH or {NSs}), which can then be thought of as target or transition state in the geometry  evolution as regulated by the (first law of classic) BH thermodynamics.
Killing bottlenecks appear in certain {NSs}  as restriction of the Killing throat  in the associated light surfaces' analysis. These   structures (in general close to the extreme Kerr or {KN} solutions) were seen also as ``horizons remnants" in {NSs} \cite{renmants,ergon,observers,Pugliese:2019rfz,Pugliese:2019efv}  and   appear also connected with the concept of pre-horizon regime introduced in \cite{de-Felice1-frirdtforstati,de-FelicefirstKerr,de-Felice3,de-Felice4-overspinning,de-Felice-mass,de-FeliceKerr,de-Felice-anceKerr}. The pre-horizon was analyzed in \cite{de-Felice1-frirdtforstati,de-FelicefirstKerr,de-Felice3,de-Felice4-overspinning,de-Felice-mass,de-FeliceKerr,de-Felice-anceKerr}. In these analyses, it was concluded that a gyroscope would conserve a memory of the static
or stationary initial state, leading to the gravitational collapse of a mass distribution \cite{de-Felice1-frirdtforstati,de-FelicefirstKerr,de-Felice3,de-Felice4-overspinning,de-Felice-mass,de-FeliceKerr,de-Felice-anceKerr,Chakraborty:2016mhx,Zaslavskii:2018kix,Zaslavskii:2019pdc,Tanatarov:2016mcs}.

{The article plan:}
This article is structured as follows: In Section (\ref{Sec:metric-bundles}), we discuss  the  main properties of the {LQG} metric and metric bundles: the black hole solutions are introduced in Section (\ref{Sec:metric}), metric bundles are discussed in Section (\ref{Sec:litig-colo}) and  we  construct  the extended plane for these solutions  in Section (\ref{Sec:ext-plane}).  The comparison with  the case of the  Reissner--Norstr\"om geometry is addressed in Section (\ref{Sec:RN}).
{Metric bundles of the {LBHs}} are the focus of Section (\ref{Sec:wordsee-h}).
In Section (\ref{Sec:termo}).
We review some aspects of the BH thermodynamics exploring in Section (\ref{Sec:termo-1}) the {BHs} surfaces' gravity, the luminosity and the temperature in terms of the loop model parameters, then these quantities are  considered  on metric bundles.
In Section (\ref{Sec:conlc}), we summarize the main steps of this analysis, concluding this article.

Throughout this work, we introduced a  number of symbols and notations  necessary to
explain all the results obtained for these recently introduced objects; however, there is in fact a relatively small set of objects  constituting  a core of the {MBs} we analyze along this analysis,  and are  listed for reference in Table \il\ref{Table:pol-cy-multi}.
\begin{table}[H]
\caption{Lookup table with the main symbols and relevant notations  used throughout the article with a brief description and reference to the first introduction of the term.  Links to associated sections, definitions, and figures are also listed.  Notation {RN}  refers to Reissner--Norstr\"om   geometry.  We specify that $\sigma\equiv\sin^2\theta$  and $H(r)=g_{\phi\phi}/\sigma$. In general, we  adopt notation $\mathcal{Q}_{\bullet}\equiv \mathcal{Q}(r_{\bullet})$ for any quantity $\mathcal{Q}$ evaluated  on  a general radius $r_{\bullet}$. A notable  example concerns the case of quantities $\mathcal{Q}_{\pm}\equiv \mathcal{Q}(r_{\pm})$  evaluated on horizons $r_{\pm}$  where we use superscript (occasionally subscript where necessary)  $\pm$ respectively and, for convenience with the common use in literature and specified in the text, we use $\mathcal{Q}_{H}\equiv \mathcal{Q}_{+}$ for quantities evaluated  on the outer horizon  $r_+$. The frequency notation is excluded from this rule:  $\omega_{\pm}$ are limiting photon orbital frequencies which on  the horizons of the spherically symmetric  geometries considered here are clearly null or $\omega_{\pm}(r_{\pm}=0)$.}
\label{Table:pol-cy-multi}
\centering
\resizebox{.99\textwidth}{!}{%
\begin{tabular}{lll}
 \hline \hline
 \\
  $(\xi_t,\xi_{\phi})$ &  Killing fields  of the geometry&
Equation~(\ref{Eq:ass})--Section (\ref{Sec:metric})--Section (\ref{Sec:wordsee-h})
\\
$a_o= A_{min}/8\pi$,&   the area parameter&Equation~(\ref{Ep:espi-p})--Section (\ref{Sec:metric})
\\
 $A_{min}$&  minimum area gap of {LQG}& Equation~(\ref{Ep:espi-p})--Section (\ref{Sec:metric})
\\
  $P$&   metric polymeric parameter&Equation~(\ref{Ep:espi-p})-- Figures (\ref{Fig:colorPP},\ref{Fig:vengplre})--Section (\ref{Sec:metric})
\\
$\epsilon=\gamma _{BI} \delta$& $\delta$=metric polymeric parameter, $\gamma _{BI}$=Barbero--Immirzi parameter&Equation~(\ref{Ep:espi-p})- Figures (\ref{Fig:colorPP},\ref{Fig:vengplre})-Section (\ref{Sec:metric})
  \\
$M$&   ADM mass in the Schwarzschild limit &
Equation~(\ref{Ep:espi-p})--Section (\ref{Sec:metric})
\\
 $m$& mass  polymeric parameter function &  Equation~(\ref{Ep:espi-p})-- Figures (\ref{Fig:colorPP},\ref{Fig:vengplre})--Section (\ref{Sec:metric})
\\
 $r_\pm$& horizons &   Equation~(\ref{Eq:horizons})--
  -- Figures (\ref{Fig:colorPP},\ref{Fig:vengplre})--Section (\ref{Sec:metric})
  \\
  $\epsilon_{\pm}$ &horizons  in $\epsilon$-loop parameter&
Equation~(\ref{Eq:pok-fac})--
Figure\il (\ref{Fig:vengplre})
\\
$P_{\pm}$ &horizons  in   $P$-loop parameter   in extended plane
&
Equation~(\ref{Eq:vengplre})
\\
$\mathcal{L}=\partial_t +\omega \partial_{\phi}$& null  Killing vector
 (generators of Killing event  horizons)& Section (\ref{Sec:litig-colo})
 \\
$\mathcal{L}_{\mathcal{N}}=0$ &  Killing vector $\mathcal{L}$ norm   $\mathbf{ g } (\mathcal{L},\mathcal{L})$   &Equation~(\ref{Eq:ass})--Section (\ref{Sec:litig-colo},\ref{Sec:wordsee-h})
\\
$ \omega_{\pm}$&   light-like   ($\mathbf{\mathcal{L_N}}=0$)  limiting frequencies  for stationary observers & Equation~(\ref{Eq:be-en})
\\
$\omega_{Sch}$ & limiting frequencies for the Schwarzschild geometry & Equation~(\ref{Eq:datitru},\ref{Eq:be-en})
\\
$\Qa_{T}$&  {RN} spacetime  "total charge"
 &
Equation~(\ref{Eq:q23d})--
Section (\ref{Sec:RN})
\\
$a_{\pm} $&
{Kerr Killing horizon curve} in the extended plane& Section (\ref{Sec:RN})
\\
 $Q_{\pm}$&
 {RN} horizon  in the extended plane& Section (\ref{Sec:RN})--Figures (\ref{Fig:vengplre})
 \\
  $Q_{\omega}$&
{RN} metric bundles &Equation~(\ref{Eq:be-en}).
\\
   $r_{a_o}^{min}=\sqrt{a_o}$ &a minimum  curve for the $\sqrt{H(r)}$ as function of $r$  ($\sqrt{H(r_{a_o}^{ min})}=2r_{a_o}^{min}$)&
    Figures (\ref{Fig:CausalP})--Section (\ref{Sec:metric})
\\
$\sigma_{\omega}$&  metric bundles:$\theta$  parametrization&
Equation~(\ref{Eq:media.-stateprima})
\\
$a_o^{\pm}(m,P)$ &metric bundles: $a_o$-parametrization&
Equation~(\ref{Eq:minsa})--Figures (\ref{Fig:SPlru8a})
\\
$r_{\tau}$& solution of $\partial_P\omega _{\pm }=0$ &Equation~(\ref{Eq:tau-def})--Figures (\ref{Fig:pcolorP1})
\\
$m_{\tau }$& solution of  $\partial _m\omega _{\pm }=0
$
&Equation~(\ref{Eq:mtau})
\\
$\kappa: \nabla^\alpha\mathcal{L}=-2\kappa \mathcal{L}^\alpha$ &
(acceleration) on  $r_{\pm}$, $\kappa_{\pm}$ define  BH surface gravity&Equation~(\ref{Eq:surface})--Section (\ref{Sec:litig-colo},\ref{Sec:termo-1})
\\
 $T_{\mathbf{BH}}$&  BH temperature &Section (\ref{Sec:litig-colo},\ref{Sec:termo-1})
\\
$A_{\mathbf{BH}}$ &  BH areas &Equation~(\ref{Eq:BHarea-poer})--Figures (\ref{Fig:pax12})--Section (\ref{Sec:litig-colo},\ref{Sec:termo-1})
\\
$L (m)$&
Luminosity&Equation~(\ref{Eq:luci})--Figures (\ref{Fig:cononapo16})--Section (\ref{Sec:termo-1})
\\
 \hline\hline
\end{tabular}}
\end{table}

\section{On LBHs  and  the Metric  Bundles}\label{Sec:metric-bundles}
In this section, we discuss the  main properties of the {LQG} metric  and the metric bundles. We introduce the {LBHs} solutions in Section (\ref{Sec:metric}). In Section (\ref{Sec:litig-colo}), we discuss some general properties of metric bundles.  Constructing the extended plane for these solutions, we  compare with  the case of the  Reissner--Norstr\"om geometry in Section (\ref{Sec:RN}).
More information on the  issues related to {BHs} solutions in {LQG} can, for example, be found in \cite{Perez:2017cmj,Barrau:2018rts,Barrau:2019swg,Haggard}, while definition of metric-bundles in the context of  geometries with  Killing horizons that is  the other aspect underlying this analysis  we refer to \cite{renmants,observers,ergon}.
 \subsection{The Metric}\label{Sec:metric}
We  consider    the  static spherical     loop BH ({LBH}) solution  derived as {LQG} approximation  as

\bea \label{Eq:dom-gionr}
&& ds^2 = - G(r) dt^2 + \frac{dr^2}{F(r)} + H(r) d\Omega^2, \quad \mbox{where}
\quad   d\Omega^2\equiv d\theta^2+\sigma d\phi^2, \quad \sigma\equiv \sin^2\theta,
 \\
 &&\nonumber G(r)\equiv\frac{(r-r_+)(r-r_-)(r+ r_{*})^2}{r^4 +a_o^2},  \quad  H(r) \equiv r^2 + \frac{a_o^2}{r^2},\quad
 F(r) \equiv \frac{(r-r_+)(r-r_-) r^4}{(r+ r_{*})^2 (r^4 +a_o^2)}
  \quad\mbox{and}
 \\\label{Ep:espi-p}
 &&r_* \equiv  \sqrt{r_+ r_-} = 2mP,\quad P(\epsilon) \equiv  \frac{\sqrt{1+\epsilon^2} -1}{\sqrt{1+\epsilon^2} +1},\quad M = m (1+P)^2\quad a_o=\frac{A_{min}}{8 \pi}.
\eea
In the metric,  $P$ is  the polymeric parameter, where generally $P\ll 1$ so $(r_- ,r_*)$  vanishes and the metric can be considered approaching a Schwarzschild limit.
$a_o$  is the area parameter,  equal to $ A_{min}/8\pi$,  $A_{min}$ being the minimum area appearing in {LQG} (minimum area gap of {LQG}),  which can be  related to  parameter
$\epsilon=\gamma _{BI} \delta$, and  $\gamma _{BI}$ is  the Barbero--Immirzi parameter.
$M$ is the  ADM mass in the Schwarzschild limit, i.e., the mass
 for an observer at  asymptotically flat  infinity;  vice versa,
 the $m$ parameter depends  on  the polymeric function $P$ and $M$.

\begin{figure}[H]
\centering
  \includegraphics[width=5cm]{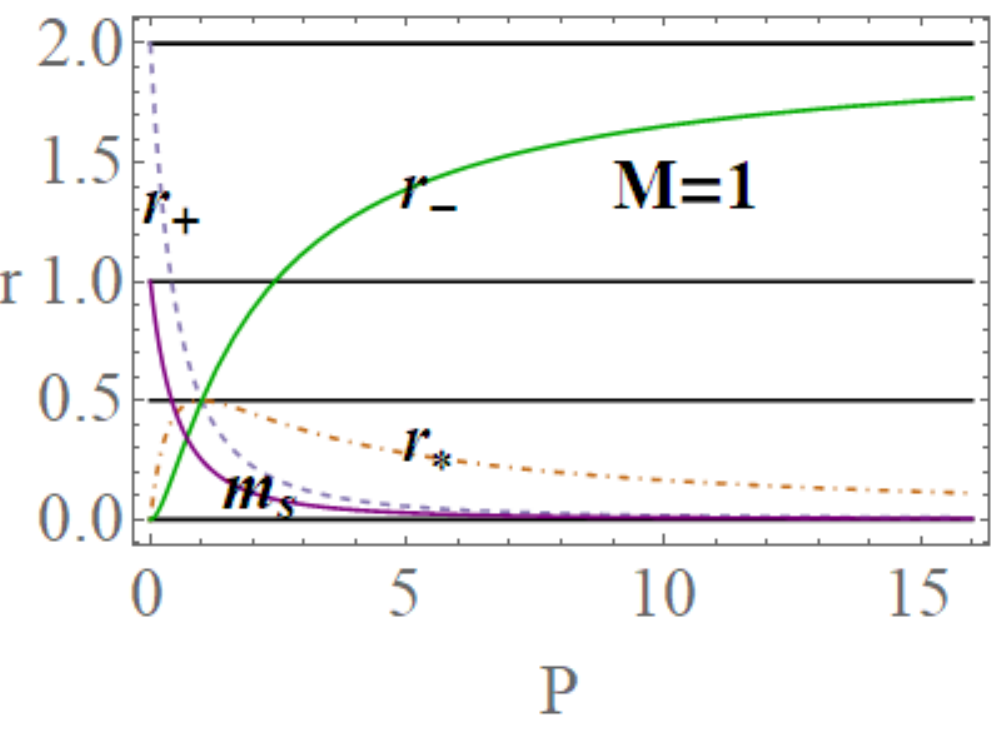}
  \includegraphics[width=4.85cm]{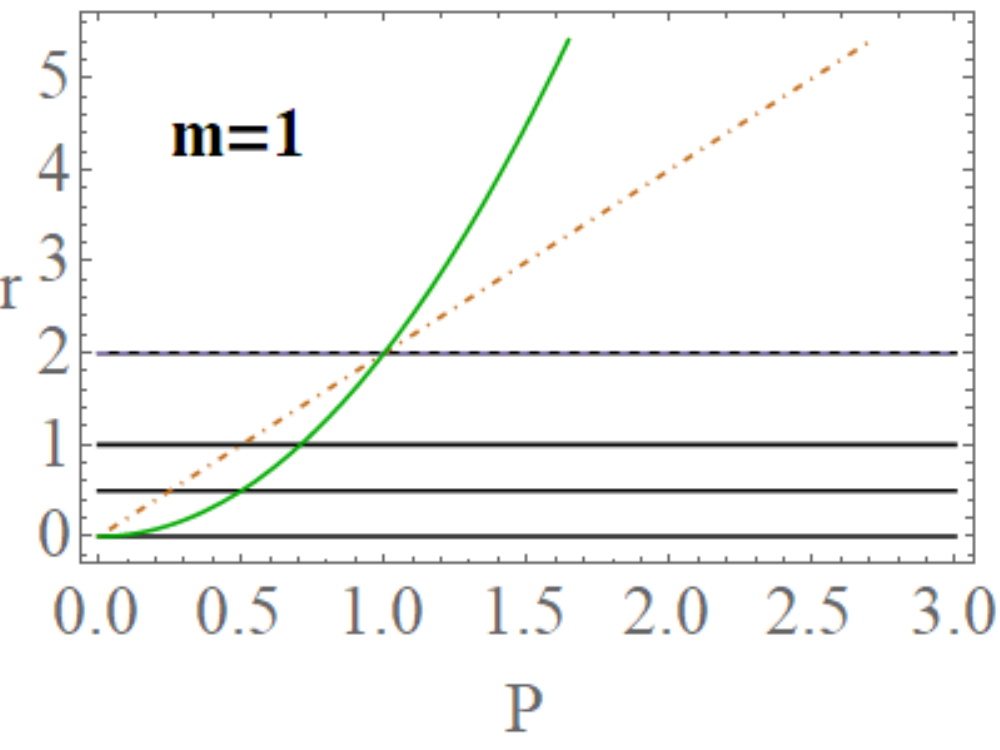}
                      \includegraphics[width=5.3cm]{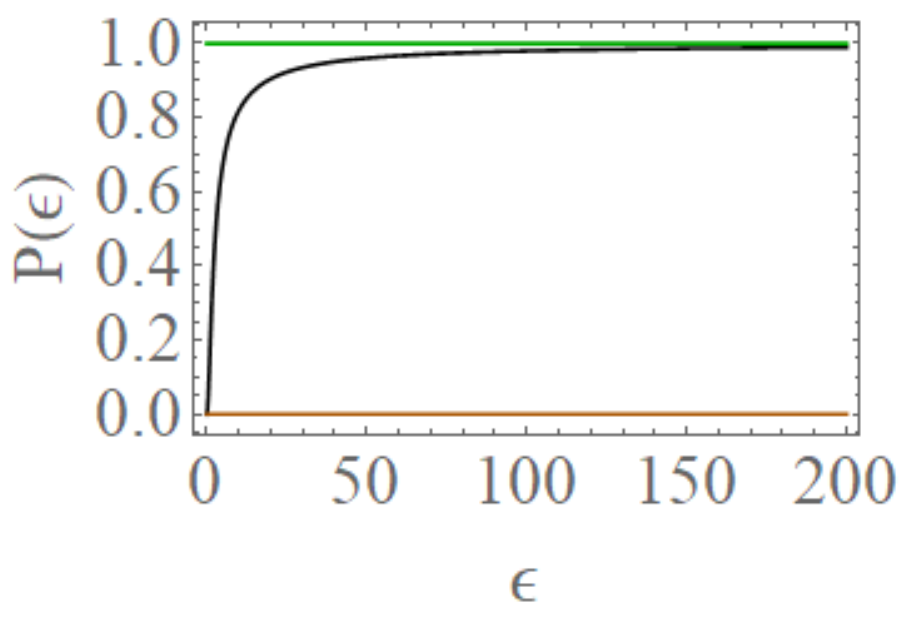}\\
          \includegraphics[width=5cm]{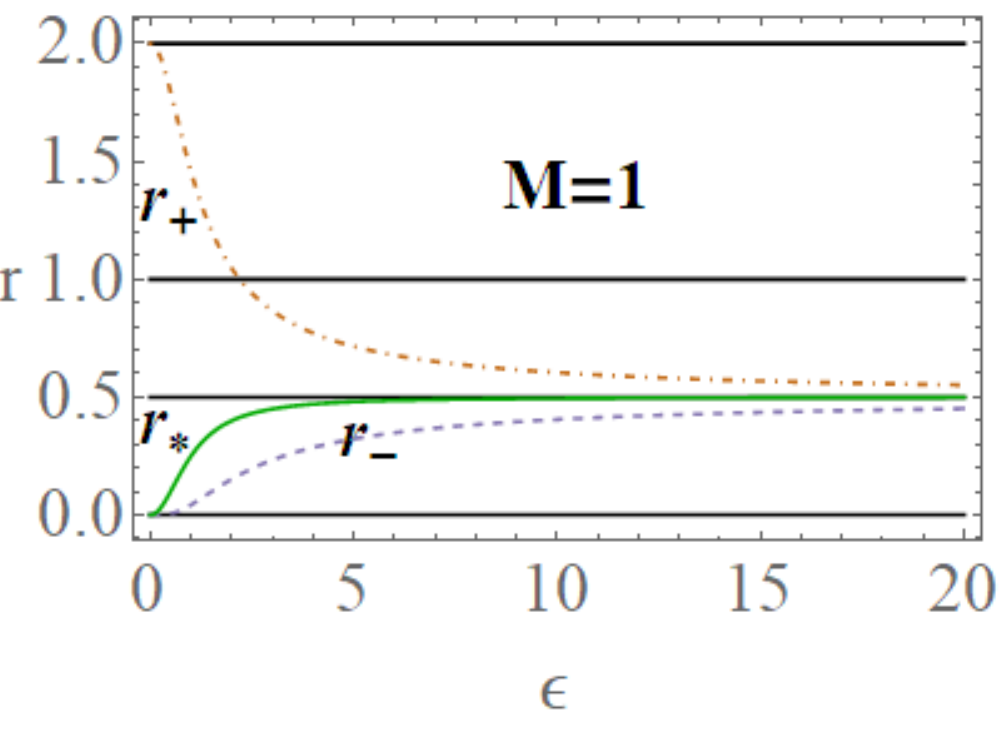}
                   \includegraphics[width=5cm]{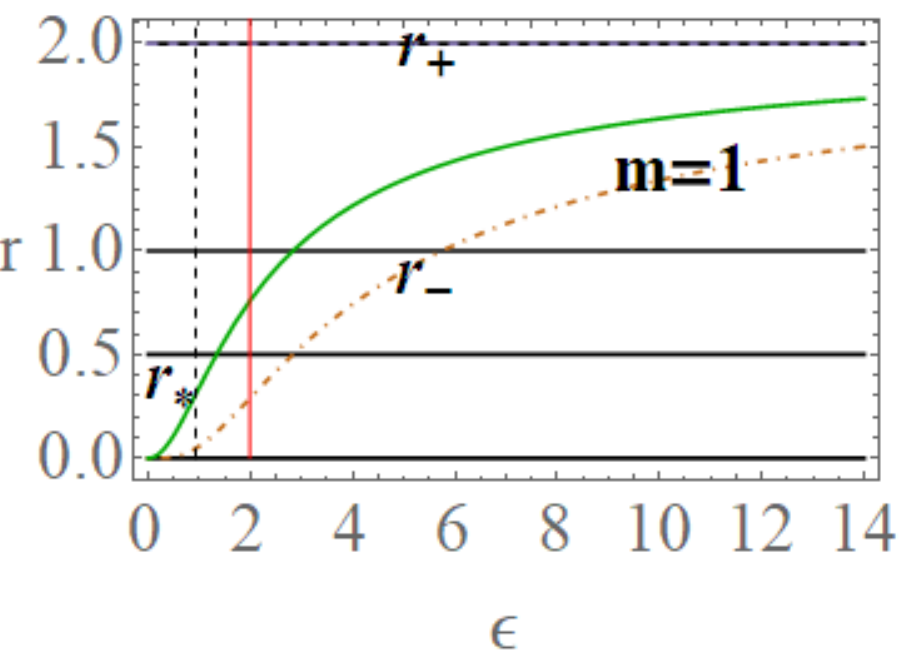}
  \caption{Upper  left panel:    horizons $r_{\pm}$ and radius $r_*$ of Equation~(\ref{Eq:horizons})  function of the $P$ polymeric metric parameter in the terms of the LQG mass parameter $m$ considered as a function of $P$ (thus the notation $m_s$),  it is here $M=1$ ($M$ is the ADM mass in the Schwarzschild limit). Upper  center panel:     $r_{\pm}$ and  $r_*$ function of the $P$  for $m=1$, colors notation follow correspondent left panel.  Upper right  panel: $P$ as a function of $\epsilon$ (a metric polymeric parameter) as in Equation~(\ref{Ep:espi-p}).   Bottom panels: horizons and radius $r_{*}$ for $M=1$  and $m=1$ respectively as functions of $\epsilon$.
 Notes on notation can be  found in Table (\ref{Table:pol-cy-multi}). }\label{Fig:colorPP}
\end{figure}

There are  two horizons $r_+$ and $r_-$ respectively:
\bea
&&\label{Eq:horizons}
 \mbox{{Horizons:}}\quad r_+ = 2m,\quad r_-= 2 m P^2.
\eea
In the following, we consider the two Killing fields  of the geometry $(\xi_t,\xi_{\phi})$.
Here, we adopt two approaches considering{ \textbf{(1)}} $m=1$ (loop BH
mass equal one) or {\textbf{(2)} } ADM mass parametrization: $M=1$.
Clearly, $M=m$ and $r_+=2m=2M$ in the  (Schwarzschild) limit of $P\approx 0$.
When considering the    \textbf{(2)}-asset, we shall use  explicitly $m(M,P)$ to signify the dependence on $P$ and $M$.
Condition $\mathbf{(1})$ implies $P=+\sqrt{M}-1$, condition \textbf{(2)} implies $m=1/(1+P)^2$, see Figures (\ref{Fig:colorPP}).
More generally, we consider $\mathcal{P}=\{m,P,\epsilon,a_o\}$ as the geometry parameters, considering differently related {(The issue of the} independence of the parameters $\mathcal{P}=\{P,m,a_o\}$  is deep and concerns the specific approximated  {BHs} model as well as the {LQG}.   Being our analysis adapted to the context of the applications of the {MBs}, we have adopted a more general approach, using Equation~(\ref{Ep:espi-p}) in fact as particular cases.
The use of $m=1$ implies a re-parametrization in terms of $M$ which here sets the scales in some analysis. However,  within the condition $m=1$, there is  $P\to \sqrt{M}-1$, and $r_+=2$ (in mass units) see Figures (\ref{Fig:colorPP})---see, for example, \cite{Aleshi2012zz}. We consider different approaches and conditions to evaluate the MBs for this geometry.).
For very large $\epsilon$, considering  function $P(\epsilon)$ of  Equation~(\ref{Ep:espi-p}), there is  $r_{\pm}=r_*=M/2$ (and  $2m$) for very small $\epsilon$, $r_+=2M(2m)$ while $r_-=r_*=0$---see Figures (\ref{Fig:colorPP}) and Figures (\ref{Fig:vengplre}).

Note that here  $r$
is only asymptotically the usual radial coordinate since
$H(r)$ is not  $r^2$.
 Figures (\ref{Fig:CausalP}) show the curves  $\sqrt{H(r)}=$constant in the plane $(r/M,a_o)$ and the extreme  curve for the $\sqrt{H(r)}$ as a function of $r$.  The function  of $r$ has an  extreme $r_{a_o}^{min}$, a minimum, notably  equal to the length from the minimal loop area i.e.,  $r_{a_o}^{min}=\sqrt{a_o}$, where $\sqrt{H(r_{a_o}^{ min})}=2r_{a_o}^{min}$.
Concerning the structure of the geometry and singularity nature and relative discussion of the  Penrose diagram, we refer to \cite{Aleshi2012zz,Modesto}: geometry at $r=0$ has no  singularity, but, in the limiting Schwarzschild case, and in fact this is a regular BH solution. The analysis of Penrose diagram shows that there is  another asymptotically flat Schwarzschild region, i.e., there are  two horizons and two pairs of asymptotically flat regions.
\subsubsection{On the MBs, Horizons, and Observers}\label{Sec:litig-colo}
The analysis of the metric bundles and the geometry properties  with  MBs  focus on the properties an observer  could measure  in the region outside the (outer) horizon $r_+$  in the BH spacetime. The observer could extract information (locally) of  the region close to the (inner and outer) horizons $r_{\pm}$---and connecting different geometries of the metric family considering local properties of causal structure with the analysis of photon-like orbits.
In this sense, the horizon's confinement and the \emph{\textbf{horizon's replicas}}.
\medskip

\begin{itemize}
\item\textbf{---Definitions of horizon's replicas and  confinement}

Considering  a generic property $\wp_{\pm}$  of the horizon as distinguished in the extended plane, as the horizon  frequency $\omega$ for  the spinning BH horizons, there is a \emph{replica} of the horizon, in the same spacetime when there is an orbit (radius) $r_x>r_\bullet$  such that
$\wp(r_\bullet)\equiv \wp_\bullet= \wp (r_x)$, where $r_{\bullet}$ is a point of the horizon curve in the extended plane.
From MB definition,
 there are  horizons replicas in different  geometries,  i.e.,  there are a $p\neq p_x$ and a $r_x>r_+$, where $p$ and $p_x$ are  values of   the extended plane parameter $\mathcal{P}$,  corresponding to two different geometries (distinguished with two  horizontal lines of the extended plane)  such that:
$\wp(r_\bullet(p),p)\equiv \wp_\bullet^p=\wp (r_x(p_x),p_x)$. In
 both points, $(r_\bullet,r_x)$, there is equal  light-like orbital  frequency.
Vice versa, the (MBs') \textbf{\emph{horizon confinement}}  is  interpreted as the presence of a  "\emph{local  causal ball}" in the extended plane, which is  a region of the extended plane $\mathcal{P}-r$, where MBs are entirely confined, this means that there are no horizons replicas  in any other region of the extended plane, in any other geometry, although we can be interested in specifying this definition  to confinement  of the $\wp$ property in  the same geometry.   Typically, for the Kerr spacetime, the  causal ball is a  region upper bounded in the extended plane by the a portion of the horizon curve corresponding to the a set of the inner horizon BHs---\cite{renmants,Pugliese:2019rfz,
Pugliese:2019efv}.
The analysis of self-intersections of the bundles curves on the extended plane, in the same geometry (horizon confinement)  or
intersection of bundles curves  in different geometries is therefore an important point of the MBs analysis.

 (It is obvious that, in the spherically symmetric spacetime, the definition of replica is adapted to the frame of the MBs  approximation  to the horizon curve in the extended pane, i.e., $r_{\bullet}\approx r_{\pm}$).

We precise the definition of the   MBs by considering  explicitly the definition for  the Kerr spacetimes; in this discussion, it is easier to consider explicitly the definition for the metric bundles adapted to the  more general axially symmetric case as in \cite{renmants,ergon,observers,Pugliese:2019rfz,
Pugliese:2019efv}).
Therefore, the Kerr  horizons are  {null} surfaces, $\mathcal{S}_0$,
whose {null} generators coincide with the orbits of a
one-parameter group of isometries;  thus, there exists a Killing field $\mathcal{L}$ that is normal to $\mathcal{S}_0$.
MBs satisfy   the condition $\mathcal{L}_{\mathcal{N}}\equiv\mathcal{L}\cdot\mathcal{L}=0$,
where $\mathcal{L}$ is a  Killing field of the geometry   $\mathcal{L}\equiv \partial_t +\omega \partial_{\phi}$.
In  BH spacetimes, this Killing vector defines also the thermodynamic variables and the Killing horizons.
Therefore, \emph{metric bundles} are solutions of the zero-norm condition $\mathcal{L_N}(\omega_\bullet)=0$ ($\omega_+=\omega_{\bullet}$ for the outer horizon $r_+$).
The condition  $\mathcal{L_{N}}=0$ is  related to the definition of  stationary observers,  characterized by a four-velocity of the form
 $
u^\alpha\propto\mathcal{L}^{\alpha}$.
The spacetime  causal structure of the Kerr geometry can be then   studied by  considering also   stationary  observers
 \cite{malament}:  timelike stationary observers have orbital frequencies (from now on simply called frequencies) in the interval
$\omega\in]\omega_-,\omega_+[ $ having
limiting orbital frequencies, which are the photon orbital frequencies  $\omega_{\pm}$,  which, evaluated on the Kerr horizons  $r_{\pm}$,  provide the frequencies $\omega_{\pm}$ of the Killing horizons.
In general, a Killing horizon is a light-like hypersurface (generated by the flow of a Killing vector),
where the norm of a Killing vector is null.
 The event horizons  of a spinning BH  are  therefore  Killing horizons   with respect to  the Killing field
$\mathcal{L}_H\equiv \partial_t +\omega_H \partial_{\phi}$, where  $\omega_H$ is in general angular velocity of the horizons. (The event horizon of a stationary asymptotically flat solution with matter satisfying suitable hyperbolic equations  is a Killing horizon). Conditions on $\omega_H=$constant represent  the BH rigid rotation.
For  static (and spherically symmetric) BH spacetimes, the
event, apparent, and Killing horizons  with respect to the  Killing field   $\xi_t$ coincide.
In the limiting case of the static Schwarzschild spacetime or the Reissner Nordstr\"om spacetime,
 the event horizons are  Killing horizons with {respect} to the  Killing vector $\partial_t$.
\\
\item
\textbf{---MBs and thermodynamics: }
In this work, we also investigate some  BHs thermodynamics properties of the LBHs  in the extended plane through the analysis of MBs.
 The BH Killing horizons of
stationary  solutions
have  constant surface gravity (zeroth BH  law-area theorem):
the norm $\mathcal{L_N}$  of  $\mathcal{L}$ is  constant on the
BH horizon.
Moreover, the BH surface gravity, which is a  conformal invariant of the metric,
may be defined as the  {rate} at which the norm  $\mathcal{L}_{\mathcal{N}}$   of  the Killing vector $\mathcal{L}$  vanishes from
outside ($r>r_+$).
 For a Kerr spacetime,
the surface gravity re-scales with the conformal Killing vector, i.e., it  is not the same on all generators, but,  because of the symmetries,  it is constant along one specific generator.
More precisely: the constant $\kappa: \nabla^\alpha\mathcal{L}=-2\kappa \mathcal{L}^\alpha$,
 evaluated on the \emph{outer} horizon $r_+$, defines the BH surface gravity, i.e.,
$\kappa_+\equiv \kappa(r_+)=$constant on the orbits of $\mathcal{L}$
 (equivalently,  we can write
  $\mathcal{L}^\beta\nabla_\alpha \mathcal{L}_\beta=-\kappa_+ \mathcal{L}_\alpha$ and  $L_{\mathcal{L}}\kappa_+=0$, where $L_{\mathcal{L}}$ is the Lie derivative---therefore defining a non-affine geodesic equation).
The BH  surface area
is non-decreasing (second BH  law); consequently, the impossibility  to achieve
by a physical
process a BH state with zero surface gravity.
More precisely,  non-extremal
BH cannot reach an extremal case in a finite number of steps---third BH law:
  at the extreme case for the Kerr geometry $a=M$, the maximum of the horizon curve in the extended plane,  where  $r_{\pm}=M$, the surface gravity  is zero and, consequently, the   temperature  is  $T_{\mathbf{BH}} = 0$,  but not its    entropy (and therefore the BH area).
(This fact poses constraints also with respect to the stability
 against Hawking radiation)
The mass variation,  the surface gravity, and the horizons frequencies are related by the
 first law of BH thermodynamics, which can be written as
$\delta M = (1/8\pi)\kappa_+ \delta A_{\mathbf{BH}} + \omega_H \delta J$, where there is
the variation of the BH
mass, the horizon area and angular momentum $J$,  for the Kerr (BH),  representing the  ``work term'', $A_{\mathbf{BH}}$ is the BH area .
\end{itemize}
\subsubsection{The Extended Plane}\label{Sec:ext-plane}

\begin{figure}[H]
\centering
 \includegraphics[width=16cm]{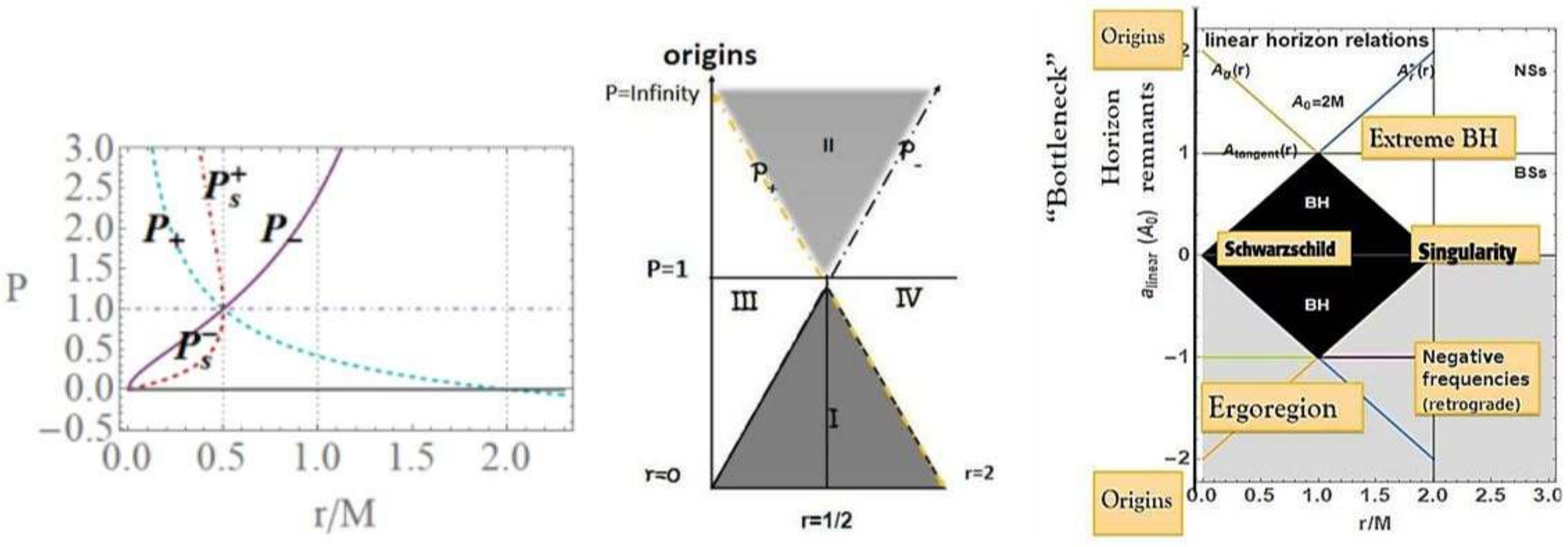}
  \caption{Extended planes in the Kerr geometries and { LQG BHs}. Left panel: $P_{\pm}$ are the horizon curves in the extended plane considered in Equation~(\ref{Eq:vengplre}).   Center panel: extended plane of the LBH geometry in the $P$-parametrization ($P$ is the polymeric parameter). Details are in Section (\ref{Sec:ext-plane}). Right panel: extended plane of the Kerr geometry, details are in \cite{renmants} and Section (\ref{Sec:ext-plane}); here, we point out the analogies with the extended plane structures in the two planes---see also Figures (\ref{Fig:vengplrea}) and (\ref{Fig:CausalP}), and Table~(\ref{Table:pol-cy-multi}).}\label{Fig:vengplre}
\end{figure}

\begin{figure}[H]
\centering
       \includegraphics[width=12cm]{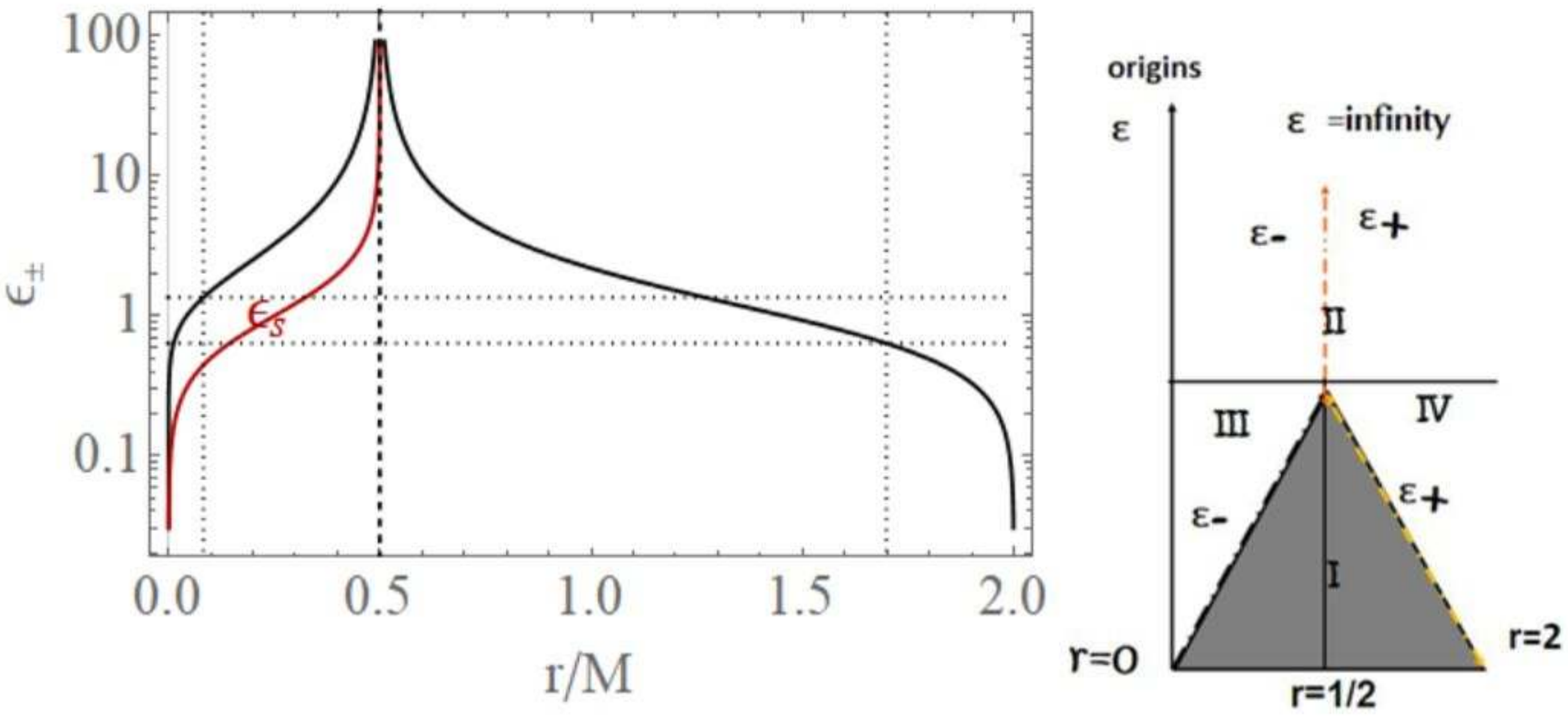}
  \caption{
Extended plane in the $\epsilon-r$ parametrization.
     The horizon's curve $\epsilon_{\pm}$  of Equation~(\ref{Eq:pok-fac}) in the extended plane $\epsilon-r$ is also represented together with the asymptote $r=0.5M$. Saddle points are horizontal  dotted lines.  $P_s^{\pm}$ curves of Equation~(\ref{Eq:superrad}) are also shown ($M$ is the ADM Schwarzschild mass, $\epsilon$, and $P$ are polymeric metric parameters). Details are in  Section (\ref{Sec:ext-plane})---see also  Figures (\ref{Fig:vengplre}) and Table (\ref{Table:pol-cy-multi}).}\label{Fig:vengplrea}
\end{figure}

In Figures (\ref{Fig:vengplre})--(\ref{Fig:vengplrea}), two realizations of the  extended plane of the regular LBH geometry  of Equation~(\ref{Eq:dom-gionr}) are represented. The first step to construct the extended plane $\mathcal{P}-r$ where the MBs are defined, is  to individuate the more convenient   parameter  $\mathcal{P}$; the second step in this construction  is  to represent the horizon curve in the plane. (It is clear that here a plane is briefly intended as a flat, two--dimensional surface, as in  Figures (\ref{Fig:vengplre})).
The main convenient parametrization for the extended plane as emerging from  the analysis of MBs in Section (\ref{Sec:wordsee-h}) is $P$-parametrization or equivalently  $\epsilon$-parametrization, obtained by using Equation~(\ref{Ep:espi-p}) and here showed in Figures (\ref{Fig:vengplre}).
(More generally, here we note that, as discussed in \cite{renmants,Pugliese:2019rfz,
Pugliese:2019efv}, the first step to explore the metric bundles is to distinguish a leading parameter for the metric bundles, which in the Kerr spacetime was the dimensionless spin $a/M$ of the singularity (a BH or NS), as clearly connected to the characteristic frequency. This is obviously determined by the Killing  horizons' definitions. This choice is, however, not  immediate. In  the case of Kerr-Newman {(KN)} spacetimes for example, as discussed in \cite{renmants}, one could choose spin $a/M$ or  the dimensionless electric change $Q/M$ or, for example, the "total charge" $\Qa_{T}\equiv \sqrt{(Q/M)^2+(a/M)^2}$).


\begin{figure}[H]
\centering
      \includegraphics[width=5cm]{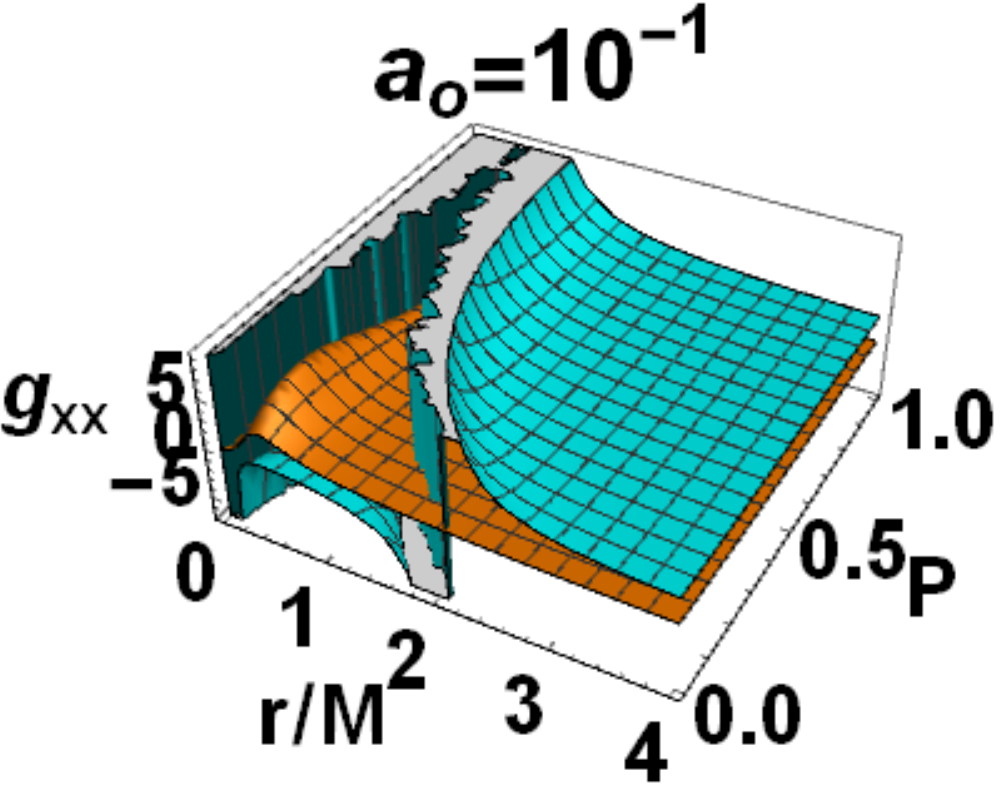}
          \includegraphics[width=5cm]{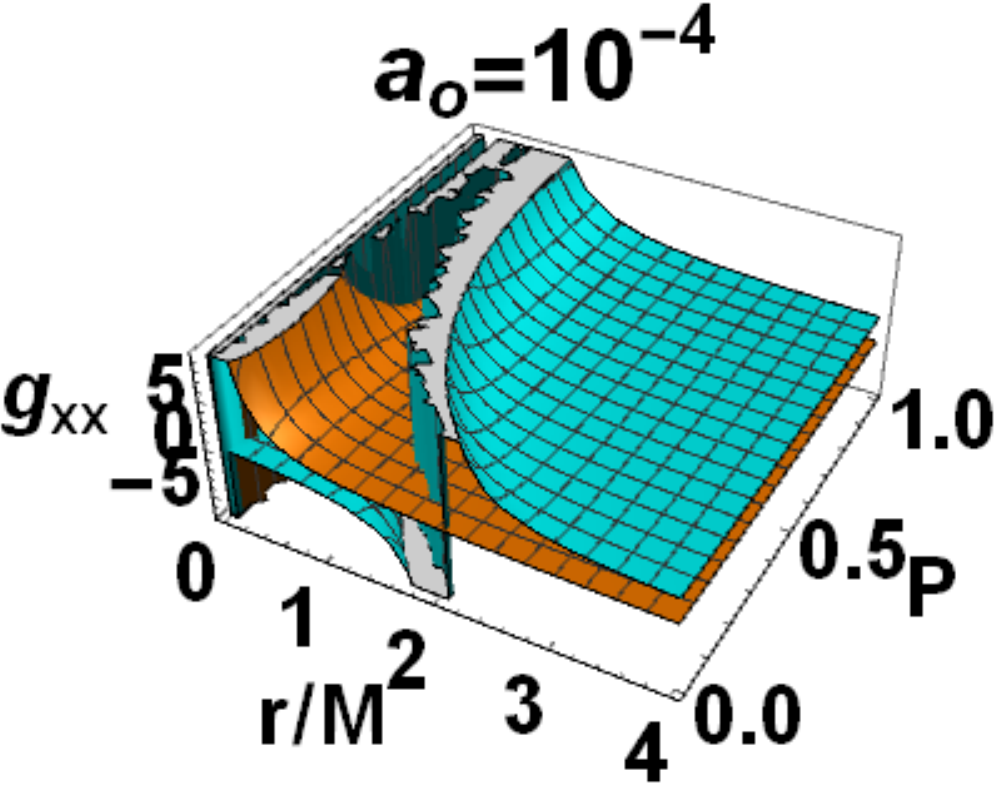}
             \includegraphics[width=5cm]{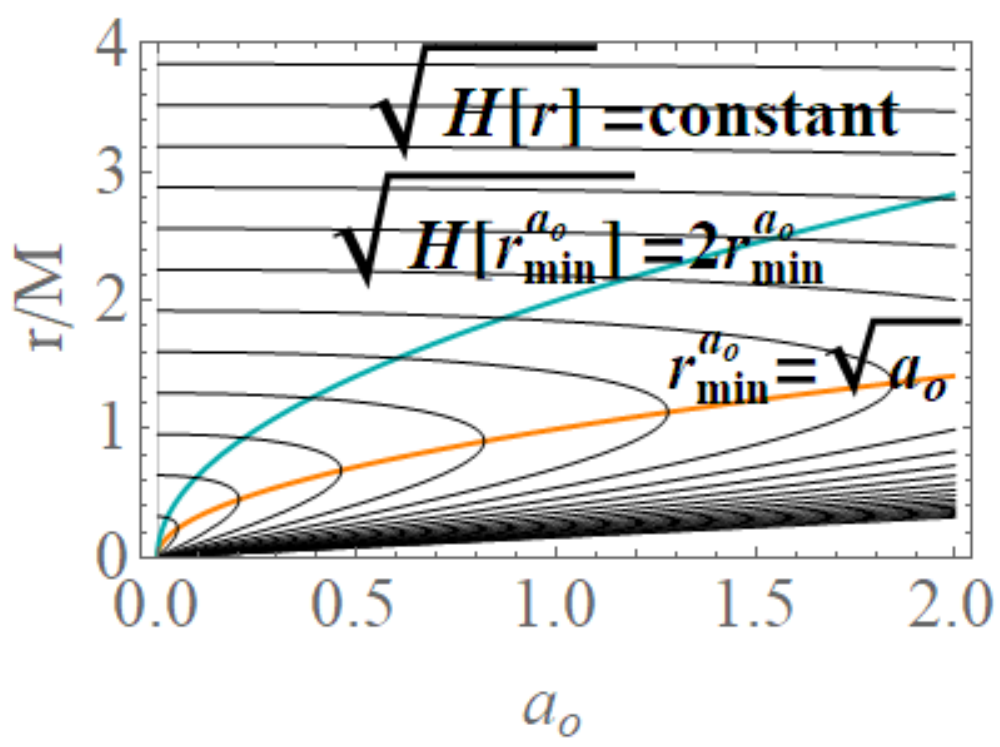}
          \\
  \includegraphics[width=5cm]{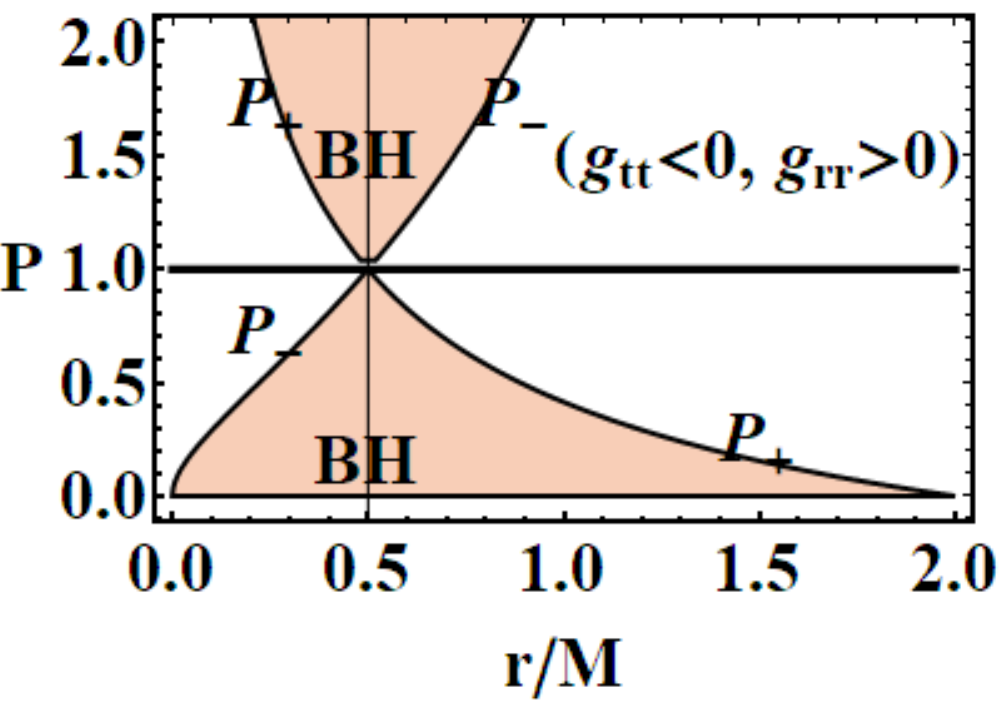}
      \includegraphics[width=5cm]{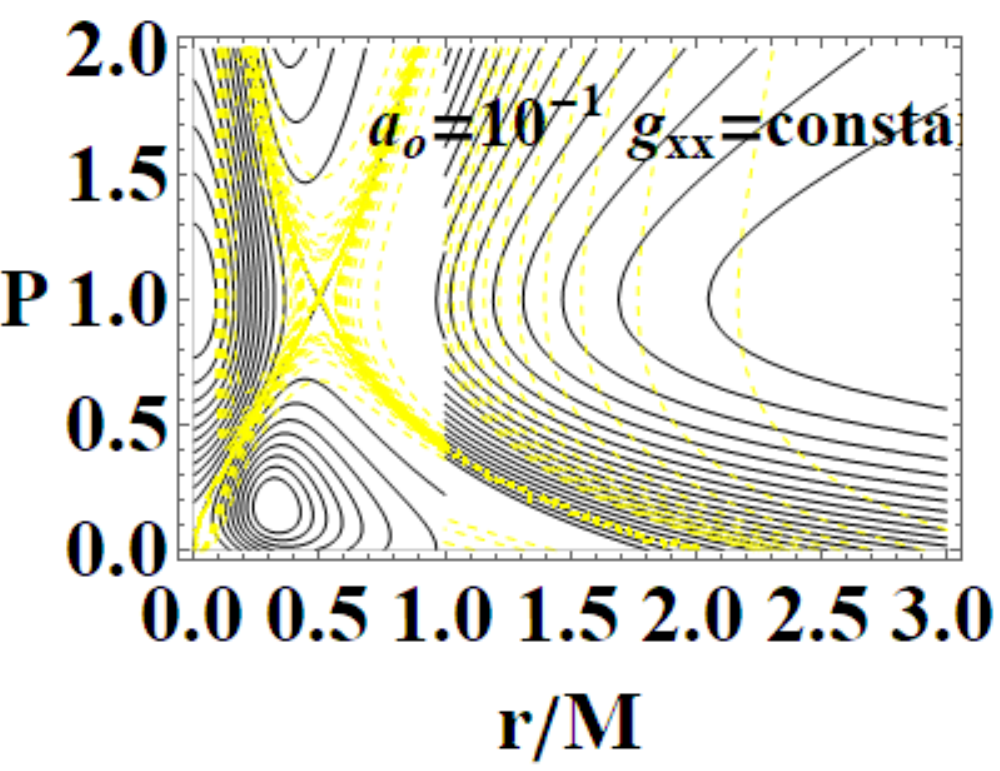}
    \includegraphics[width=5cm]{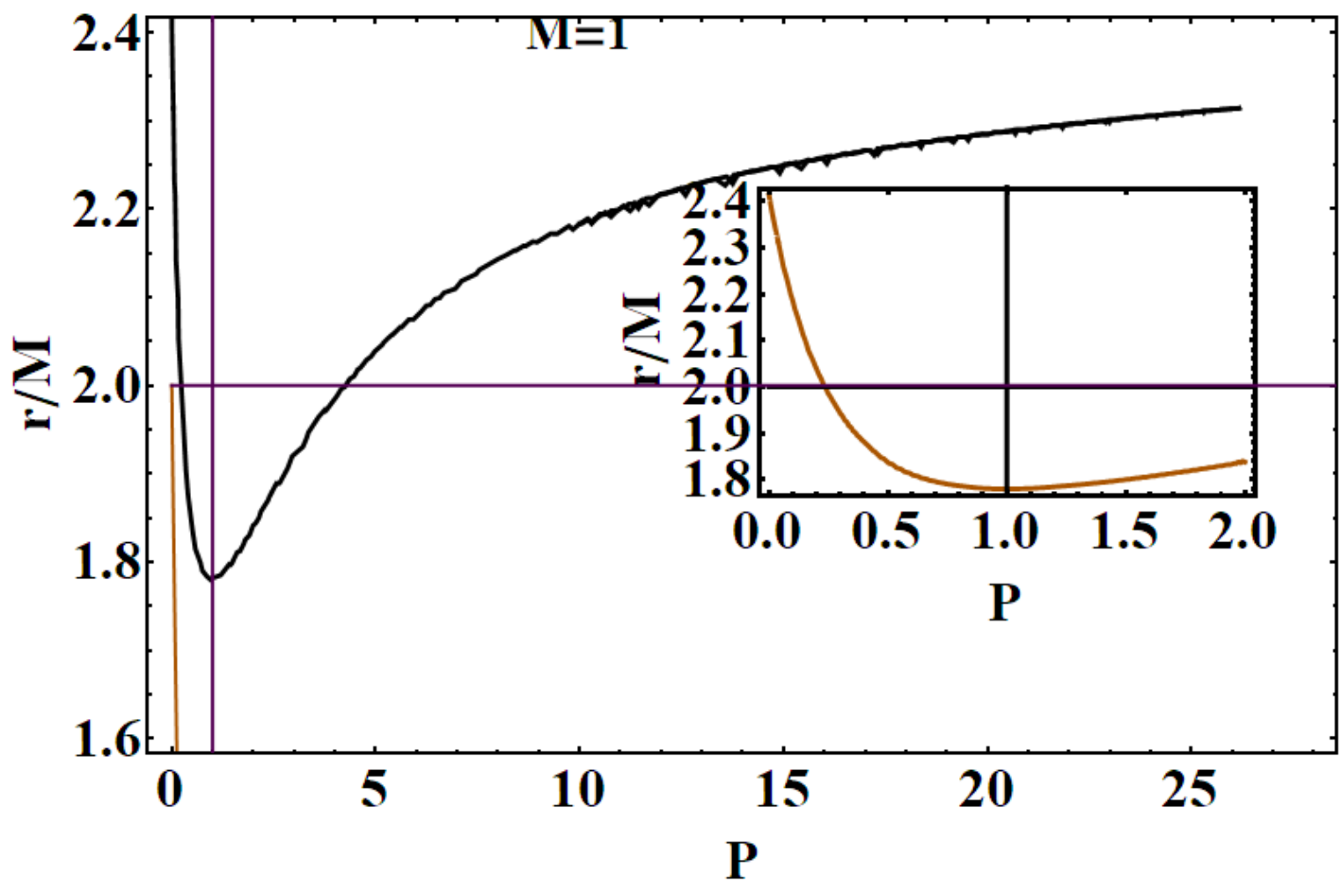}
  \caption{Analysis of metric (\ref{Eq:dom-gionr}) and metric bundles.  3D Plots show $g_{xx}$ for $x\in\{t,r\}$ as a function of $r/M$ (ADM Schwarzschild  mass $M=1$)  and  polymeric metric parameter $P$, for different values of $a_o$ (length from the minimal LQG area). Orange surface is $g_{tt}$. The Upper Right panel shows the  metric component  $\sqrt{H(r)}=$constant in the plane $(r/M,a_o)$, $H(r)=g_{\phi\phi}/\sigma$ in Equation~(\ref{Eq:horizons}), $\sigma\equiv\sin^2\theta$. Curve $r_{a_o}^{min}$ is the extreme  curve for the $\sqrt{H(r)}$ as a function of $r$. Note that the extreme, a minimum, is equal to the length from the minimal area $r_{a_o}^{min}=\sqrt{a_o}$, and the function $\sqrt{H(r_{a_o}^{ min})}=2r_{a_o}^{min}$. Bottom Left  panel:
extended plane $P-r$.
Region $g_{tt}>0$ and $g_{rr}<0$ (pink-BH region, we do not consider the region $P>1$) and outer region is  $g_{tt}<0$ and $g_{rr}>0$.
Horizons $P_{\pm}$ is shown. Regions do not depend on a $a_o$ parameter.
Line $r=1/2$ is also shown; this is an $M=1$ approach.  Bottom center panel: curves  $g_{xx}=$constant for $x\in\{t,r\}$.
Bottom right panel:
The curve  $r(P)$ such that $g_{tt}=c_t=$constant and $g_{rr}=c_r=$constant; in other words, the families (in terms of $P$ parameters) of metric solutions having equal $g_{tt}$ and $g_{rr}$ in the \emph{same} point $r$. The inside plot is a zoom in  the region $P\in[0,1]$.
 See also Table (\ref{Table:pol-cy-multi}) for further details  on notation.}\label{Fig:CausalP}
\end{figure}

A second  issue in the plane construction is the choice of  the axis $r$, which  is here the radial coordinate $r$,  especially relevant in the case of LBH of Equation~(\ref{Eq:dom-gionr}).
{As we discussed in Section (\ref{Sec:metric}),
concerning the interpretation of the radial coordinate  $r$ for metric  Equation~(\ref{Eq:dom-gionr}), the radius $\tilde{r}\equiv\sqrt{H[r]}$ only asymptotically approaches the standard radial coordinate, i.e., $r$
is only asymptotically the usual radial coordinate {(
For this reason}, one could think of selecting $\tilde{r}$ instead of the asymptotic standard  Schwarzschild radial coordinate $r$ which is the  circumferential
radial coordinate of the Schwarzschild geometry (from the integration around a full circle at radius $r$, we
get a circumference of $2\pi r$, i.e.,
the surfaces at fixed $t$ and $r$ appear in GR as round spheres where $ d s^2=d \Omega^2$ is the  standard Riemannian metric on the (unit radius) two sphere or
$d \Omega^2$  is an interval of spherical solid angle in  standard spherical coordinates $(\theta,\phi)$---
for a fixed $r$, the surface area the circle $4 \pi r^2$, and the associated  sphere with
Gaussian curvature$ 1/r^2$).
The line element on an (equatorial) circle is, in the GR  geometry,
$ds^2=r^2d\phi^2$, vice versa  in our case, metric (\ref{Eq:dom-gionr}), we shall have  $ds^2=\tilde{r}^2d\phi^2$.
In this regard, we also note here that,
   as in general relativity, we could adopt the Eddington--Finkelstein coordinates adapted to radial null geodesics. However,
we  can explicitly use MBs, based on null circular orbits, to rewrite the line element.) with respect to the reference GR solution,  as
$H(r)$ is not just $r^2$. In this respect,
in Figures (\ref{Fig:CausalP}), we represent the $\sqrt{H(r)}$ asymptotical behavior for large  distance from $r=0$  (the line  defining the bundles origins) and particularly the value $\sqrt{H(r)}=r$. We have represented both the asymptotic behavior and the curves   $\sqrt{H(r)}=$constant (versus $r=$constant) in the plane $(r/M,a_o)$ and the extreme  points curve for the $\sqrt{H(r)}$ as a function of $r$. In fact,  $\sqrt{H(r)}\equiv \tilde{r}$  remarkably is not monotone in $r$ but has a minimum,  $r_{a_o}^{min}=\sqrt{a_o}$, where $\sqrt{H(r_{a_o}^{ min})}=2r_{a_o}^{min}$. Furthermore, in Figures (\ref{Fig:CausalP}), we can see the situation for  different  geometries (line $r=$constant) and at fixed geometry  ($a_o=$constant), where there are two  orbits  $r$ with equal values of $\tilde{r}$.
 }
The main point of the construction of the extended plane consists in the fact that  it allows for  considering the properties of a parameterized family of geometries  in a "global" prospective, by  considering  different features as seen at variation of  the metric family  parameter $\mathcal{P}$. This  turns out, therefore, to also be relevant  for the exploration of the transformations leading from one solution to another of the family, as, for example,  after a   (dimensionless) spin shift of Kerr BH, a shift of the dimensionless electric charge in the RN metric, or eventually a variation of loop parameters $\mathcal{P}$ for the  LBHs.  Thus, in this   "global" frame, we also search for a replica of the horizons of one geometry in different solutions, investigating the presence of  horizons' characteristics  of different spacetimes of the family (or MBs).
In this way, we also relate different geometries through their local causal properties and also {BHs'} thermodynamical characteristics. These aspects are here explored in dependence on the model parameters $\mathcal{P}$  and evaluated on the metric bundles in Section (\ref{Sec:termo}).

Metric (\ref{Eq:dom-gionr}) shares several similarities with the Reissner--Nordstr\"om geometry, considered here in Section (\ref{Sec:RN}). To discuss some properties of the extended plane and the MBs, it is convenient, however, to refer first to the representation of the  extended plane  of the axially symmetric, stationary, vacuum Kerr solution.   The Kerr geometry is a well-known, exact, asymptotically flat  solution of Einstein equations---\cite{Kerr,Kerr-Newman,ergon,renmants,observers} . In Figures (\ref{Fig:vengplre}), the panel on the Kerr extended plane shows
the negative region corresponding to counter-rotating orbits i.e.,  to photon orbital frequency equal to  the  horizon frequencies in magnitude. In the spherically symmetric cases, we can restrict our analysis,  without loss of generality, to the  upper part of the plane,  related to the  positive frequencies.
Line $\la\equiv a\sqrt{\sigma}=0$, corresponding to line $\mathcal{P}=0$ of the  plane $\mathcal{P}-r$, is the Schwarzschild limiting  case,  where $a$ is the dimensionless spin, $\sigma\equiv \sin^2\theta$ in Boyer--Lindquist coordinates.
The origin (\emph{line of bundles origins}) is here $\la_0\equiv a_0 \sqrt{\sigma}$, corresponding to $\mathcal{P}$- line  ($r=0$) of the $\mathcal{P}-r$ plane. The Kerr geometry ergoregion (in the extended plane) is the strip  $r<2M$ (the ergoregion of the Kerr geometry is bounded by and outer ergo-surface $r_{\epsilon}^+\geq r_+$,  where $r_+(a)<2M$ is the outer horizon and the inner ergo-surface $r_{\epsilon}^-(0)\leq r_-$, $r_-$ is the inner horizon, where only on the equatorial plane is there $r_\epsilon^+=2M$ independently from the spin $a$).
Extreme  Kerr  BH is for $\la=1$, in  panel regions for Kerr  BHs or Kerr naked singularities  ({NSs}) are also reported. For the Kerr MBs, it is relevant to consider the polar angle $\theta$, in $\sigma\equiv \sin^2\theta$; in the spherical symmetric case considered here, we can consider $\sigma=1$.
The extended plane of the Kerr geometry is constructed considering different functions (including the tangent curves to the horizon),
$\{A_x\}_x$,  which are  given in \cite{renmants} (they correspond also to the linearized horizons relations  $r_{+}(r_-)$ in an equivalent extended plane).
The extended plane of Kerr geometries and LBHs  in Figures (\ref{Fig:vengplre})  show clear analogies.
 We consider an  extended plane realization in Figure~(\ref{Fig:colorPP}) for the LBHs. We can  consider the vertical and horizontal lines and the horizons curves  $P_{\pm}$ in Figures (\ref{Fig:vengplre}) as a function of $r$ (for $M=1$).
In the $P-r$  plane ($M=1$), the horizons are  determined by the polymeric function  as
\bea\label{Eq:vengplre}
&&
P_{+}\equiv \frac{\sqrt{2}}{\sqrt{r}}-1, \quad P_-=\frac{1}{P_+},\quad P_+P_-=1,
\\ \nonumber &&(P_{\pm}=1 \quad r=\frac{1}{2}),\quad\lim_{r\rightarrow0} P_{\pm}=\begin{pmatrix}
                                                           +\infty \\
                                                           0
 \end{pmatrix},
  \quad
 \lim_{r\rightarrow2} P_{\pm}=\begin{pmatrix}
  0 \\
   +\infty
    \end{pmatrix},
\eea
where we adopted a shortened notation for the limits of the horizons $P_{\pm}$ of Equation~(\ref{Eq:vengplre}).
In the LBHs $P$-parametrization, we note the
intersection of $P_-$  curve as  "inner horizon" and  $P_+$ as "outer horizon" and the values  $r=1/2$, $r=2$,  and the limiting  $r=0$ correspondent to the singularity in the Schwarzschild  limit.
A schematic representation of the extended plane  is  rendered in the right panel, enlightening  the BH regions
(where $P_+=P^{-1}_-)$.
The  LQG  extended plane shows  limiting  points
$P_+=0$ for  $r=0$, and
$ P_+$ for  $r=1/2$ (note the analogy with  the  maximum  of the horizon curve  of the Kerr geometries for $a=M$   correspondent to the extreme  Kerr BH).
The limiting values $P_{\pm}=1/2$  are clear; this is an extreme point where $r_+=r_-$.
To complete the analysis of  plane in terms of the horizons $r_{\pm}$,  we can consider the  curves $P_s^{\pm}$ of the extended plane:
\bea\label{Eq:superrad}
P_s^{\pm}\equiv\frac{1-r\pm\sqrt{1-2 r}}{r}:\quad r=r_*^{\pm},
\eea
where $P_s^{+} P_{s}^{-}=1$.
 Alternately, we consider the horizons in the $\epsilon-r$  plane, making explicit the dependence of $P$ in terms of $\epsilon$-loop parameter having the horizon:
\bea &&\label{Eq:pok-fac}
\epsilon_{\pm}\equiv\frac{\sqrt{-4 r^2+6 r+2 \sqrt{2} \sqrt{r}}}{\sqrt{(1-2 r)^2}},\quad \lim_{r\rightarrow\wp} \epsilon_{\pm}=0,\quad \wp\in \{0,2\},\\
                                                         && \lim_{r\rightarrow1/2} \epsilon_{\pm}=+\infty
,\quad
\epsilon_s\equiv\frac{\sqrt{r}}{\sqrt{\frac{1}{2}-r}}: \quad P_s^{\pm}(\epsilon)=P({\epsilon}).
\eea
In $\epsilon_s$ of Equation~(\ref{Eq:pok-fac}), we consider   Equation~(\ref{Eq:superrad}).
The choice of the $\epsilon$--representation has several advantages, evident  from the horizons representations  in the
$\epsilon-r$ extended plane.
   The two limiting points $r =
 0 $ and $r =
  2 $ also are connected to  the value  $\epsilon=0$,  and the limit $r =
   1/2$ is a vertical asymptote for  $\epsilon\to+\infty$.
   On the other hand, there is $\partial_ {\epsilon}^2\
r_* (\epsilon)= 0$ on  $\epsilon_ {\pm} =  {1}/{\sqrt {3}}$ for $r_* = {1}/{8}$
---Figures (\ref{Fig:vengplre}).
It is clear from Figure \il (\ref{Fig:vengplre})  $\epsilon_s$  is bounded in the range $r\in[0,1/2]$.

We stress  that, in the metric bundles analysis  and the geometry properties distinguished with MBs, we are  concerned with  the properties an observer  could measure  in the region outside the outer horizon $r_+$; in this context, therefore, we  consider   observers at infinity, adopting  an  adapted frame. This obviously compels a reinterpretation of the metric bundles in the region close to the horizons $r_{\pm}$---see Figures (\ref{Fig:CausalP}).  In this framework, the horizons confinement in the MBs sense may be  interpreted as the presence of a  "local  causal ball" in the extended plane, which is a region where  MBs are entirely confined in---i.e., no horizons' replicas can be found in the other regions of the extended plane.

\begin{figure*}
\centering
            \includegraphics[width=5cm]{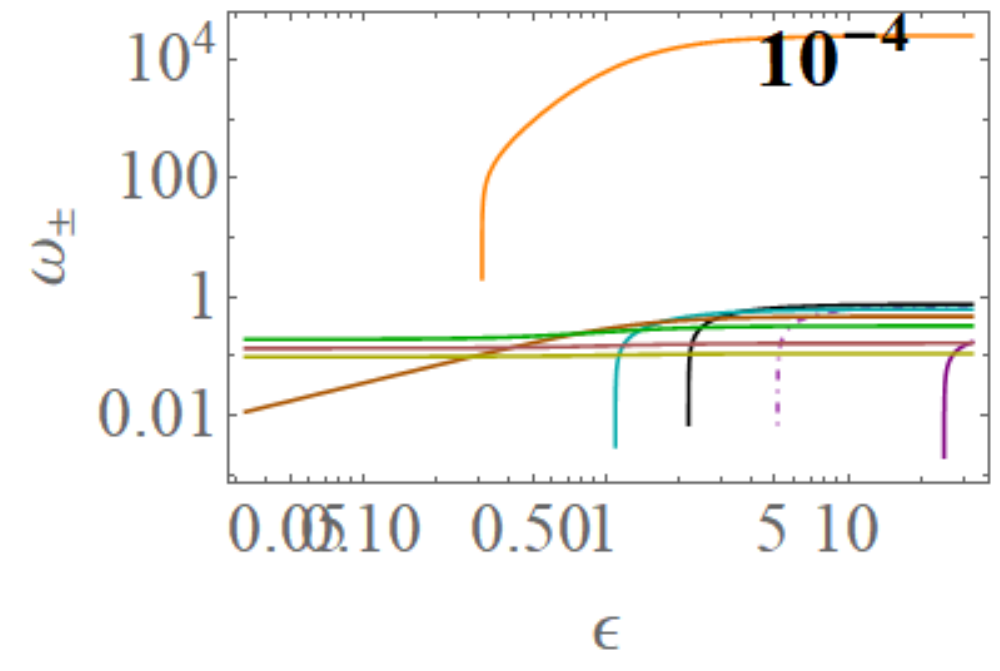}
              \includegraphics[width=5cm]{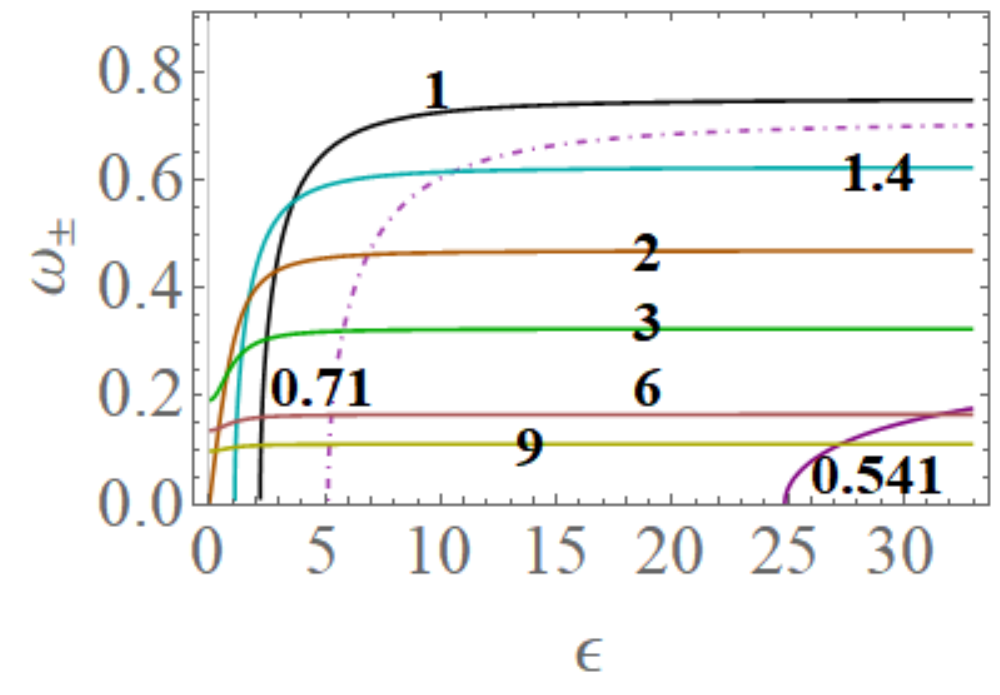}
            \includegraphics[width=4cm]{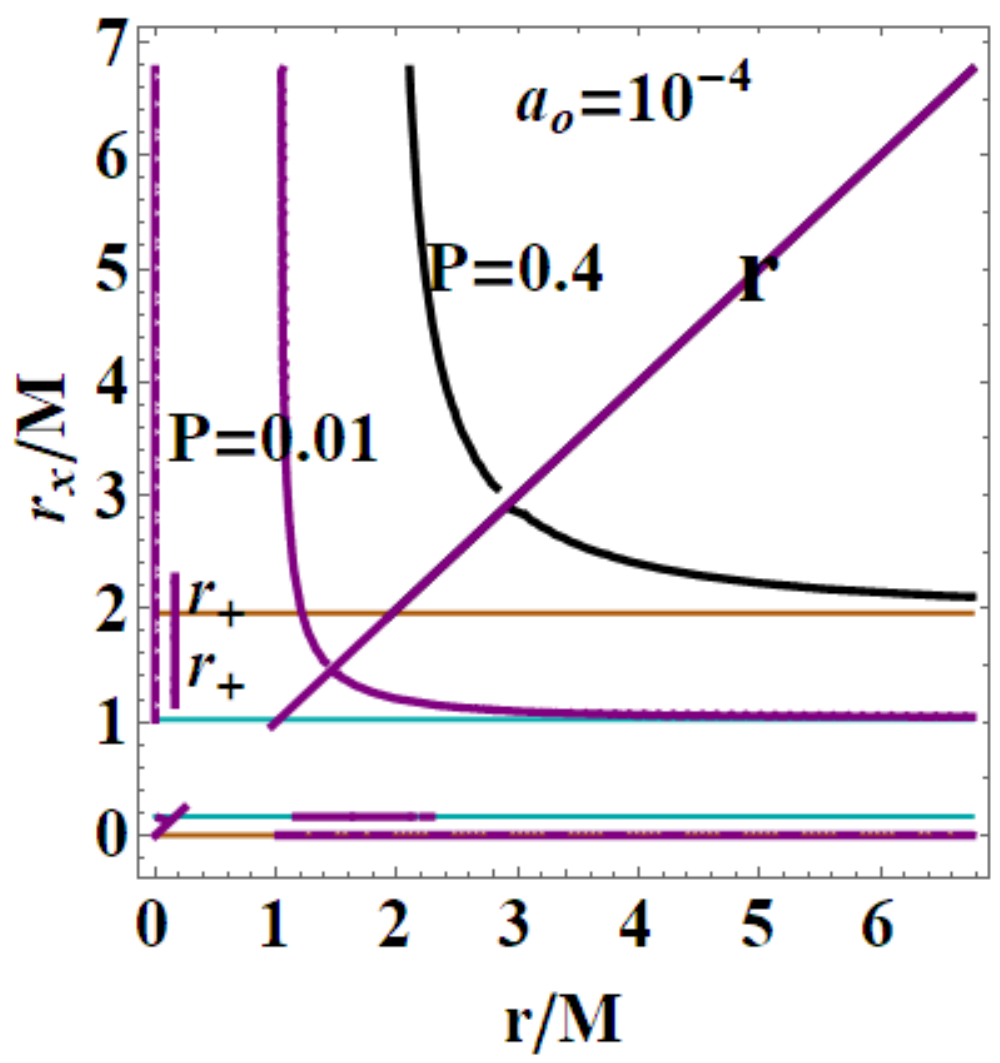}
  \caption{Left and center panels show the vertical lines of the extended plane; in other words, the MB intersections with the curves $r=$constant for different values of $r$ signed in pictures, exploring  different regions of the  $\omega_{\pm}$ values of limiting photon orbital frequency.
 Further notes on notation are in Table (\ref{Table:pol-cy-multi}).
The right panel shows the solutions of the problem $\omega(r_x)=\omega_{\pm}(r)$ ("horizons"' replicas in this spherically symmetric geometry); in other words, the horizontal lines in the extended plane for different  polymeric metric parameters $P$ and for a selected LQG area parameter  $a_o$ ($M$ is the ADM mass and $r_{\pm}$ are the BH horizons). }\label{Fig:govmentr}
\end{figure*}

{Figures} (\ref{Fig:govmentr}) represent location  of orbits   $r<r_x$ at equal frequency $\omega$ (therefore belonging to the same bundle) in two cases. In the first case, we  consider the  same geometry, with $P=P_x$ corresponding to a horizontal line of the extended plane (implying clearly  certain conditions on the  MBs curvature in the $\mathcal{P}-r$ plane).
 Clearly, we look particularly to the couple of radii  $r$ and  $r_x$  such that the two radii are in an inner region of the extended plane, upper bounded of the  horizon $P_-$, and the second outside the region bounded by the horizon curve $P_+$.

Detecting self-intersections of the bundles curves on the extended plane, in the same geometry (horizon confinement)  or
intersection of bundles curves  in different geometries, is a crucial point in the MBs analysis, related to the issue of confinement and causal balls.  There are no MB curves' self-intersections, on equal  $P$ and $r$ (a part some special cases such as  extreme  Kerr and RN BH spacetime); in other words, there are no MB knots.
  Figures (\ref{Fig:CausalP}) show on the other hand an analysis that is  particularly interesting  for the metric bundle approach  where  the curve   $r(P)$, a solution of the problem  $g_{tt}=c_t=$constant \emph{and} $g_{rr}=c_r=$constant, is showed. This curve  collects  the  metric solutions  (in terms of $P$ parameter) having equal $g_{tt}$ and $g_{rr}$, on the same radius $r$; clearly, we are interested particularly to consider the values of $P$ close to the maximum $P=1$ or  for  $P<1$.
\begin{figure*}
\centering
  \includegraphics[width=7cm]{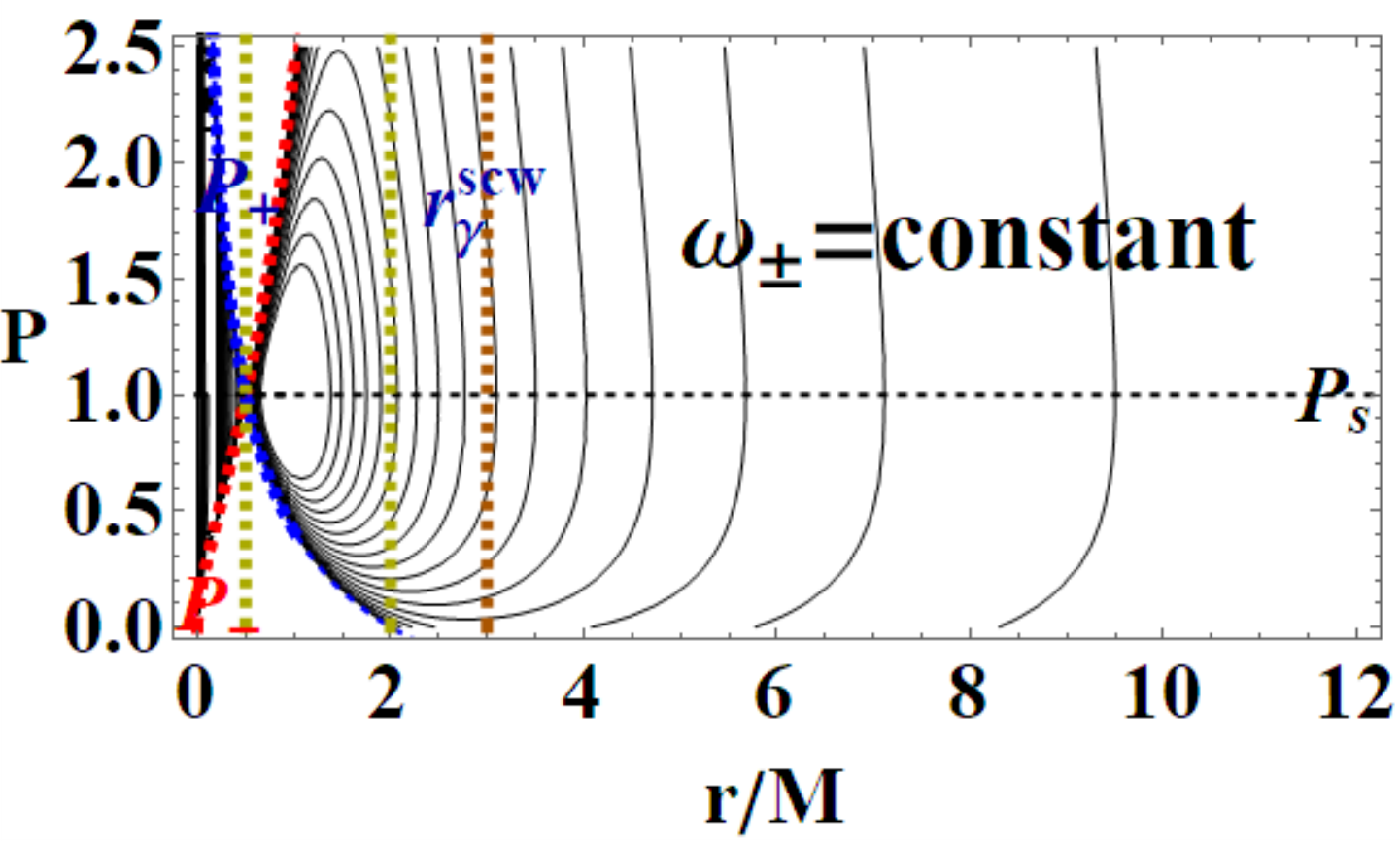}
            \includegraphics[width=7cm]{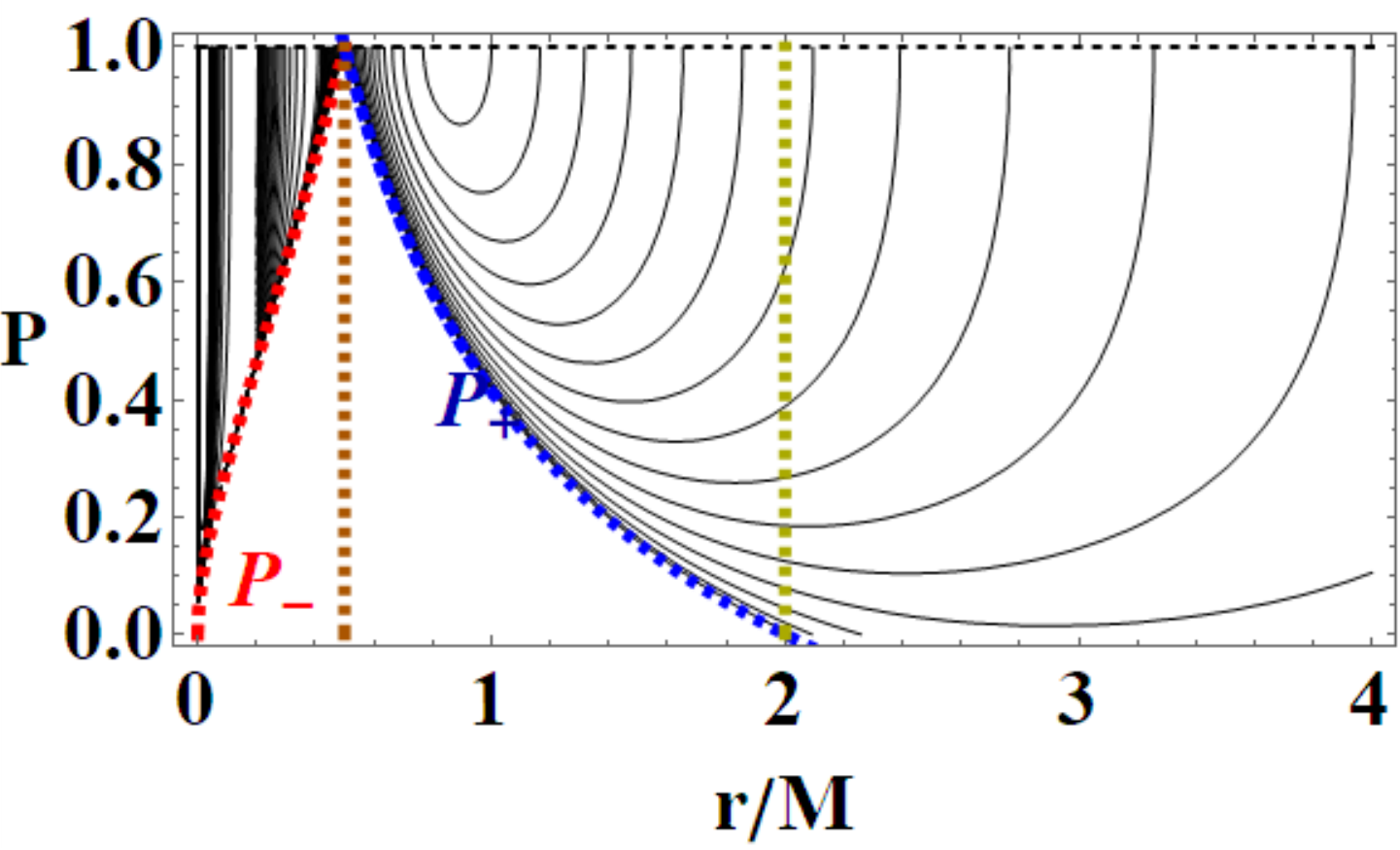}\\
                  \includegraphics[width=7cm]{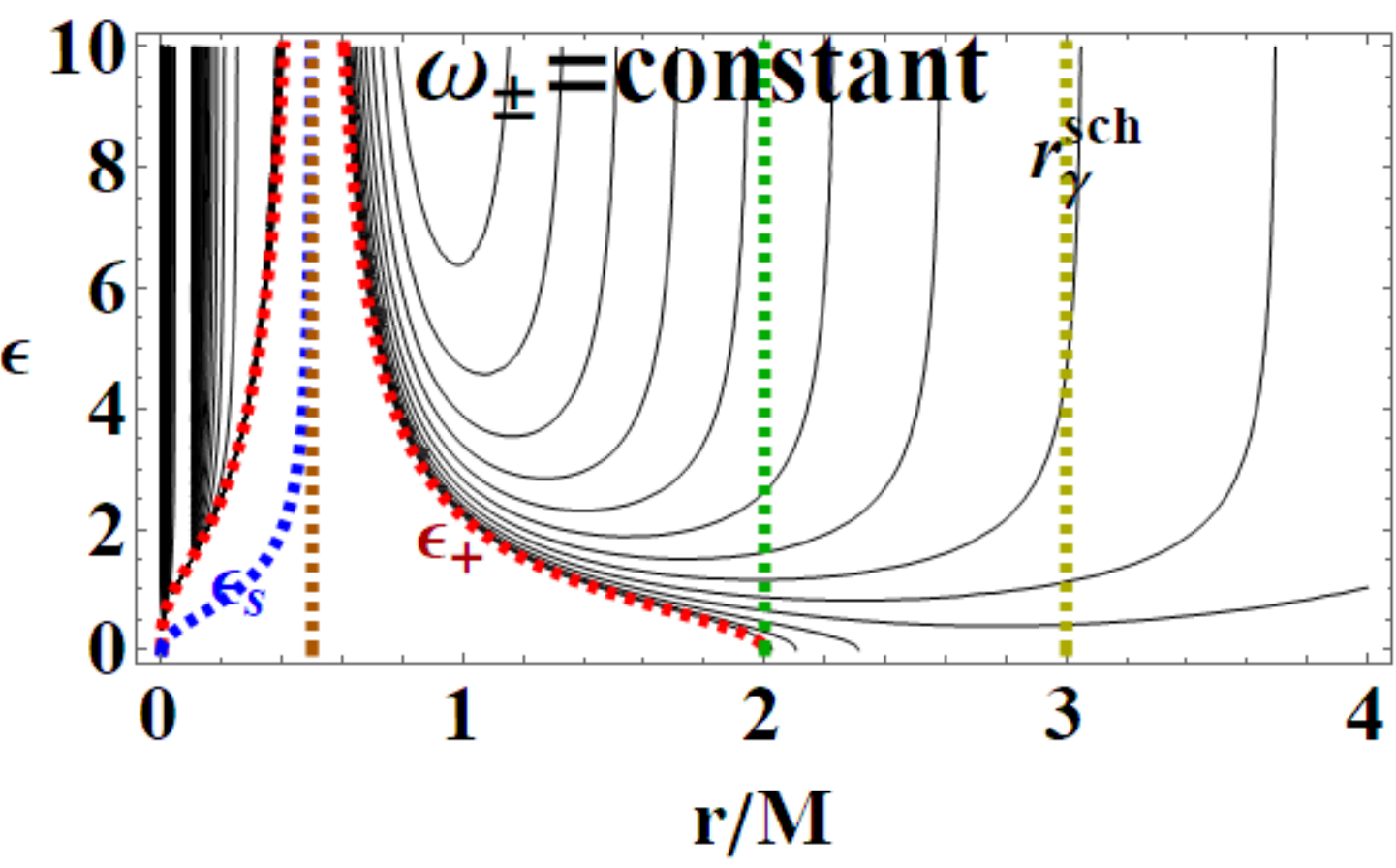}
                             \includegraphics[width=7cm]{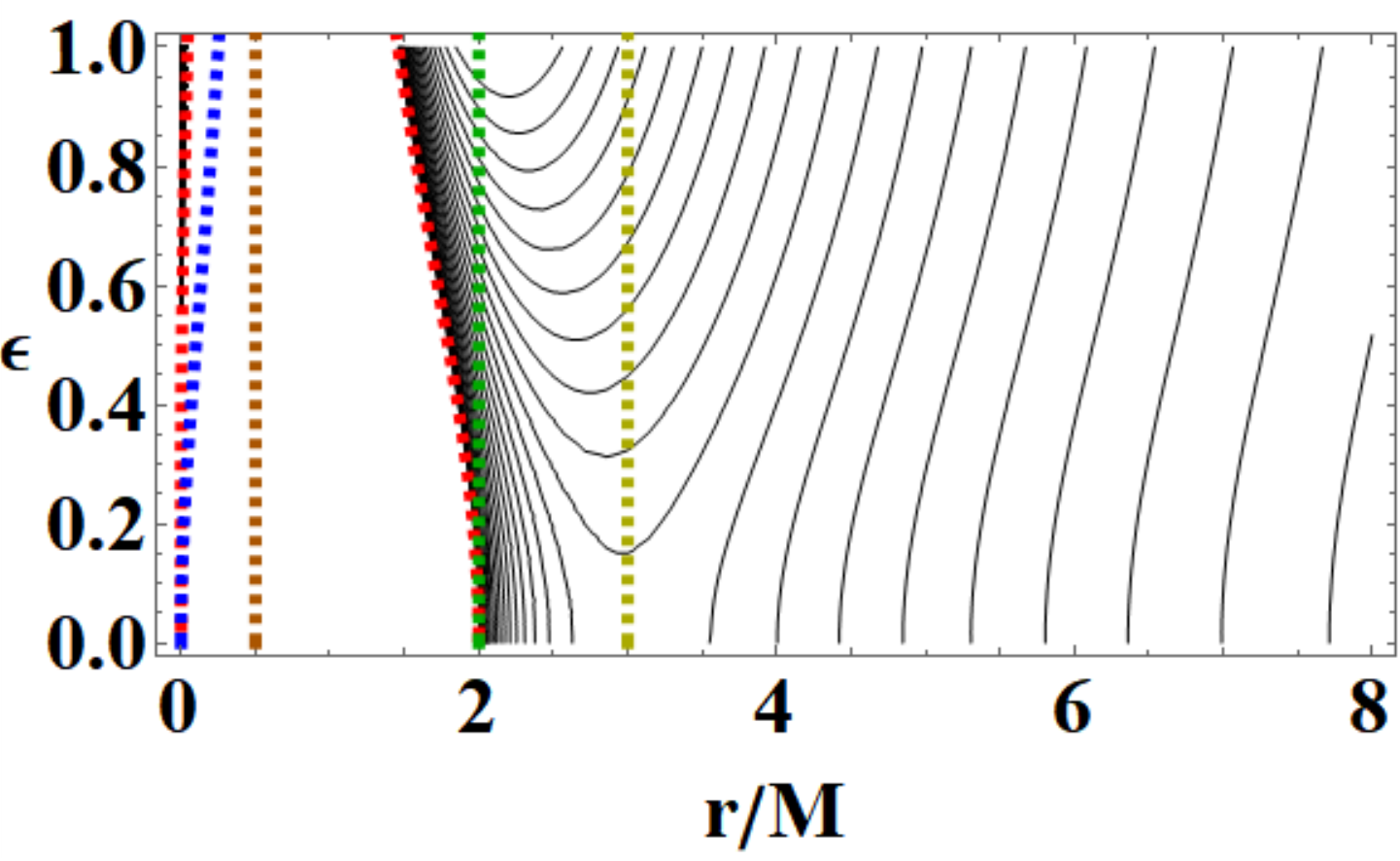}
  \caption{{Metric bundles}  (MBs) in the extended plane $P-r$ (upper panels) and $\epsilon-r$ (bottom panels) for ADM mass  $ M=1$. ($P$ and $\epsilon$ are polymeric parameters).
MB curves are the bundles at equal limiting photon orbital frequency $\omega_{\pm}$.
Radius $r=0.5$ is  relevant for the analysis of the horizons
and $r=2$  is the horizon in the Schwarzschild limit, $r=3$  is the photon circular orbit  in the Schwarzschild limit (a geodesic in this spacetime).
Curves $P_{\pm}$ of the horizons,
$P_s=P_+ P_-=1$ is also shown.
 The Schwarzschild limit is for $ P_-=0$.
We note the presence of curves at $r<2$ for very small $P$; the role of $r=3M$ is the photon orbit in the Schwarzschild limit ($M$ is the Schwarzschild ADM mass).
In the bottom panels, MBs are also shown in the plane  $\epsilon-r$,
red curves are the horizons  $\epsilon_{\pm}$, curve $\epsilon_s: r_*=r$ is also shown.
Further notes on notation are in  Table (\ref{Table:pol-cy-multi}).}\label{Fig:govmentra}
\end{figure*}
Figures (\ref{Fig:govmentr}) and Figures (\ref{Fig:govmentra})   show the
 metric bundles and  vertical and horizontal lines   in the  $P-r$ and $\epsilon-r$   extended plane. The approximations of the bundles to the horizons curves and the  study of horizons' replicas for  different values of the parameter are clear.  In Section (\ref{Sec:wordsee-h}), we also consider the explicit expression  of the metric bundles, as solutions $\mathcal{L}_{\mathcal{N}}=0$, adopting different parameterizations .
\subsection{Comparison with the Reissner--Norstr\"om Geometry}\label{Sec:RN}
 The Reissner--Norstr\"om  (RN) metric  is a well-known  spherically symmetric (and static) electro-vacuum solution of Einstein equations, with inner $r_-$ and outer $r_+$  Killing horizons (with Killing vector $\xi_t$):
\bea&&\label{Eq:q23d}
r_-\equiv M-\sqrt{M^2-Q^2};
                                                         \quad r_+\equiv M+\sqrt{M^2-Q^2};
\eea
(here, we adopt usual spherical symmetric coordinates $(t,r,\theta,\phi)$, where also  $\sigma\equiv \sin^2\theta$).
There is a naked  singularity  for $Q>M$, the extreme {RN-BH} geometry, where $r_{\pm}=M$ occurs in the limit $Q=M$. We construct the metric bundles $Q_{\omega}$, introducing the Killing field $\laa$ and the zero-quantity $\laa_{\mathcal{N}}$ as follows:
\bea\label{Eq:be-en}
&&
\laa\equiv \xi_t+\omega\xi_{\phi};\quad  \laa_{\mathcal{N}}\equiv\textbf{g}(\laa,\laa)=g_{tt}+g_{\phi\phi}\omega^2=\sqrt{r} \sqrt{r^3 \sigma  \omega ^2-r+2}=0,\\
                                                         &&  \omega_{\pm}=\pm\frac{-g_{tt}}{g_{\phi\phi}}={\pm} \frac{\sqrt{Q^2+(r-2) r}}{r^2},\quad
Q_{\omega}\equiv\sqrt{r} \sqrt{r^3 \sigma  \omega ^2-r+2},\\&&\label{Eq:be-en0}
Q_{\omega}=0,\quad \mbox{for}\quad \omega=\omega_{Sch}\equiv\frac{\sqrt{r-2}}{r^{3/2}}, \quad or \quad r=0
\eea
$\omega_{Sch}$ is the  frequency  $\omega_{\pm}$ in the Schwarzschild geometry. In the RN geometry, to make it easier to read, we use geometric units where $M=1$  in many quantities. Here, and in the following, since the metric is spherically symmetric,  we can use, where more convenient,  $\sigma=1$,  i.e., we  fix an arbitrary equatorial plane without loss of generality. On  the other hand,  the re-parametrization $\sigma  \omega ^2\rightarrow\omega^2$ is an important  definition adopted also in the axially symmetric case of the Kerr MBs.

\begin{figure}
  \centering
  \includegraphics[width=5cm]{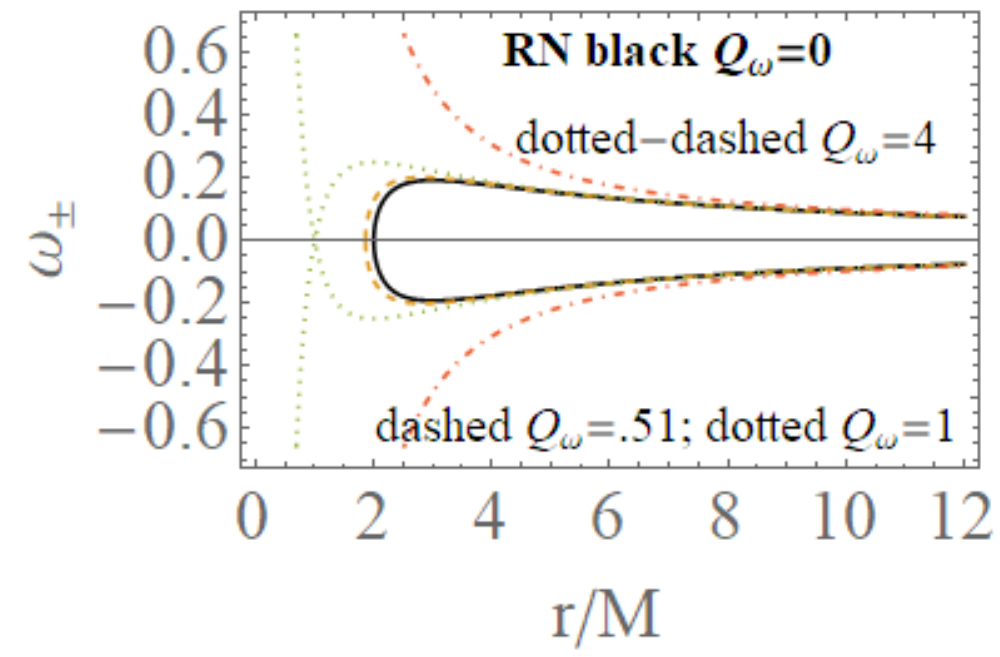}
   \includegraphics[width=5cm]{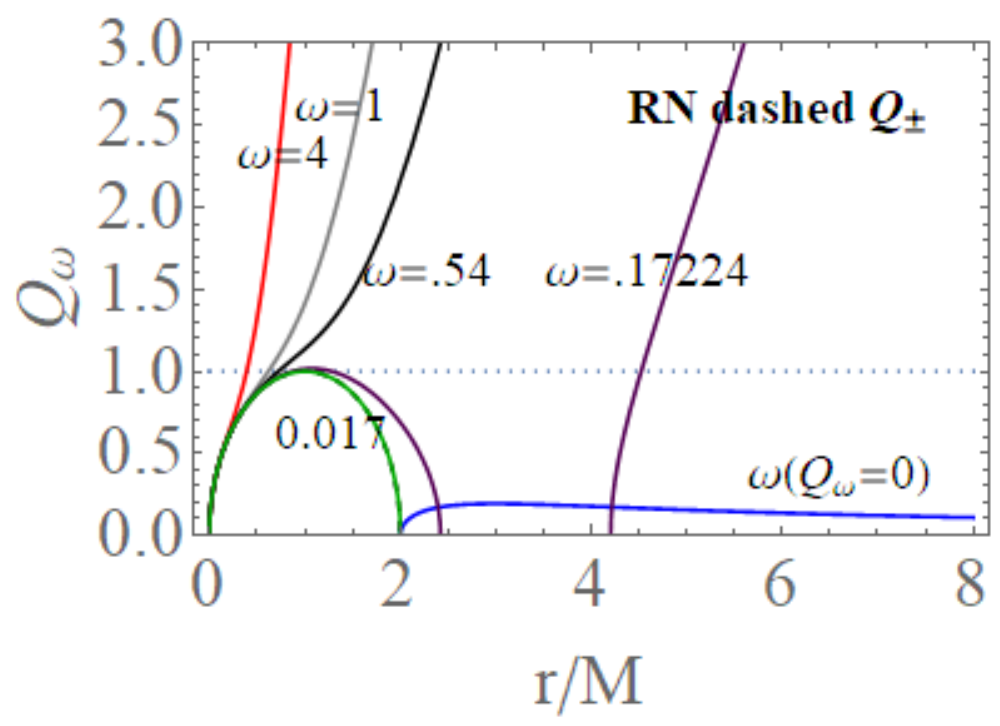}
    \includegraphics[width=5cm]{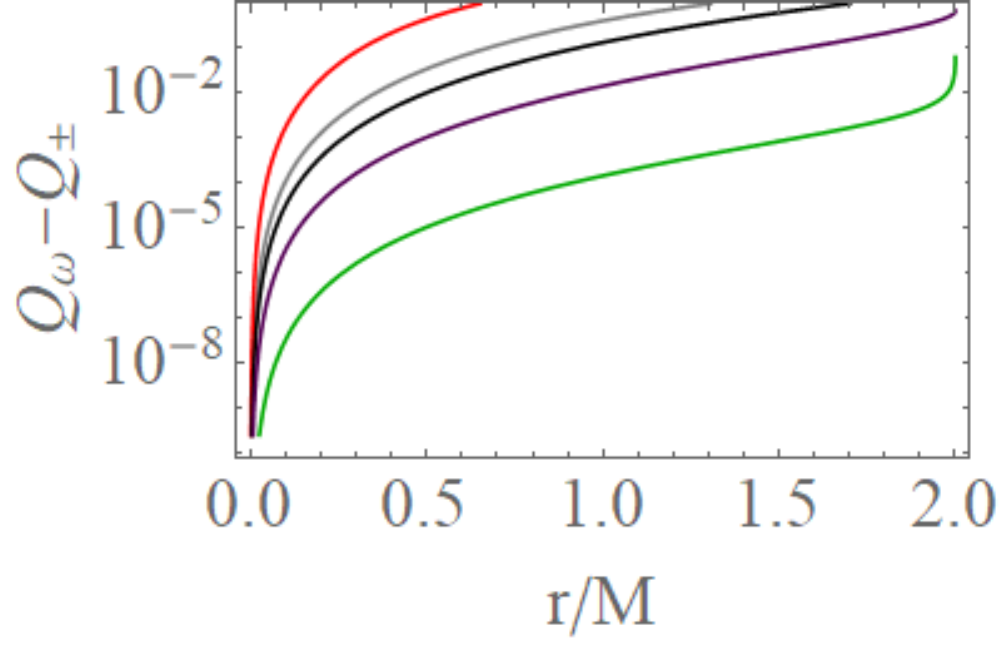}
  \caption{{Reissner--Nors}tr\"om  (RN) analysis: $Q_{\omega}$ are the metric bundles of Equation~(\ref{Eq:be-en}),  $\omega$ is the bundle frequencies, $Q_{\pm}$ is the horizon curve in the extended plane, frequency solution of $Q_{\omega}=0$ is in Equation~(\ref{Eq:be-en0}). (Here, $M$ is the metric mass parameter of the  Reissner--Norstr\"om  line element).  The right panel is the difference $Q_{\omega}-Q_{\pm}$ versus $r/M$, for the frequency values as in the central panel. The left panel shows the frequencies $\omega_{\pm}$ of Equation~(\ref{Eq:be-en}) versus $r/M$ as function of different $Q$ from BH to NS (naked singularities)---see also Table (\ref{Table:pol-cy-multi}).}\label{Fig:colorP1}
\end{figure}

Functions $\omega_{\pm}$ are
limiting light-like particles  orbital frequencies  showed in Figures (\ref{Fig:pcolorP1});  they  constitute the limiting conditions for the measure of the stationary light-like observing orbital frequencies.
The RN horizon  $Q_{\pm}=\pm\sqrt{r(2-r)}$  in the extended plane has a similar form to the Kerr geometry  horizon  $a_{\pm}=\pm\sqrt{r(2-r)}$---see Figures (\ref{Fig:vengplre})---where $\Upsilon\in\{|Q_{\pm}|,|a_{\pm}|\}$ has values  in $[0,M]$, where  $\Upsilon=0$ corresponds to the Schwarzschild geometry  and $\Upsilon=M$ is for the extreme RN  $(Q=M)$ or extreme Kerr BH case, respectively. For both geometries, this is a maximum of the horizon curve in the extended plane corresponding to the extreme BH solution. The horizon curve is closed and bounded in the range $r\in[0,2M]$, where $r=0$ is the Schwarzschild, Kerr, or  RN central singularity, $r=2M$ is the horizon in the Schwarzschild limit of the RN and  Kerr geometry as well as in the  LBHs .
The analysis of the {RN  MBs} in Figures (\ref{Fig:colorP1}) shows   that   it is  $Q_{\pm}=Q_{\omega}$ only for $\sigma=0$ or $\omega=0$, or $r=0$.  In fact, the absence of a tangency condition of the bundles with the horizon curves  is obviously an expression of the spherical symmetry. On the other hand, the approaching of MB curves to the horizons and the role of  electric charges are shown in Figures (\ref{Fig:colorP1}).
There is also $\omega=0$ only for bundles confined in the region  $r\leq 2$.
The choice of bundle parametrization for the RN case  is clear, being  related to the horizons' definitions. The MB introduction in the RN geometries  clarifies  some aspects of the  NSs, and  aspects of (local) causal structure on the bases of the horizons properties, and individuates the bottleneck region  typical of certain  NSs.
The precise definition of bottleneck goes far from the goals of the present analysis of the regular LBH solutions;  however,
the bottleneck is a restriction of the surfaces defined by the functions $\omega_{\pm}$  in the plane $r-\omega$, as a frequency tunnel (or equivalently the light-surfaces (LSs)) for some solutions   of  NSs close, in the extended plane, to the extreme BH solution.
The right panel of Figure~(\ref{Fig:colorP1}) shows the approximation of the bundles to the horizon curves in the extended plane typical of a spherically symmetric case. It is clear that the distance  between the MBs and the horizons' curve  decreases with the characteristic frequencies and increases with $r\in [0,2M]$. The bundles zeros curve in the central panel
$\omega(Q=0)$ is the limiting Schwarzschild frequencies. The left panel shows the frequencies $\omega_{\pm}(r)$ (or the LSs in the plane $r-\omega$) making evident the symmetries for negative and positive frequency values (in the case of spinning BHs, this symmetry with respect to positive and negative frequencies is broken).  Increasing the value of the electric charge $Q$ from  the Schwarzschild $Q=0$ to NSs  $Q>M$, the frequency curves are very different from the BH  with the self intersecting curve of the extreme {RN--BH }, which is, however, a regular maximum point  of the horizon curve in the extended plane. The self-intersections of the MBs is in fact a relevant aspect of the MB features.
\section{Metric Killing Bundles of the LBHs}\label{Sec:wordsee-h}
The four-velocity of the stationary observers are adapted to the Killing field $\laa=\xi_t+\omega\xi_{\phi}$ in the limiting Schwarzschild geometry and considered here in metric (\ref{Eq:dom-gionr}). We consider the null-like condition on the norm  $\laa_{\mathcal{N}}$
\bea&&\label{Eq:ass}
\laa_{\mathcal{N}}=\textbf{g}(\laa,\laa)=\frac{\sigma  \omega ^2 \left(a_o^2+r^4\right)}{r^2}-\frac{(r-2 m) (2 m P+r)^2 \left(r-2 m P^2\right)}{a_o^2+r^4}=0.
\eea
It is obvious that $\laa_{\mathcal{N}}$ is zero on the horizons only for $\omega=0$ ($\mathbf{g}(\xi_t,\xi_t)=0$---static observers are defined by the particular solution $\omega=0$; these observers, for example, cannot exist in the ergoregion of a Kerr geometry).
Note that we  can use also an adapted parametrization as
\bea
&&
\laa_{\mathcal{N}}=\frac{(R-2) \left(R-2 P^2\right) (2 P+R)^2}{A_o^2+R^4}+\frac{W^2 \left(A_o^2+R^4\right)}{R^2};
\\
&& \mbox{where}\quad\{r\rightarrow m R, a_o\rightarrow A_o m^2, \omega\rightarrow{W}/({m \sqrt{\sigma }})\}.
\eea
 In the plane  $(P,R)$, the outer horizon is
$
P_+\equiv\sqrt{{1}/{2R}}
$.
(We should also note that, in this way, $W$, $R$, and the length $A_o$ depend  on the ADM mass $M$ and the polymeric function $P$. The frequency  $\omega \sqrt{\sigma}$ re-parametrization  is instead a typical property of the MBs, which is generally related, for static as well as axially symmetric  solutions   to the MBs origin $r=0$ properties).

We can explore the spherical symmetry in the context of the metric bundles; it is clear that we can take advantage of this symmetries by considering  $\sigma=1$, i.e., an {(Schwarzschild BH)} equatorial plane, without loss of generality. However,  on $r_{\pm}$ and $r_s$, there is
\bea&&
\laa_{\mathcal{N}} (r_+)=\frac{\sigma  \omega ^2 \left(a_o^2+16 m^4\right)}{4 m^2},\\
&&\laa_{\mathcal{N}}(r_-)=\frac{\sigma  \omega ^2 \left(a_o^2+16 m^4 P^8\right)}{4 m^2 P^4},
\\
&&
\laa_{\mathcal{N}} (r_*)=\frac{64 m^4 (P-1)^2 P^3}{a_o^2+16 m^4 P^4}+\frac{\sigma  \omega ^2 \left(a_o^2+16 m^4 P^4\right)}{4 m^2 P^2}.
\eea
Therefore, more precisely:
\bea
&&
(r_-):\quad\laa_{\mathcal{N}}(r_-)=0,\quad \sigma =0,\quad \omega>0,\quad P\in]0,1],\quad \omega=0,\quad P\in]0,1],
\\
&&(r_+):\quad \laa_{\mathcal{N}}(r_+)=0,\quad \sigma =0,\quad \omega>0,\quad P\in[0,1],\quad \omega=0,\quad P\in[0,1],
\\
&&
(r_*):\quad
\laa_{\mathcal{N}}(r_*)=0,\quad \sigma =0,\quad \omega>0,\quad P=1,\quad \omega=0,\quad P=1.
\eea

\subsection{Light Surfaces (LS) Frequencies }

\begin{figure*}
\centering
  \includegraphics[width=5cm]{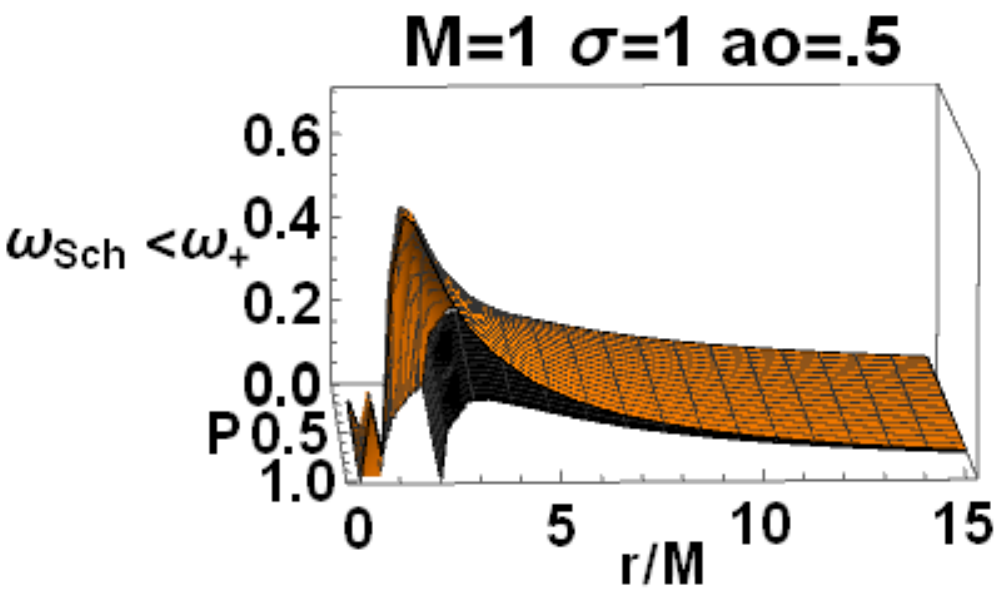}
    \includegraphics[width=5cm]{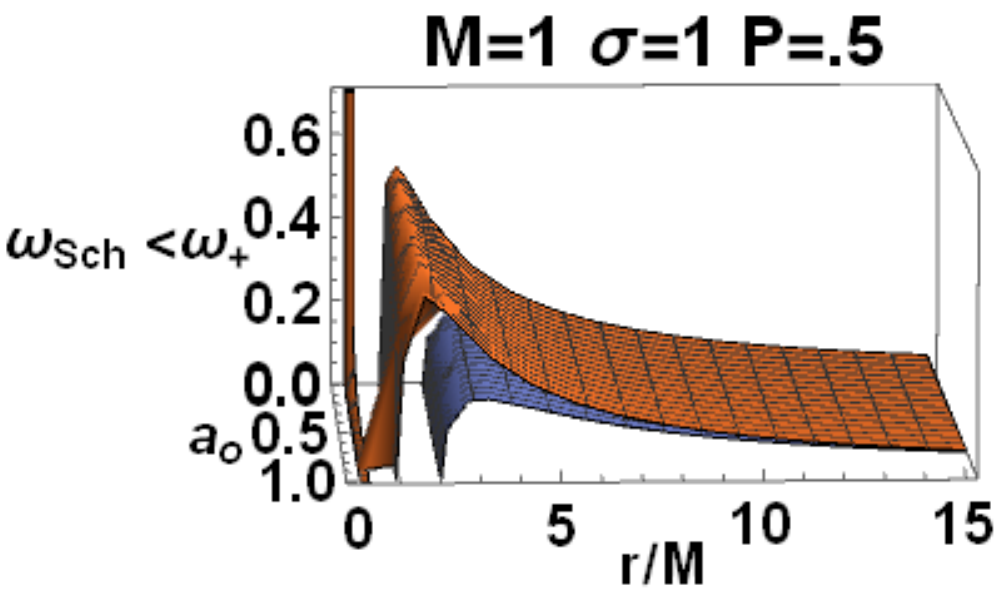}
      \includegraphics[width=5cm]{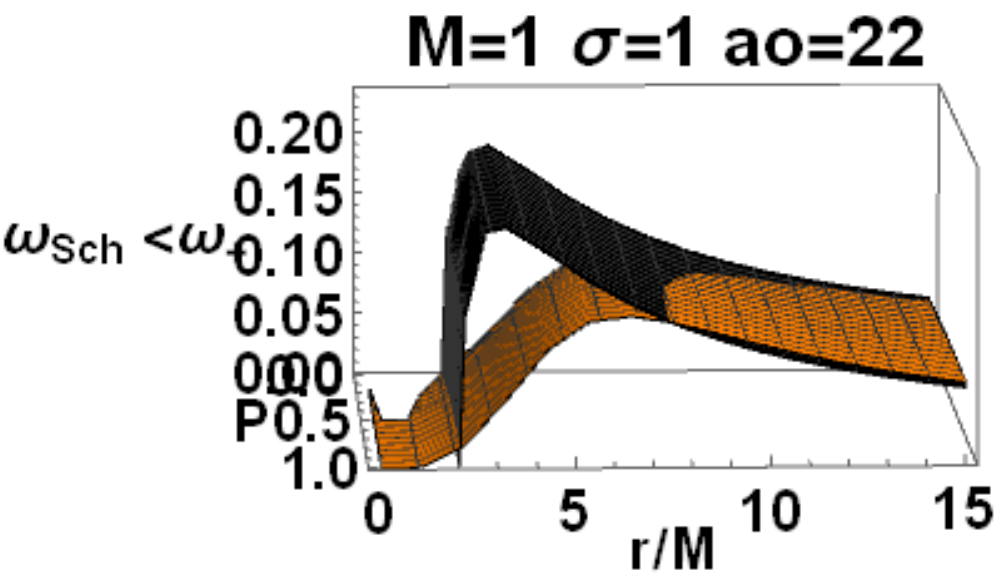}
        \includegraphics[width=5cm]{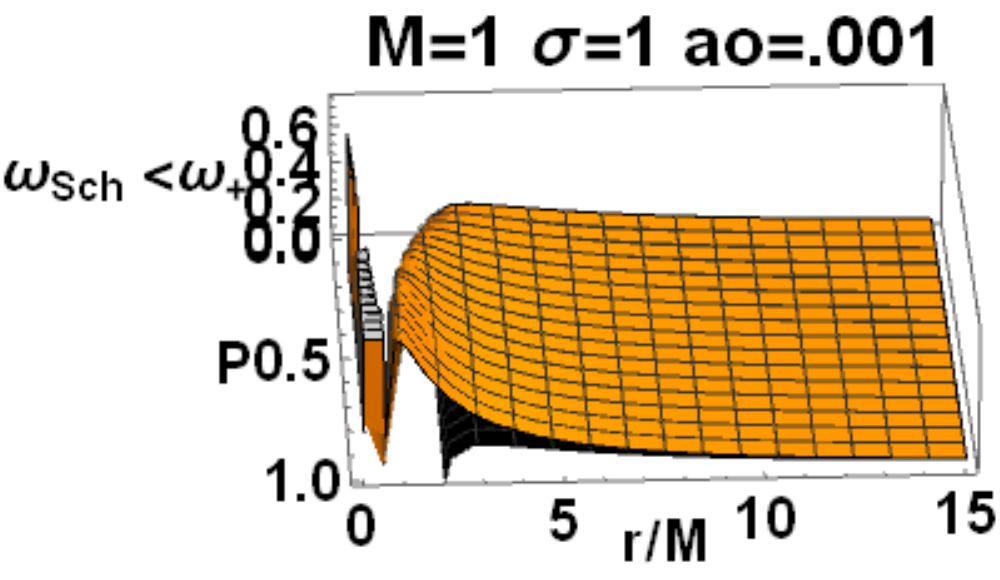}
          \includegraphics[width=5cm]{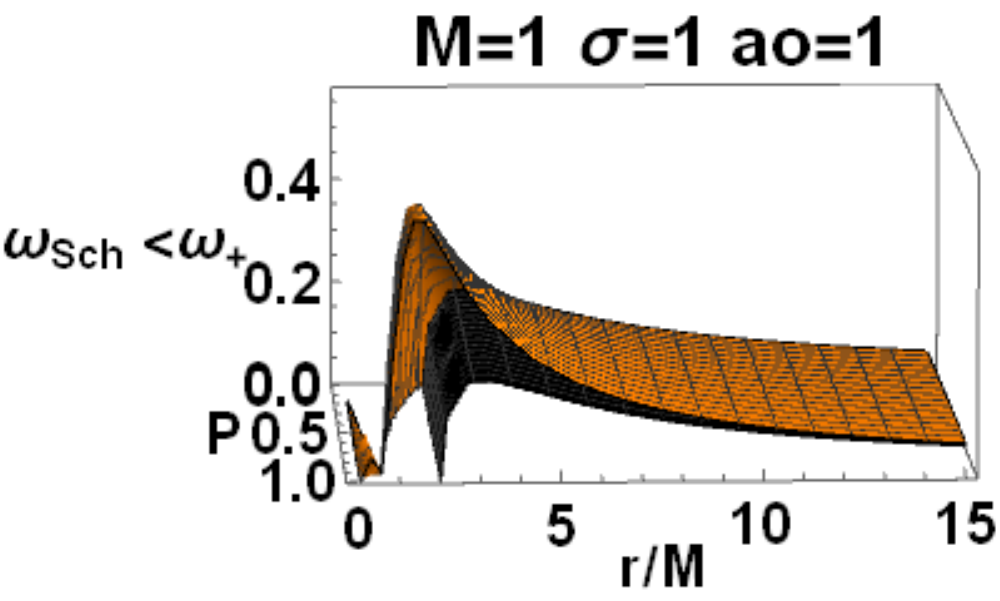}\\
            \includegraphics[width=5cm]{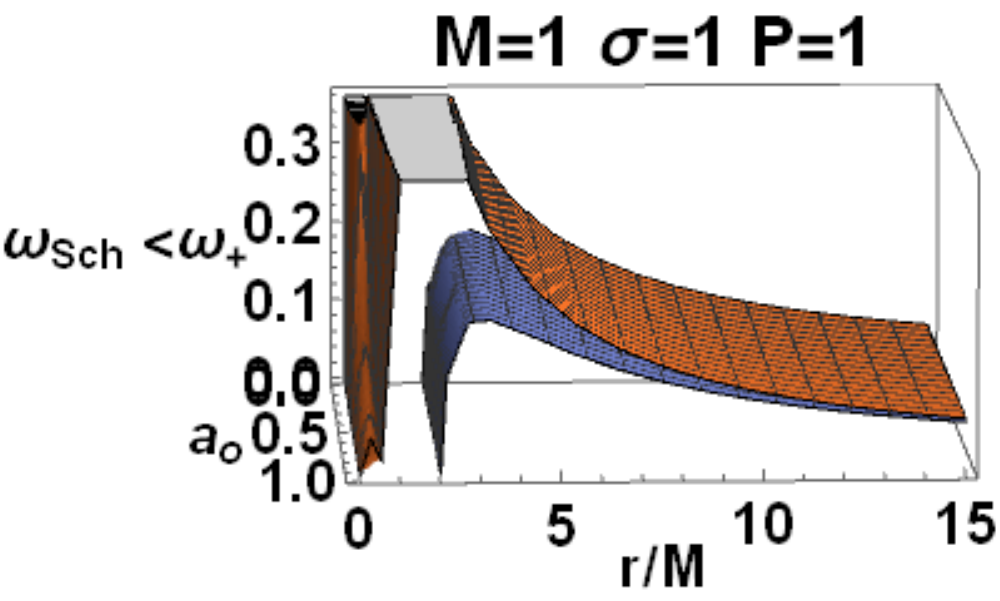}
              \includegraphics[width=5cm]{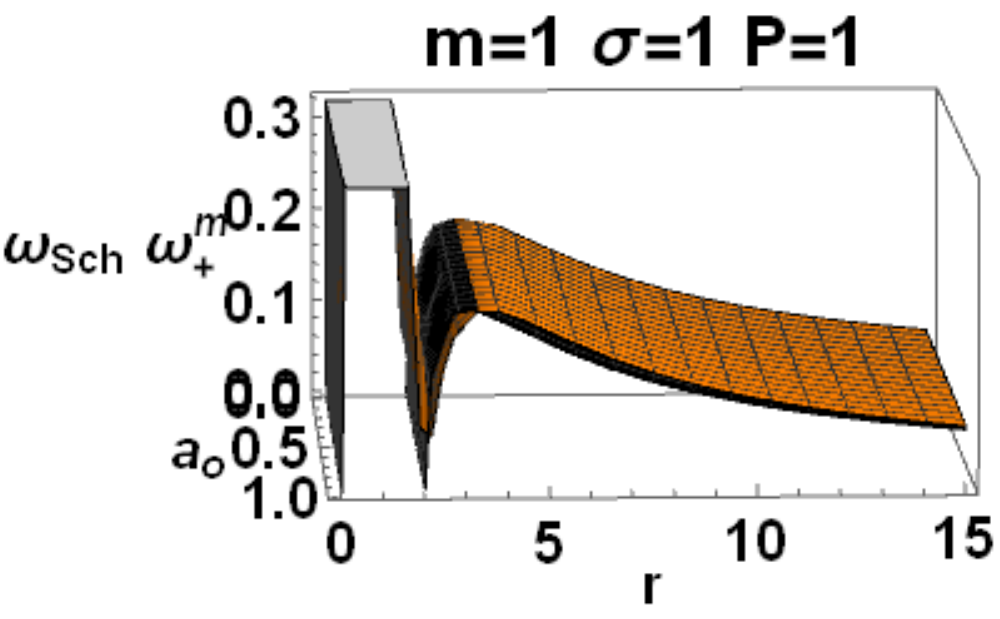}
                \includegraphics[width=5cm]{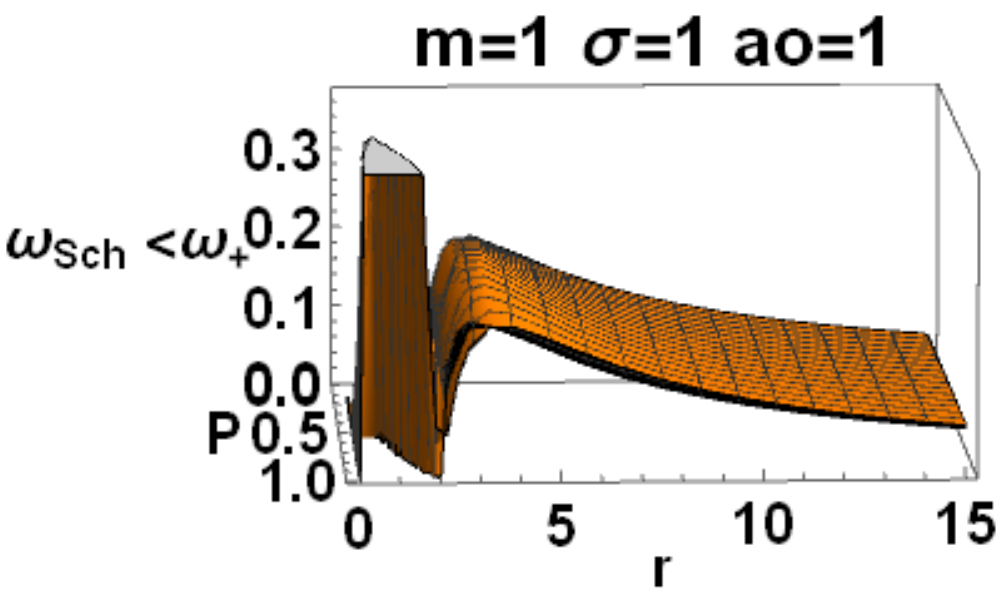}
                  \includegraphics[width=5cm]{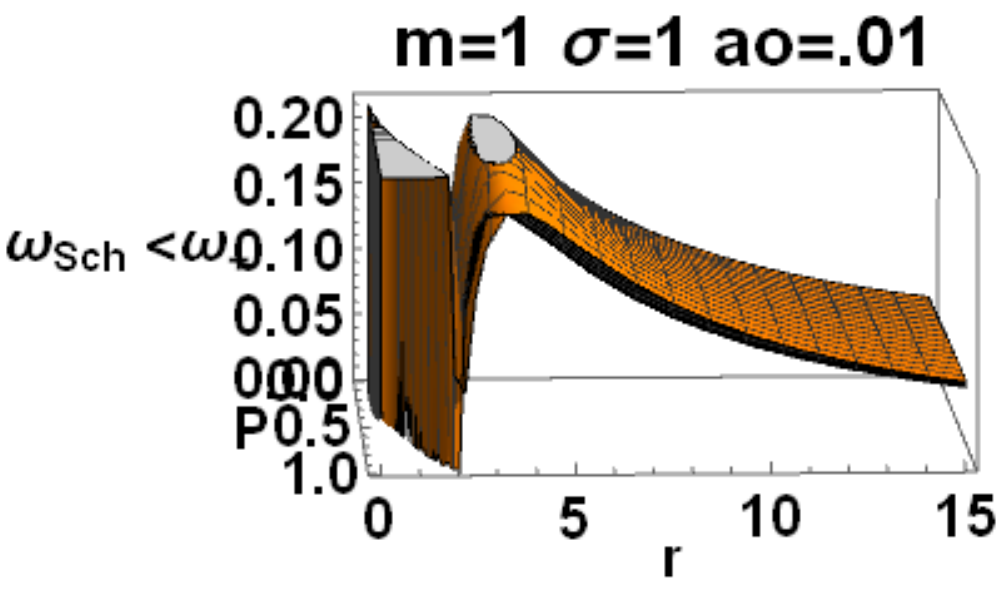}
                    \includegraphics[width=5cm]{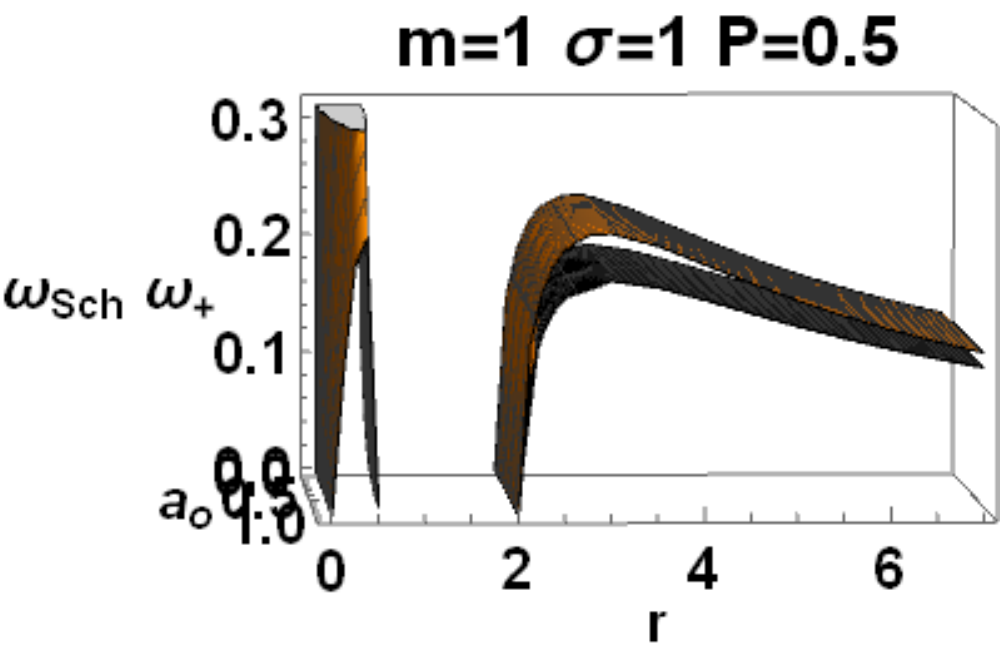}
  \caption{Limiting light-like frequencies $ \omega_\pm({\mathbf{x}})$ (orange) with  $\mathbf{x}=\{m,M\}$ ($M$ is the ADM mass, $m$ is the polymeric mass), solutions of $\laa_{\mathcal{N}}=0$ and $ \omega_{Sch}$  the   light-like frequencies of the  Schwarzschild geometries, as functions of $r$, the LQG area parameter $a_o$, and the  metric polymeric  $P$ and different values of $P$, and $a_o$, respectively. There is $\sigma\equiv \sin^2\theta$, where $\sigma=1$ is the Schwarzschild BH equatorial plane---see Equation~(\ref{Eq:scrib})
 and  Table (\ref{Table:pol-cy-multi}).}\label{Fig:fut10}
\end{figure*}

\begin{figure*}
\centering
    \includegraphics[width=3.74cm]{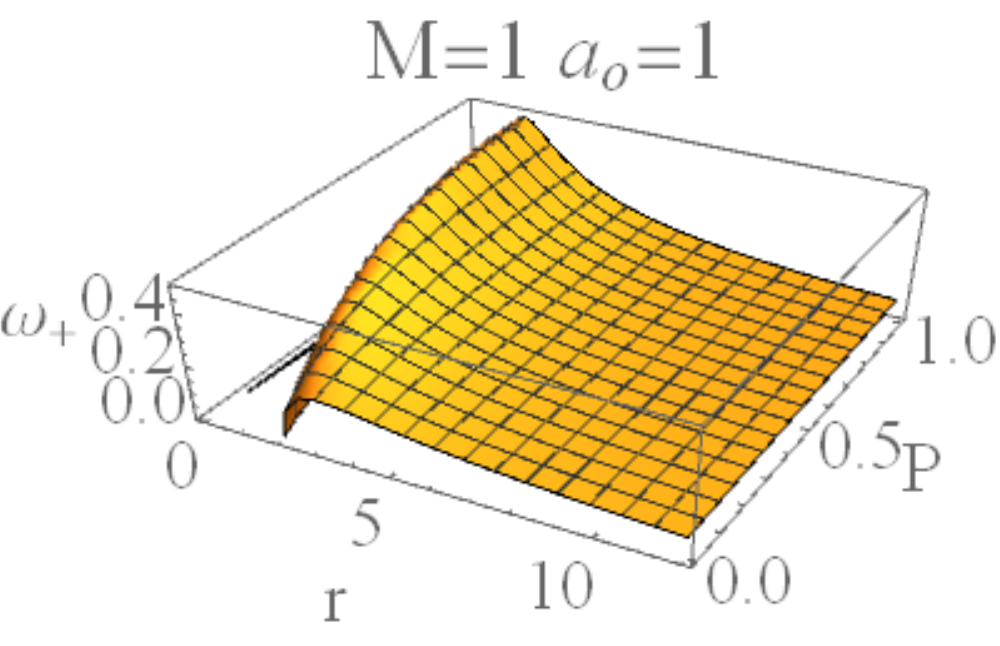}
  \includegraphics[width=3.74cm]{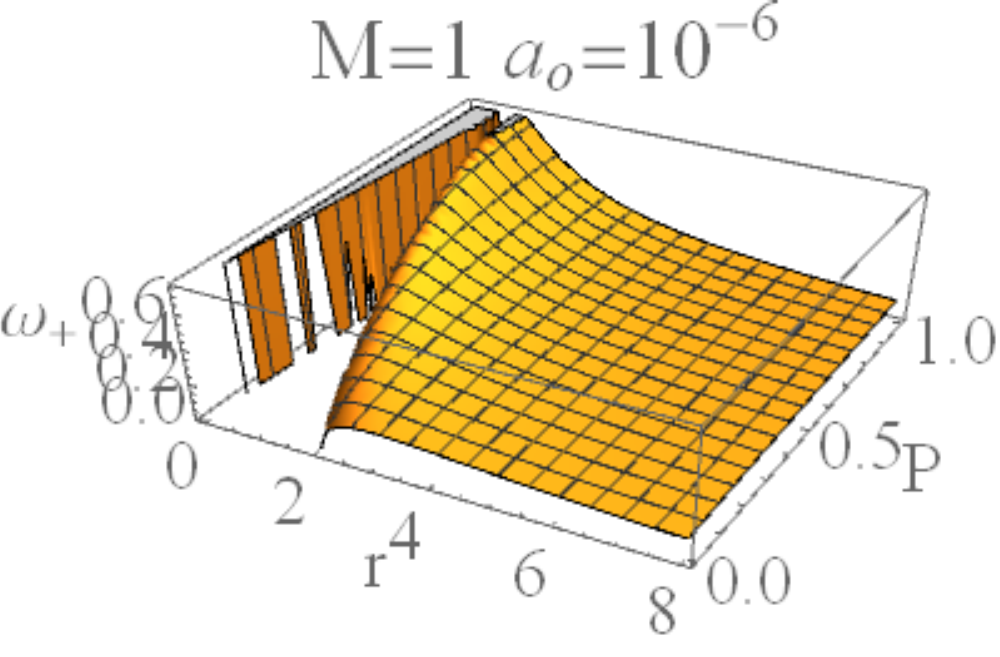}
  \includegraphics[width=3.74cm]{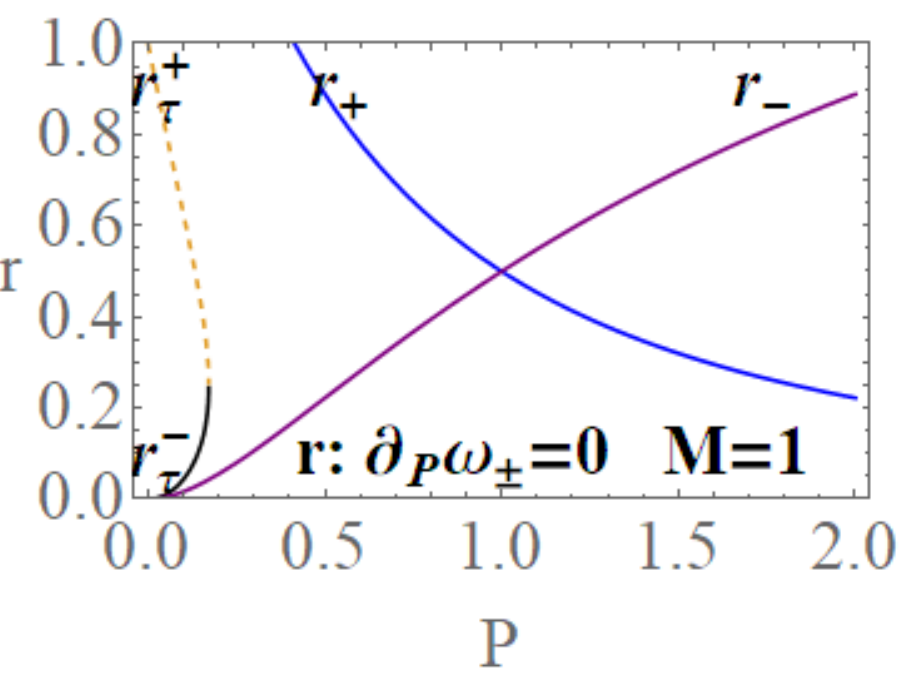}
  \includegraphics[width=3.74cm]{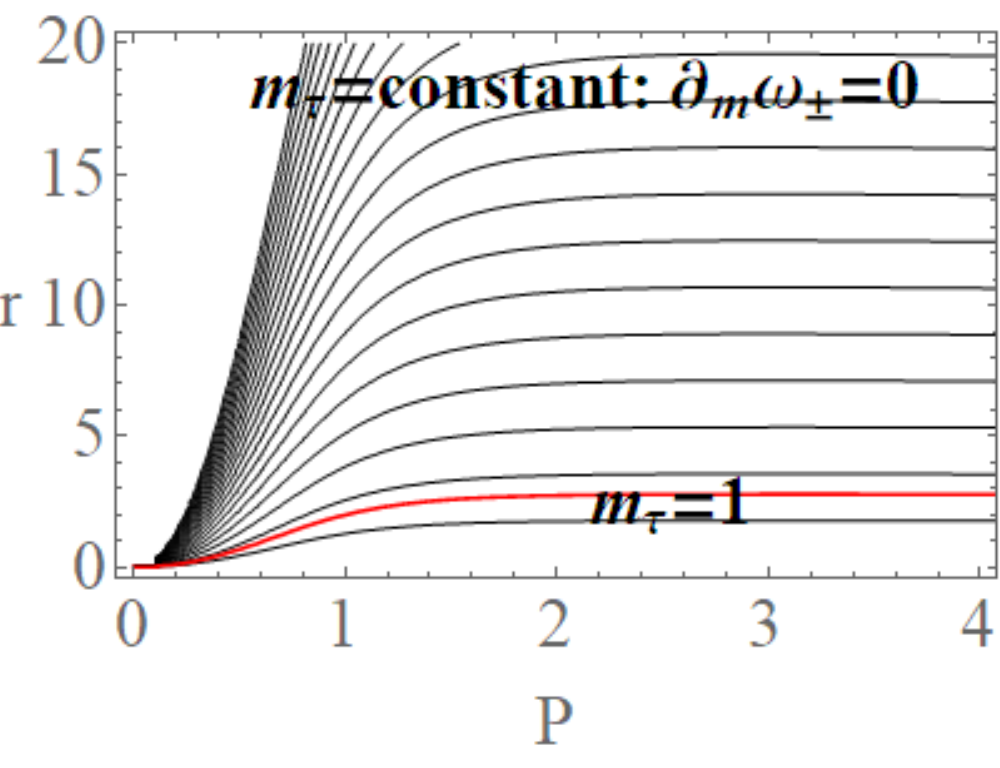}
  \caption{3D plots
show the limiting frequencies $\omega_{\pm}$ of stationary observers in the spacetime (\ref{Eq:tau-def})
for light surfaces as a function of $r$ and $ P$ (polymeric metric parameter) for different LQG length (area)
parameter $a_o$,  frequencies are in Equation~(\ref{Eq:be-en}). The extremes as function of $P$ are shown in the 2D third and fourth panels. Third panel: radii $r_{\tau}^{\pm}$ of Equation~(\ref{Eq:tau-def})
as functions of $P$ for the ADM mass $M=1$ ($r_{\pm}$ are the BH horizons).
Fourth panel: LQG mass parameter $m_{\tau}=$constant of Equation~(\ref{Eq:mtau}) in the plane $(r, P)$,
--see also Table (\ref{Table:pol-cy-multi}).
}\label{Fig:pcolorP1}
\end{figure*}

Adopting the procedure discussed in Section (\ref{Sec:RN}), we evaluate the zero-quantity $\laa_{\mathcal{N}}$, obtaining the light-like  orbital frequencies $\omega_{\pm}$
\bea\label{Eq:scrib}
&&\laa_{\mathcal{N}}=0,\quad  \omega_{\pm}(m,P)\equiv\pm\frac{-g_{tt}}{g_{\phi\phi}}=\pm\frac{G(r)}{H(r)\sigma}=\pm\frac{\sqrt{\frac{(2 m-r) \left(2 m P^2-r\right) (2 m P+r)^2}{a_o^2+r^4}}}{\sqrt{\frac{\sigma  \left(a_o^2+r^4\right)}{r^2}}},
\\&&
 \omega_\pm(M,P)= \pm\frac{\sqrt{\frac{\left(2 M-(P+1)^2 r\right) \left(2 M P^2-(P+1)^2 r\right) \left(2 M P+(P+1)^2 r\right)^2}{(P+1)^8 \left(a_{o}^2+r^4\right)}}}{\sqrt{\frac{\sigma  \left(a_{o}^2+r^4\right)}{r^2}}},
\eea
which are also the characteristic frequencies of the bundles. Note that these frequencies also define the light surfaces in many aspects of BH physics and more generally in the processes of energy extractions.
In these spherically symmetric  geometries, consider the magnitude $|\omega_{\pm}|$.
 The "Schwarzschild" limit is :
\bea\label{Eq:datitru}
&&
\omega_{Sch}\equiv\pm\frac{\sqrt{-\frac{2 M-r}{r}}}{\sqrt{r^2 \sigma }},
\\ &&\nonumber \lim_{a_o\rightarrow0}\lim_{P\rightarrow0} \omega_\pm({\mathbf{x}})=\omega_{Sch},\quad
\mathbf{x}=\{(m,P),(M,P)\}.
\eea
The two frequencies $\omega_{\pm}({\mathbf{x}})$ and $ \omega_{Sch}$ coincide   for special values of $a_o$ and $P$: there is generally $\omega_{\pm}({\mathbf{x}})\geq\omega_{Sch}$ with exceptions we show in Figure~(\ref{Fig:fut10}). On the other hand, these figures show differences between  $\omega_{\pm}({\mathbf{x}})$ considering as functions of $(m,P)$ or $(M,P)$.
Clearly at the horizons, there is $\omega_{\pm}(r_{\pm})=0$ and
$
\lim_{r\rightarrow0}\omega_{\pm}=0.
$
 The significance of these frequencies relies on the fact that they can be connected directly with the observations  of properties that are also conformal invariants  determining several properties of the geometry.
We find therefore the limiting frequencies that have an extreme for the loop mass  $m$, which is

\bea
&&\label{Eq:mtau}
m_{\tau }:\quad\partial _m\omega _{\pm }=0,
 \\
 &&m_{\tau}\equiv\frac{\sqrt{\left(-6 P^2+4 P-6\right)^2 P^2 r^2+64 \left(P^2-2 P+1\right) P^3 r^2}-P \left(-6 P^2+4 P-6\right) r}{32 P^3}
 \eea
---Figures (\ref{Fig:pcolorP1}). More precisely, limiting photon orbital frequencies have an extreme for the polymer function,
$P$  different according to  the different orbit $r$,
 \bea&&
 \label{Eq:tau-def}r_{\tau}:\quad\partial_P\omega _{\pm }=0\quad  (M=1)\\ &&\nonumber r_{\tau}^{\pm}\equiv\frac{1}{2} \left(\pm\sqrt{\frac{8}{(P+1)^2}-\frac{8}{P+1}+1}-\frac{4}{P+1}+\frac{4}{(P+1)^2}+1\right).
\eea
These radii are represented in Figures (\ref{Fig:pcolorP1}) with respect to the horizons.

The bundles are therefore a set of geometries, as defined by a leading parameter of the chosen parametrization such that all the  geometries of the bundles are only those characterized by a photon limiting orbital frequency $\omega$,  the bundle characteristics' frequency is in the Kerr applications, and the BH horizon frequency  in the extended plane.
In the spherical symmetric spacetime, we investigate in this work the frequencies connecting geometries very close to the horizon (in the extended plane) with other geometries. Mainly here we consider the $P$ parametrization. Alternately, we can consider the $\epsilon$-parametrization in the extended plane of Figures (\ref{Fig:vengplre}).
 The horizons' curves are the  functions $P_{\pm}$ of Equation~(\ref{Eq:vengplre}). Clearly, we could have used the $\epsilon_{\pm}$ representation in a plane $\epsilon-r$ as in Figures (\ref{Fig:vengplre}).
 The vertical lines in the extended plane $r=$constant represent a fixed point  in different geometries; the collections of all points on the bundles at $r=$constant provide the set of different or equal frequencies $\omega$ connecting therefore  different solutions making evident the modifications of the frequencies  due to the shift of the polymeric functions.
The horizonal lines in the extended plane, correspondent to $P=$constant, represent  one  geometry with $P=$constant, and the crossing of the horizontal line with all the bundles of the plane provides the set of light surfaces $r{\pm}(\omega)$ solutions of $\mathcal{L}_{\mathcal{N}}=0$.

\subsection{Metric Bundles Parametrization}\label{Eq:everi-chan}
In this section, we provide an explicit expression for MBs according to different parameterizations.
\begin{itemize}
\item\textbf{Metric bundles: parametrization according to $\sigma$}
As the metric is spherically symmetric,
we can consider, without loss of generality,  $\sigma=1$, i.e., a fixed  {(Schwarzschild  BH)} equatorial plane. Nevertheless, we could  consider explicitly a parametrization according to the "poloidal" angle  $\theta$, obtaining the curves:
\bea\label{Eq:media.-stateprima}
&&
\sigma_{\omega}=\frac{r^2 (r-2 m) (2 m P+r)^2 \left(r-2 m P^2\right)}{\omega ^2 \left(a_o^2+r^4\right)^2}.
\eea
In here, metric bundles with $\omega=$constant are on the hyperplane { $\sigma_{\omega}=1$.}
Clearly, this choice is relevant for the parametrization $W=\omega\sqrt{\sigma}$.
Note that there is $\sigma_{\omega}=0$ for  $r\in\{0,r_{\pm}\}$.
Figures (\ref{Fig:SPlru8}) represent special and limiting cases of metric bundles.
\\
\item\textbf{Metric bundles: $a_o$-parametrization}

It is relevant to study a $a_o$-parametrization  to  consider the families of metrics for different minimum areas parameter $a_o$. Implementing therefore the notion of metric bundles with   the area parameter, we obtain  explicitly
\bea&&\label{Eq:minsa}
(a_o^{\pm}(m,P))^2\equiv\pm\sqrt{\pm\frac{\sqrt{r^2 \sigma  \omega ^2 (r-2 m) (2 m P+r)^2 \left(r-2 m P^2\right)}}{\sigma  \omega ^2}}-r^4,
 \\\nonumber
 &&(a_o^{\pm}(M,P))^2\equiv\frac{\sqrt{(P+1)^8 r^2 \omega ^2 \left(P^2 (r-2)+2 P r+r\right) \left((P+1)^2 r-2\right) \left((P+1)^2 r+2 P\right)^2}}{(P+1)^8 \omega ^2}-r^4
\eea
represented in Figures (\ref{Fig:SPlru8a}).  Providing constraints on the loop minimal areas,
functions  $(a_o^{\pm}(M,P)),a_o^{\pm}(m,P)))$ are not well defined on the horizons in the extended plane, and limits to $r=0$ is null.
\begin{figure*}
\centering
    \includegraphics[width=5cm]{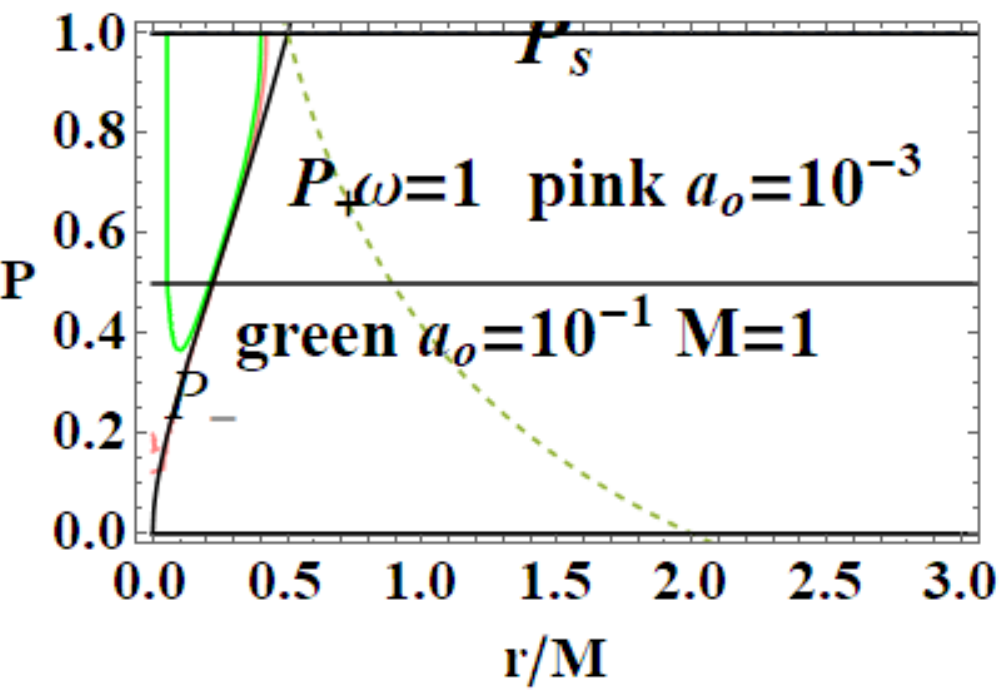}
     \includegraphics[width=5cm]{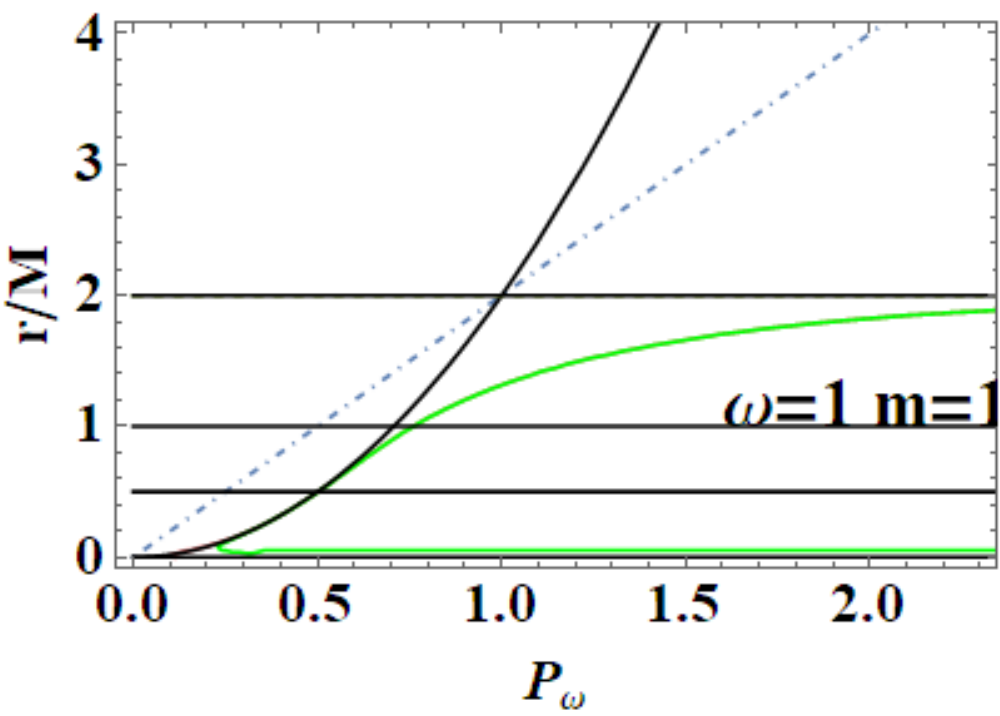}
     \includegraphics[width=5cm]{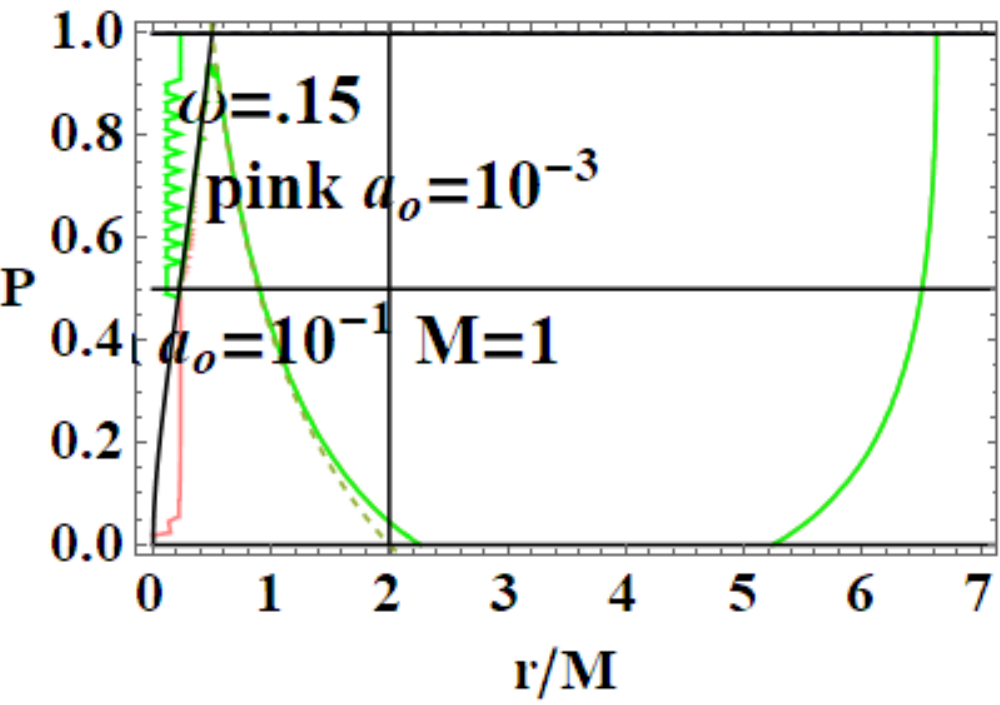}
      \includegraphics[width=5cm]{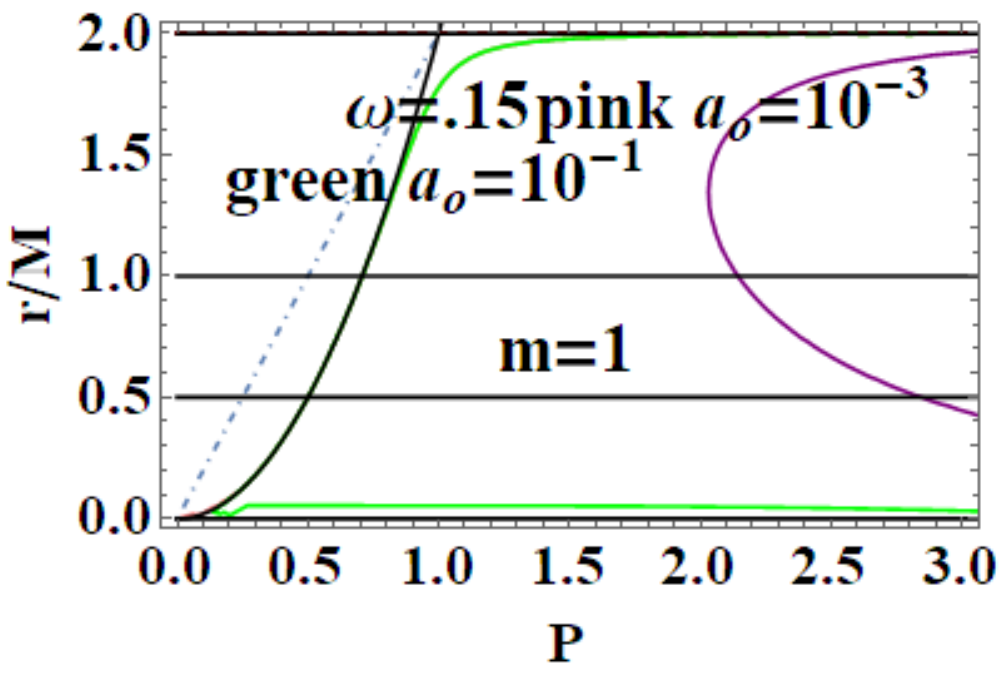}
       \includegraphics[width=5cm]{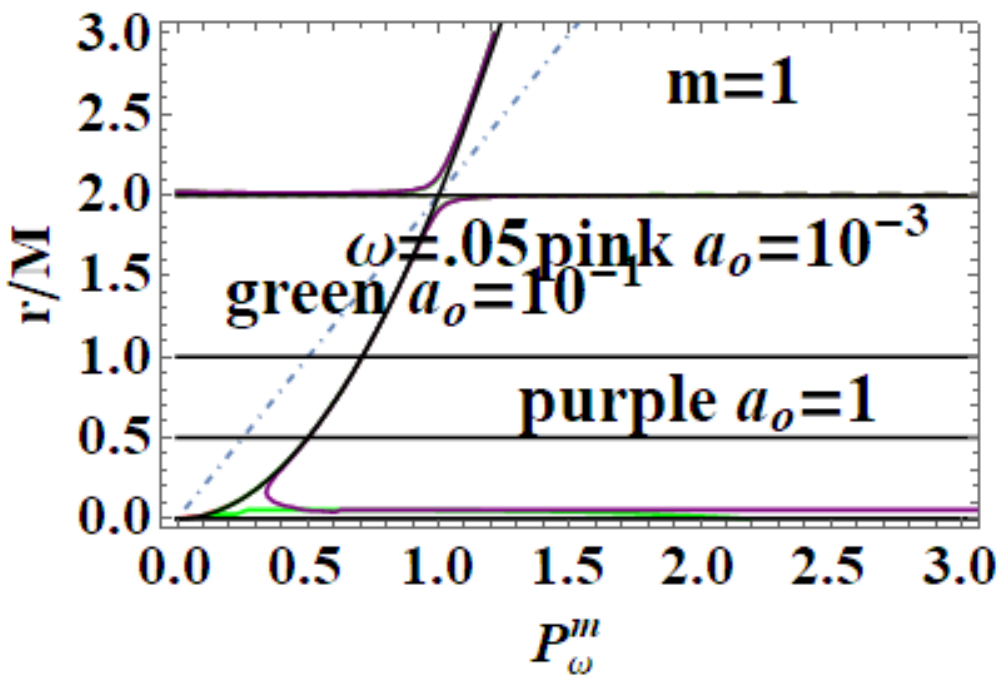}
       \includegraphics[width=5cm]{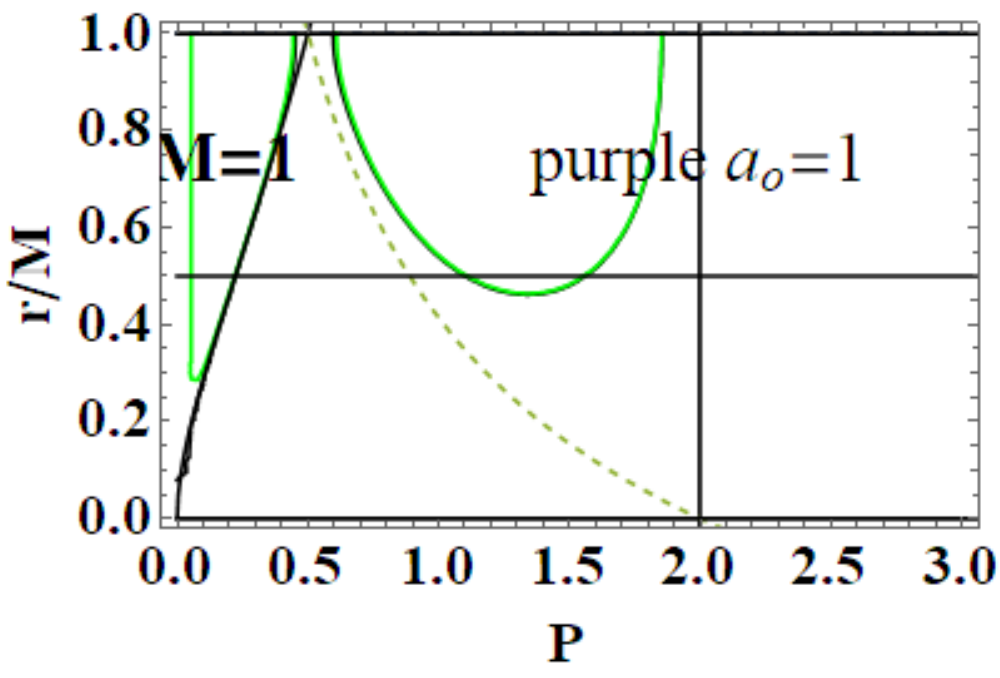}
        \includegraphics[width=5cm]{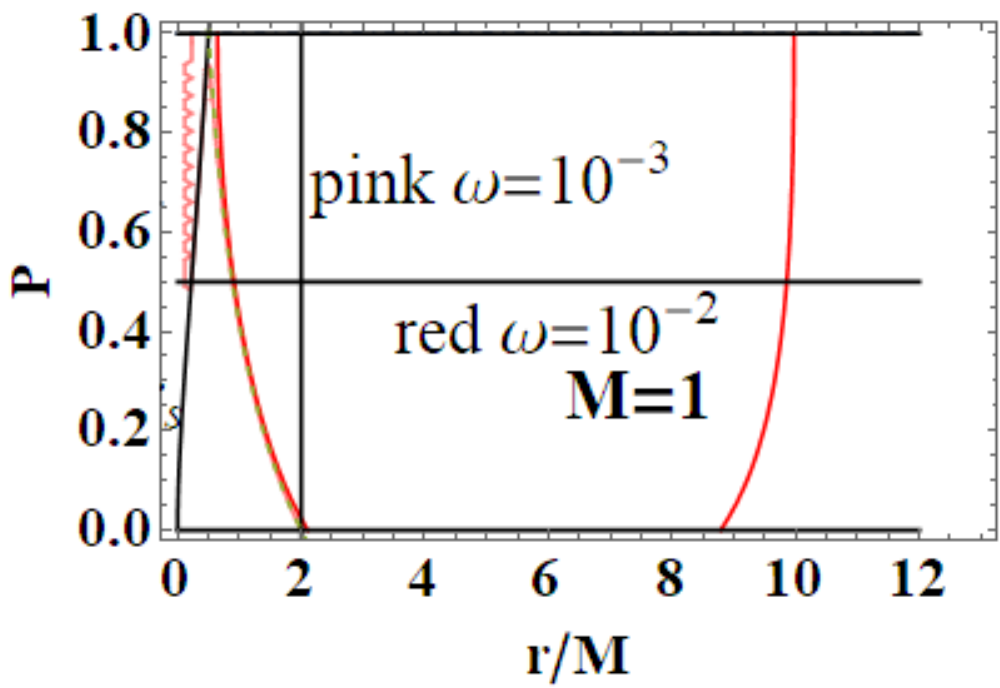}
                \includegraphics[width=5cm]{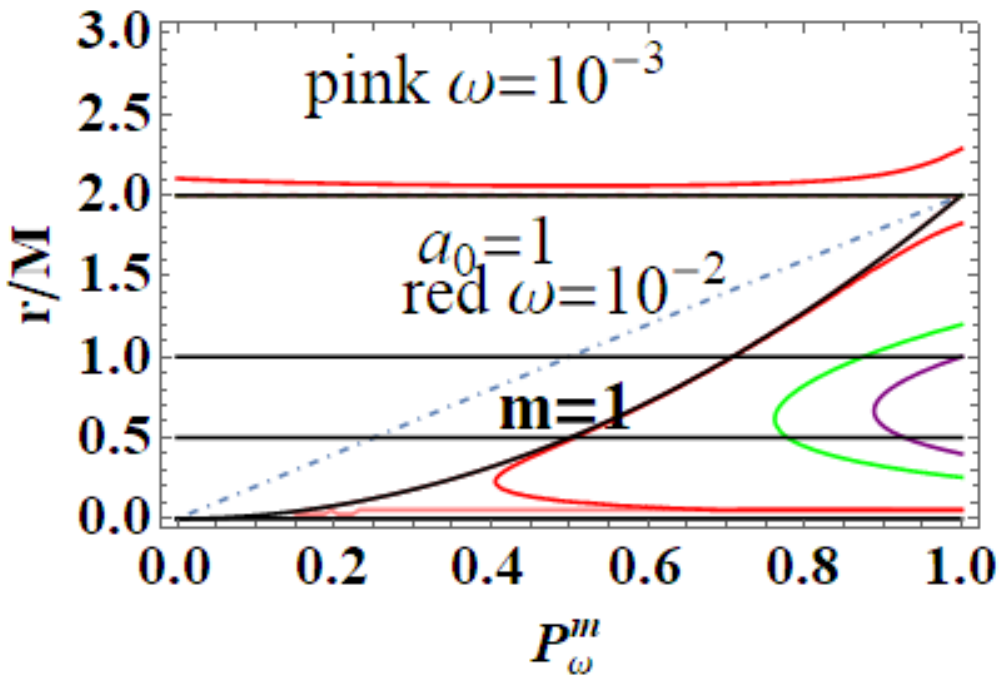}
                  \includegraphics[width=5cm]{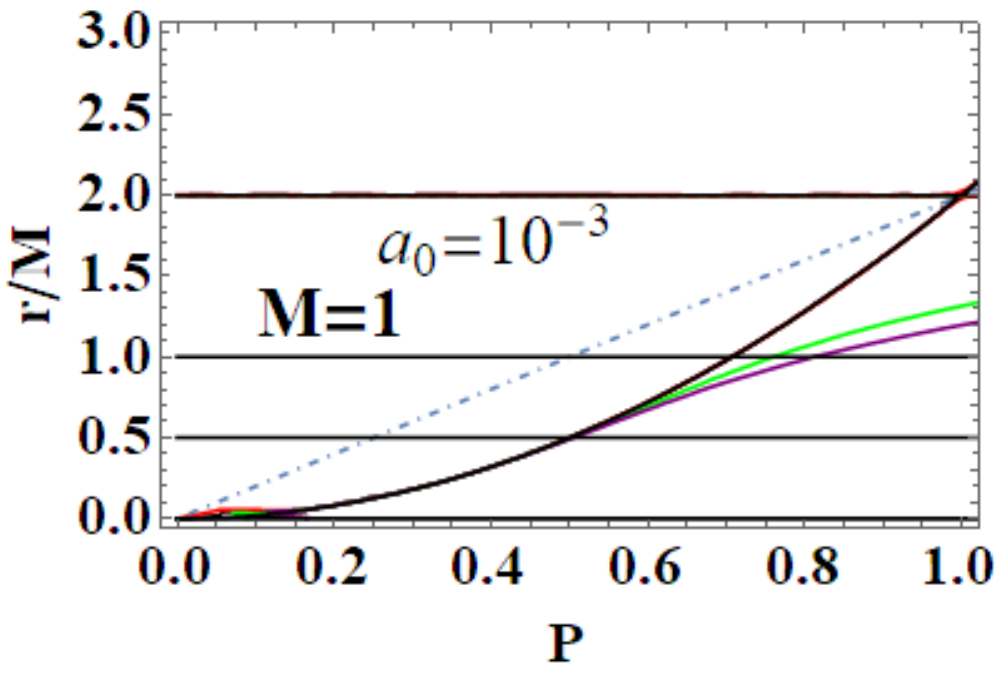}
                    \includegraphics[width=5cm]{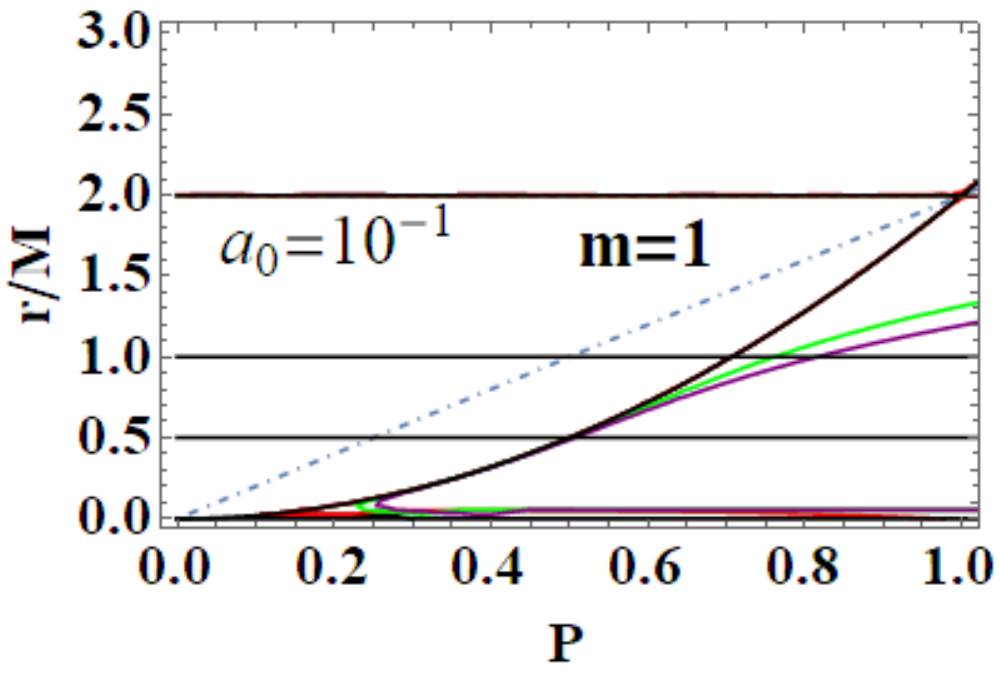}
           \includegraphics[width=5cm]{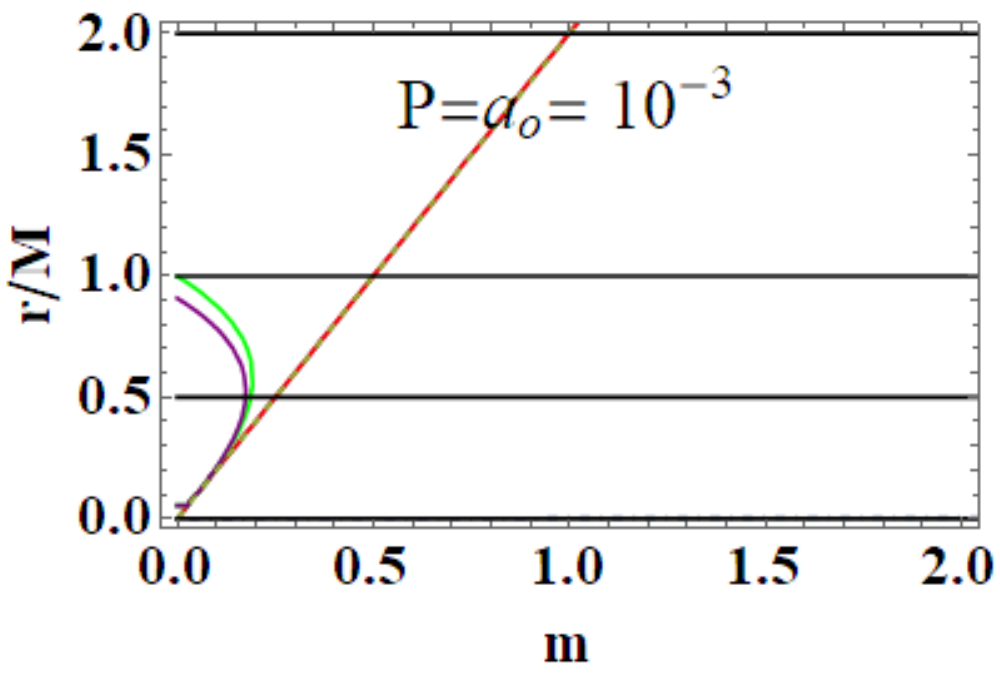}
            \includegraphics[width=5cm]{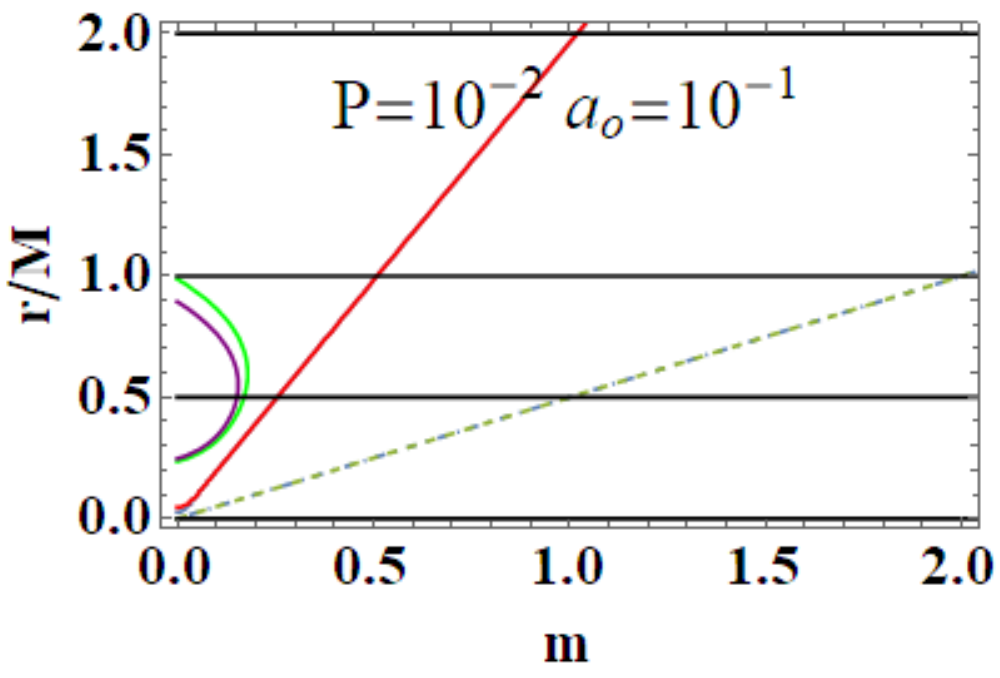}
  \caption{Metric bundles (solutions of $\laa_{\mathcal{N}}=0$, connected to the  light surfaces) for selected  values of the polymeric model parameters as functions of  the polymeric parameter $P$ or loop mass $m$ ($M$ is the ADM mass parameter) in the plane $(r,P)$ or $(r,m)$  for different bundle frequencies $\omega$ (according to colors reported in panels). $a_o$ is the LQG length parameters.  Horizons $r_{\pm}$ and radius $r_*$ are also shown---see also Table (\ref{Table:pol-cy-multi}) for details on the notation.
 }\label{Fig:SPlru8}
\end{figure*}
\begin{figure*}
\centering
  \includegraphics[width=6cm]{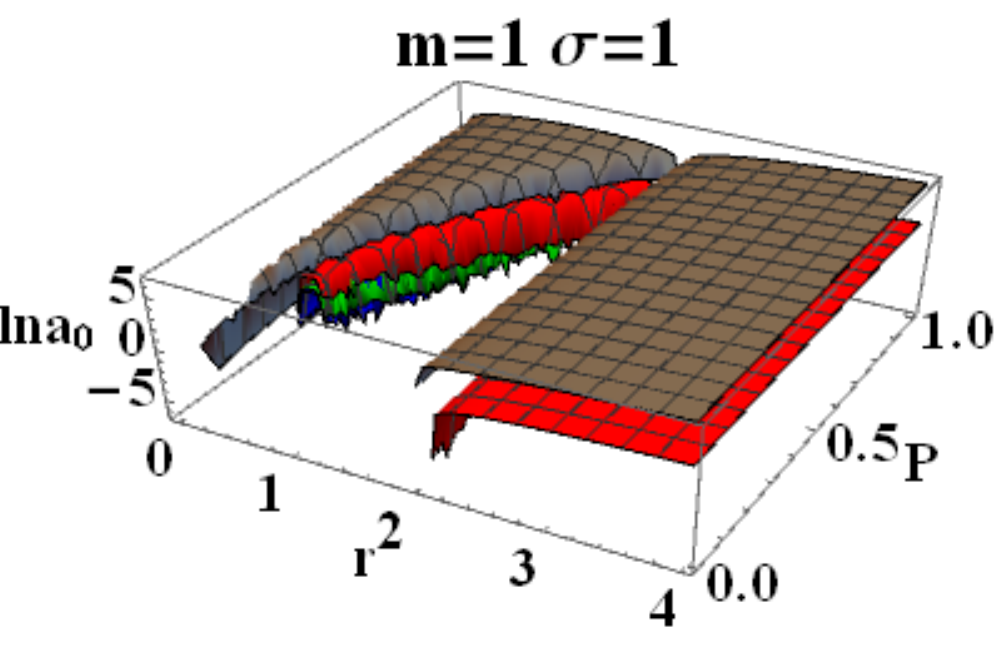}
   \includegraphics[width=6cm]{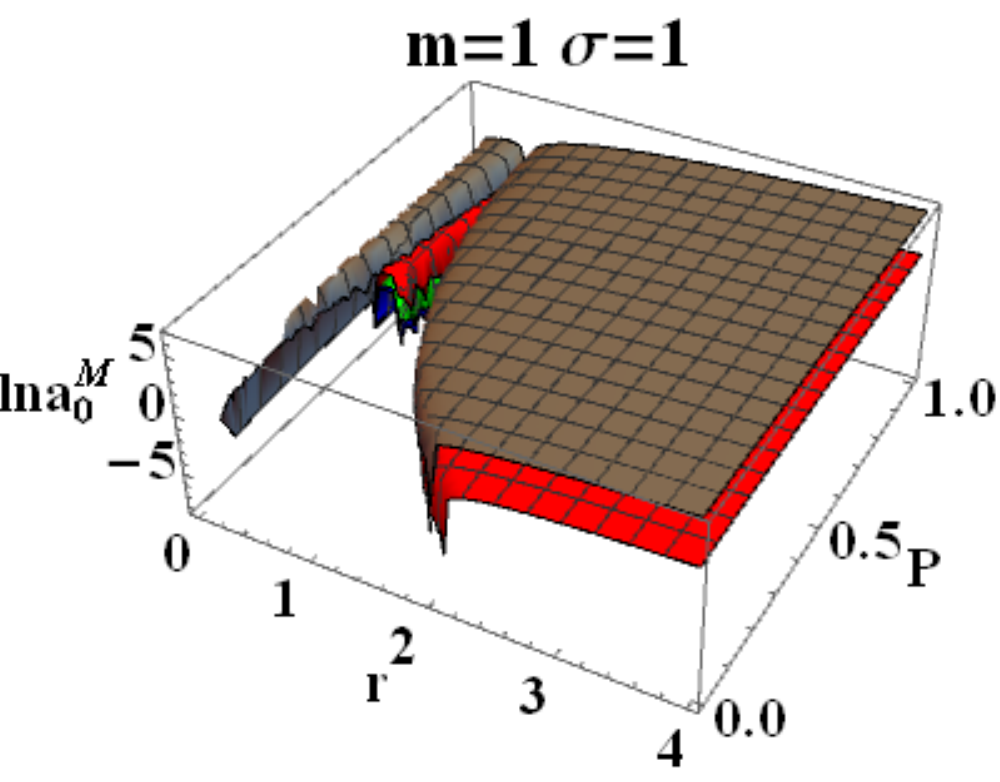}
  \caption{3D plots represent metric bundles  $a_o^{\pm}(m,P)$ for $a_o$-parametrization see
Equation~(\ref{Eq:minsa}), for
 $\omega = 10^{-4} $ (gray) $\omega = 0.1 $ (red), $\omega = 1$ (green), where $a_o$ is the LQG length parameters, $P$ is the polymeric metric parameter, $\omega$ is the light-like orbital limiting frequencies (stationary observers), $M$ is the ADM mass, $m$ is a polymeric mass, only asymptotically equivalent to the ADM mass. In the panel, we adopt the notation $a_o=a_o^{\pm}(m,P))$ and $a_o^{M}=a_o^{\pm}(M,P)$ (right panel). We also took advantage  of the symmetries  $a_o^{\pm}=\mp a_o^{\mp}$, and
$\sigma\equiv\sin^2\theta$, the (BH Schwarzschild) equatorial plane is $\sigma=1$---see also Table (\ref{Table:pol-cy-multi}) for details on the notation.
 }\label{Fig:SPlru8a}
\end{figure*}
\\
\item\textbf{Metric bundles: $P$-parametrization}

Here, we consider the   leading parameter  $P$. Metric bundles in the extended plane $P-r$ and $\epsilon-r$ are shown in Figures (\ref{Fig:govmentra}), where there is also a focus on the  vertical and horizontal lines of the extended planes and horizons' replicas at different values of the parameter.

The equation for the metric bundles according to the  $P$-parametrization $(M=1)$ is polynomial function of degree 8, $f(P;\mathbf{\zeta})=\sum_{i=0}^8P^i \zeta_i$, {where}
\bea\label{Eq:p-powe-bundlr}&&\zeta_0\equiv\omega ^2 \left(a_o^2+r^4\right)^2-(r-2) r^5;\\ &&\nonumber\zeta_1\equiv8 \omega ^2 \left(a_o^2+r^4\right)^2+8 \left(-r^2+r+1\right) r^4;
\\\nonumber
&&\zeta_2\equiv4 \left[7 \omega ^2 \left(a_o^2+r^4\right)^2+[r (r (2-7 r)+6)+2]r^3\right];\\ &&\nonumber\zeta_3\equiv8 \left[7 \omega ^2 \left(a_o^2+r^4\right)^2-r^4 \left(7 r^2+r-3\right)\right];
\\\nonumber
&&\zeta_4\equiv2 \left[35 \omega ^2 \left(a_o^2+r^4\right)^2-r^2 [r (r [5 r (7 r+2)-8]+8)+8]\right];\\ &&\nonumber\zeta_5\equiv8 \left[7 \omega ^2 \left(a_o^2+r^4\right)^2-r^4 \left(7 r^2+r-3\right)\right];
\\\nonumber
&&\zeta_6\equiv4 \left[7 \omega ^2 \left(a_o^2+r^4\right)^2+[r (r (2-7 r)+6)+2] r^3\right];\\ &&\nonumber\zeta_7\equiv8 \left[\omega ^2 \left(a_o^2+r^4\right)^2-r^6+r^5+r^4\right];\\&&\zeta_8\equiv\omega ^2 \left(a_o^2+r^4\right)^2-(r-2) r^5.
\eea
\end{itemize}
\section{The LBHs Thermodynamical Properties}\label{Sec:termo}
In this section, we review some aspects of the BH thermodynamics. In Section (\ref{Sec:termo-1}), we explore the BHs  surfaces gravity, the luminosity, and the temperature in terms of the loop model parameters; then, these quantities are  considered  on metric bundles. We discuss connections between different geometries of one bundle  considering their
  thermodynamical  properties.     In general, the Hawking emission at $r_+$  leads to the evaporation process with BH
Bekenstein--Hawking temperature. The mass loss on the process can be deduced with the luminosity ($L$).
 The focusing idea is how the BH thermodynamical properties can explain and distinguish  the construction of the underling graph in the LQG model adopted here, especially on the grounds of the MBs structures analyzing the horizons properties in regions far from the horizons---see also \cite{Aleshi2012zz,Myung:2007av}.

\subsection{BHs Thermodynamics  and  LQG Parameters}\label{Sec:termo-1}
In this section, we evaluate the BH areas and the BHs surfaces gravity and temperature in terms of the LBH parameters $\mathcal{P}$.

\begin{itemize}
\item
\textbf{BH areas}
We can evaluate the BH areas as follows:
\bea\label{Eq:BHarea-poer}
&&
(A_{\mathbf{BH}}^+):\quad
A_{\mathbf{BH}}^+(m)=\frac{\pi  \left(a_o^2+16 m^4\right)}{m^2},
\quad
 \quad A_{\mathbf{BH}}^+(M,P)=\pi  {\frac{(P+1)^4 \left(a_o^2+\frac{16 M^4}{(P+1)^8}\right)}{M^2}}.\\
&&\nonumber
(A_{\mathbf{BH}}^-):\quad
A_{\mathbf{BH}}^-(m,P)=\pi  {\frac{\left(a_{o}^2+16 m^4 P^8\right)}{m^2 P^4}}, \quad A_{\mathbf{BH}}^-(M,P)=\pi {\frac{(P+1)^4\left(a_o^2+\frac{16 M^4 P^8}{(P+1)^8}\right)}{M^2 P^4}},
\\
&& \qquad\qquad\qquad A_{\mathbf{BH}}^-(M,P(m))=
\pi  {\frac{\left(a_{o}^2+16 \left(\sqrt{m}-\sqrt{M}\right)^8\right)}{\left(\sqrt{m}-\sqrt{M}\right)^4}}.
\eea
 In  $A_{\mathbf{BH}}^-(M,P(m))$, we used the quantity $P(m)=\frac{\sqrt{M}-\sqrt{m}}{\sqrt{m}}$,
where  $A_{\mathbf{BH}}^{\pm}$ is related to the surface bounded by the outer and the inner BH horizon. (In the extended plane, it is necessary to consider horizons $r_{\pm}$). Note that $A_{\mathbf{BH}}^+$ does not depend explicitly on $P$.
When considered in the extended plane, there are some special values of the $\mathcal{P}$ parameters for which there is a coincidence of the areas $A_{\mathbf{BH}}^{\pm}$, for the  equal values of  $(m, a_o)$.
There is $
A_ {\mathbf{BH}}^-(m, a_o) =
 A_ {\mathbf{BH}}^+(m, a_o) $ (within the assumption $M = 1$) for $ m = 1/4$ or $a_o = a_ {ox}^b$,   while $A_{\mathbf{BH}}^-(P, M) =
 A_{\mathbf{BH}}^+(P, M)$ for  $a_o =
  0$, $ P = 1$ or $ a_o = a_ {ox}^a$,  where
  \bea\label{Eq:som-ox-pe}
 a_{ox}^a\equiv \sqrt{\frac{16 M^4 P^4}{(P+1)^8}},\quad a_{ox}^b\equiv\sqrt{16 \left(\sqrt{m}-1\right)^4 m^2},
  \eea
 see Figures (\ref{Fig:pax12}).
\begin{figure}
   \includegraphics[width=3.84cm]{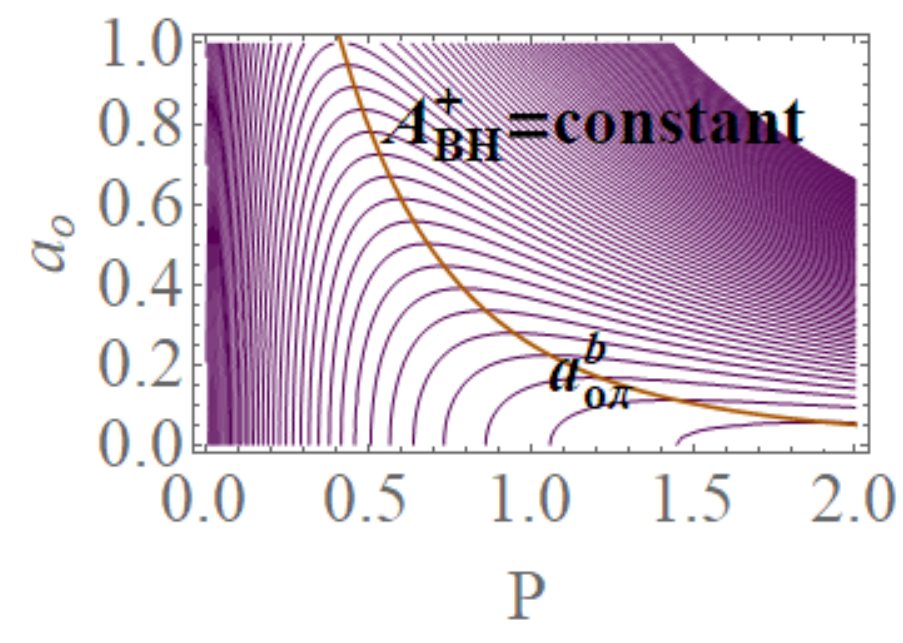}
    \includegraphics[width=3.84cm]{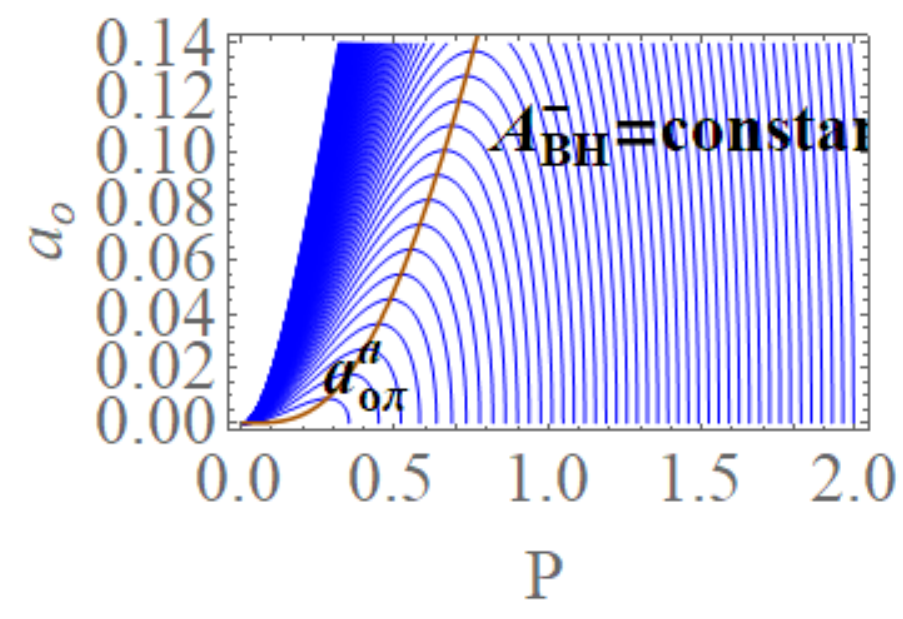}
      \includegraphics[width=3.84cm]{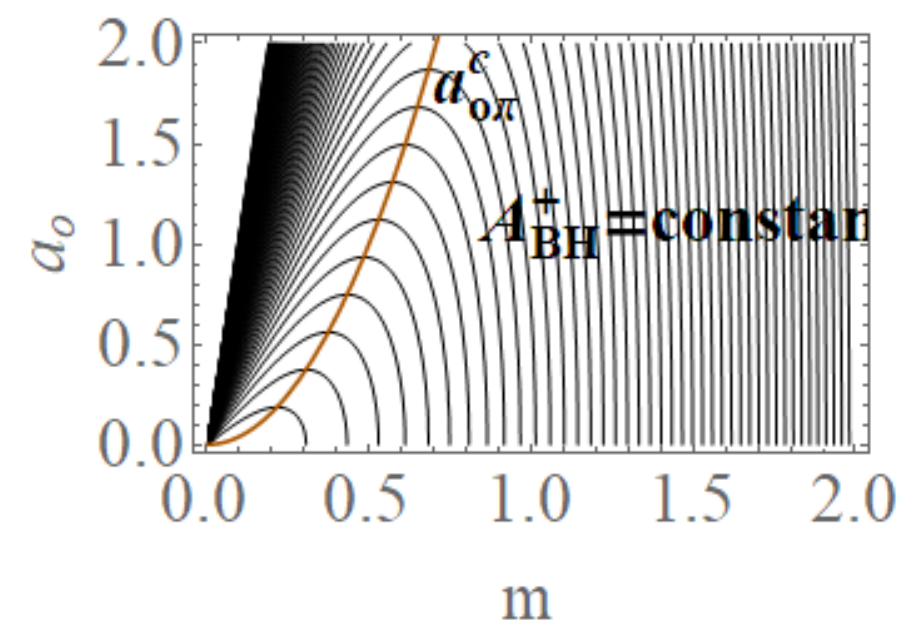}
       \includegraphics[width=3.84cm]{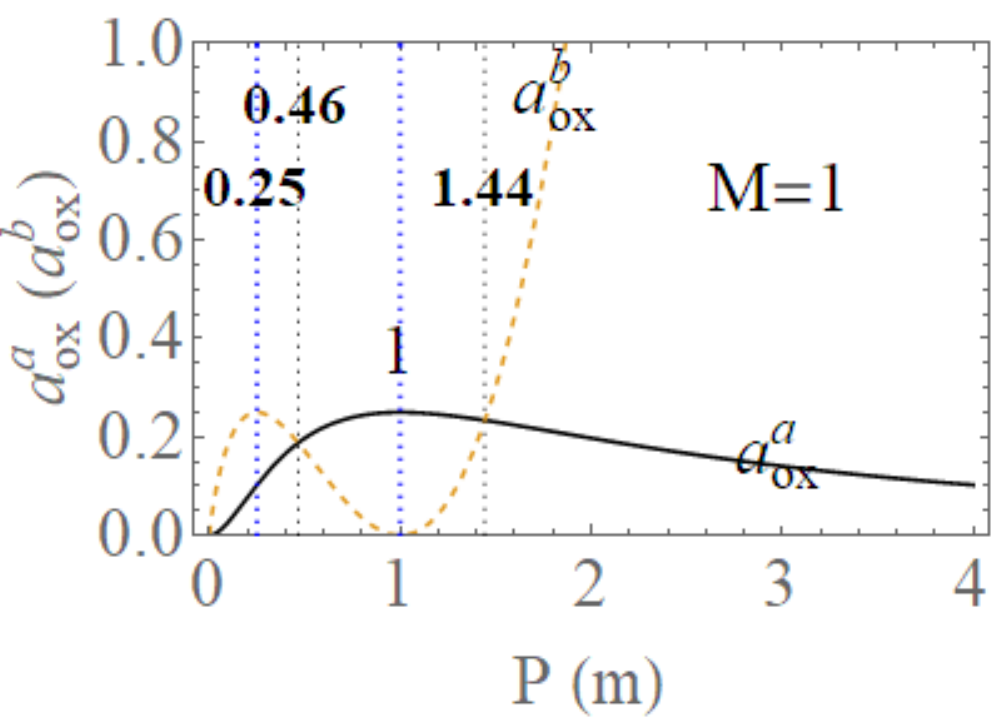}  \caption{Curves of constant BH areas $A_{\mathbf{BH}}^{\pm}$ (BH areas relatives to BH  horizons $r_{\pm}$) are shown. Left  first and second  panels: $A_{\mathbf{BH}}^{\pm}$  in the $(a_o,P)$ plane, respectively. Third panel: area  $A_{\mathbf{BH}}^{+}$  in the $(a_o,m)$  plane. Details on the notation can be found in Table (\ref{Table:pol-cy-multi}). Extreme length parameter $a_{o\pi}^{i}$ for $i\in\{a,b,c\}$ is also shown---Equations (\ref{Eq:aopi-ojh})--Equation~(\ref{Eq:mpbill}). Here, $a_o= A_{min}/8\pi$ is an area parameter where
 $A_{min}$ is   the minimum area gap of LQG,
  $P$   is a  metric polymeric parameter, and
  $M$ is the  ADM mass in the Schwarzschild limit, while
 $m$ is  a parameter depending  on  the polymeric function. Right panel: loop length curves  $a_{ox}^b$  ($a_{ox}^a$) as function of  the loop mass $m$ (polymeric parameter $P$ for $M=1$)---Equation~(\ref{Eq:som-ox-pe}). $a_{ox}^b$, $a_{ox}^a$ are solutions of  $A_{\mathbf{BH}}^-=A_{\mathbf{BH}}^+$ for the BH areas.}\label{Fig:pax12}
\end{figure}
 \begin{figure}
\centering
            \includegraphics[width=5cm]{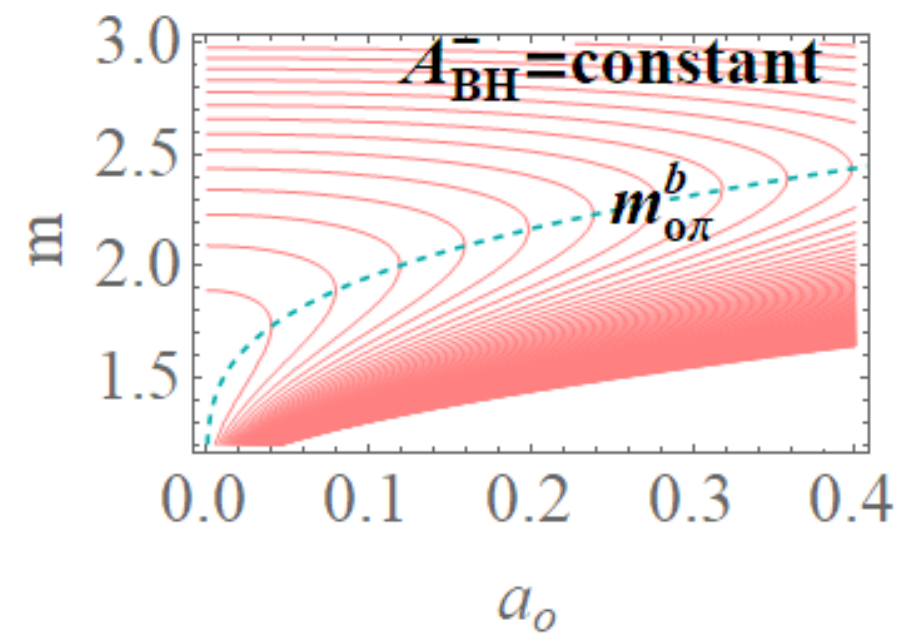}
    \includegraphics[width=5cm]{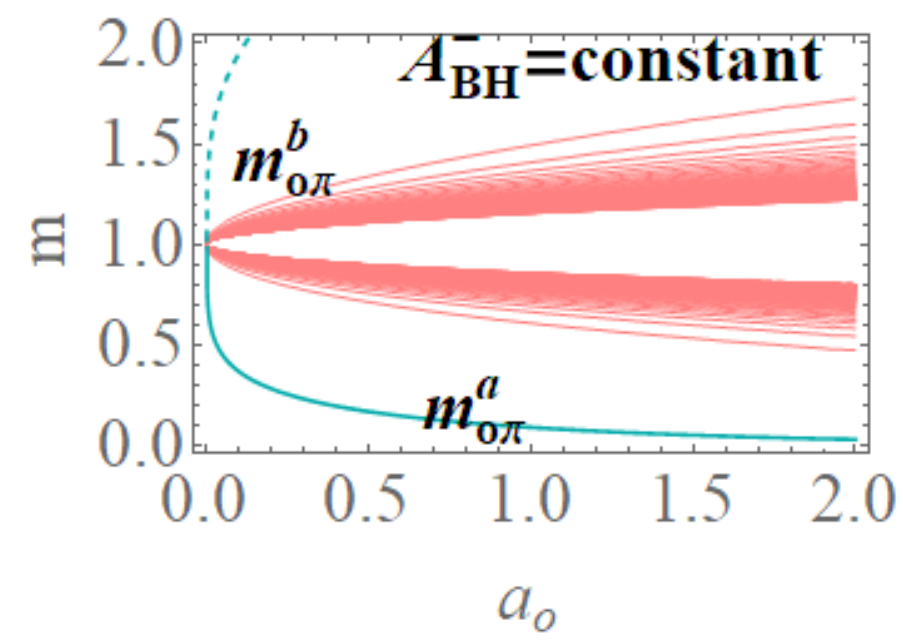}
     \includegraphics[width=5cm]{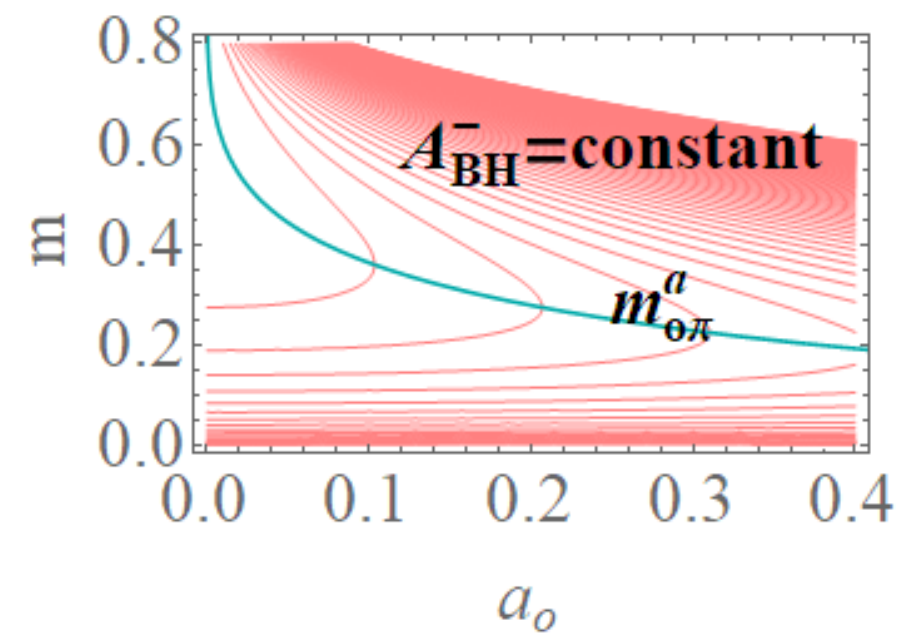}
  \caption{Curves of constant BH areas $A_{\mathbf{BH}}^{-}$ (BH area function  on $r_-$)   in $(m,a_o)$ the plane. Different panels show a focus on ranges of $m$, where $m$ is  a mass parameter depending   on  the polymeric function and the ADM mass while $a_o= A_{min}/8\pi$ is an area parameter and
 $A_{min}$ is    minimum area appearing in LQG (minimum area gap of LQG). Extreme  loop mass $m_{o\pi}^{a}$ and $m_{o\pi}^b$ curves are  shown--- Equation (\ref{Eq:aopi-ojh}) and Equation~(\ref{Eq:mpbill}).
Details on the notation can be found in Table (\ref{Table:pol-cy-multi}).}\label{Fig:pax12m}
\end{figure}
 Interestingly, however,
 the LBHs areas have extreme points: $\partial_P A_{\mathbf{BH}}^ -= 0$ for $a_ {o} =
 a_ {o\pi}^a$ and  $\partial_ {a_o} A_{\mathbf{BH}}^{\pm} = 0$ for $a_ {o} = 0$; finally,
$\partial_P A_{\mathbf{BH}}^ += 0$ for $a_ {o} = a_ {o\pi}^b$, where
 \bea\label{Eq:aopi-ojh}
 &&
a_ {o\pi}^a\equiv4 \sqrt{\frac{P^8}{(P+1)^8}},\quad a_ {o\pi}^b\equiv \frac{4}{(P+1)^4},
\eea
$\partial_M A_{\mathbf{BH}}^ -=
 0$ for ($m_ {o\pi}^{a}, m_ {o\pi}^{b})$,  ($ \partial_ {a_o} A_{\mathbf{BH}}^o = 0, \partial_ {m} A_{\mathbf{BH}}^o = 0$  for $ a_o = 0$), where
\bea\label{Eq:mpbill}
&&
m_{o\pi}^{a}\equiv\frac{1}{2}\left(-2 \sqrt{2} \sqrt[4]{a_{o}}+\sqrt{a_{o}}+2\right),\quad m_{o\pi}^{b}\equiv\sqrt{2} \sqrt[4]{a_{o}}+\frac{\sqrt{a_{o}}}{2}+1,
 \eea
 ($M=1$)
 see Figures (\ref{Fig:pax12m}), where the role of
$P=0.25$ and  $P=1$ is clear.  Note, interestingly, the presence of an extreme of the BHs areas  related to the graph loop parameters.
\\\item
\textbf{Surfaces gravity: }
We can evaluate a LBH "surface gravity" correspondent to the  outer and inner horizons $r_{\pm}$, respectively, as $\kappa_{\pm}\in \{\kappa_{\pm}(M,P),\kappa_{\pm}(m,P)\}$:
\bea&&\label{Eq:surface}
\kappa_-:\quad \kappa_-(m,P)=\frac{4 m^3 P^4 \left(1-P^2\right)}{a_o^2+16 m^4 P^8},\quad \kappa_-(M,P)=\frac{4 M^3 P^4 \left(1-P^2\right)}{(P+1)^6 \left(a_{o}^2+\frac{16 M^4 P^8}{(P+1)^8}\right)};
 \\
 &&\kappa_+:\quad \kappa_+(m,P)=\frac{4 m^3 \left(1-P^2\right)}{a_o^2+16 m^4},\quad \kappa_+(M,P)=\frac{4 M^3 \left(1-P^2\right)}{(P+1)^6 \left(a_{o}^2+\frac{16 M^4}{(P+1)^8}\right)}.
\eea
Comparing the extended planes of Kerr geometries and the regular LBH geometries of Figures (\ref{Fig:vengplre}), we expect  surface gravities to vanish in some extreme conditions on the loop graph parameters. It is then clear that $\kappa_\pm=0$ for $P=1$ and $\kappa_{\pm}>0$ for $P<1$, which is the   region of  the polymeric function  values we  explore here. Thus, there is
\bea
  \lim_{P\rightarrow 0}\kappa_-=0,\quad \kappa_+(P\approx 0)=\frac{4 m^3}{a_o^2+16 m^4}-\frac{4 m^3 P^2}{a_o^2+16 m^4}+\mathrm{O}\left(P^4\right).
  \eea
The limiting $P=1$, occurring in the extended plane of  Figures (\ref{Fig:vengplre}) at $r=1/2$,   is also an extreme  for the  $\kappa_{\pm}: \partial_{a_o}\kappa_{\pm}(M,P)=0$.
Extremes $\partial_M\kappa_{\pm}(M,P)$ are for  limiting conditions $M=0$, $P=(0,1)$, $a_o=0$,  relating the limiting values on $\mathcal{P}$ and the ADM mass. A further  extreme is for minimal area $a_o= a^1_o$ (or $M=M(a_o^1)$) for $\kappa_-(M,P)$ and $a_o=a_o^2$ for $\kappa_+(M,P)$.
For convenience, we report in  Table (\ref{Table:TolorP1})  all  the relevant minimal areas $a_o^i$ for $i\in\{1,...,7\}$,  introduced in this discussion and represented in Figures (\ref{Fig:TolorP1}).
Similarly, there is  $\partial_P\kappa_ {\pm}(M,P)=0$  for $P = 1$ and  $M = 0$,
and  also for $P\in[0, 2/3]$ with $a_o = a_o^3$ for  $\kappa_{-}(M,P)$ and $P\leq 1/2$ for $a_o = a_o^4$.

There is an extreme for  $\kappa_{\pm}(m,P)$ at $a_o\neq0$  for the  limiting condition $M=0$--(\cite{Aleshi2012zz}).
There is then
$\partial_P\kappa_ {+}(m,P)$   for $P = 0$,
  $\partial_P\kappa_ - (m,P)= 0$ for $(m = 0, P = 0)$ and  for $ P\in] 0, \sqrt {2/3}[$ and $a_o = a_o^5$.
  On the other hand, $ \partial_m\kappa_ -(m,P)= 0$, for the limiting cases  $m = 0, P = (0, 1)$ (including  $(P = 1  a_o = 0)$) and for $a= a_o^6$.  Analogously, there is
$\partial_m\kappa_{ \pm}(m,P)= 0 $, an extreme  condition  having the special solution $a_o=a_o^7$, notably independent from the polymeric parameter.
\begin{table}[h!]
 \caption{Quantities $a_o^{i}$ for $i\in\{1,...,7\}$  represented in Figures (\ref{Fig:TolorP1}). $a_o$ is the minimal loop areas, functions $a_o^{i}$ are extremes of the surfaces areas $\kappa_{\pm}$. $M$ is the ADM mass, $P$ is the polymeric function, and $m$ is the polymeric mass. }\label{Table:TolorP1}
\centering
\begin{tabular}{|l l|}
\hline
$a_o^1 \equiv\frac{4 M^2 P^4}{\sqrt{3} (P+1)^4},   $ & $
   a_o^2 \equiv\frac{4 M^2}{\sqrt{3} (P+1)^4}$\\$ a_o^3 \equiv\frac{4 M^2 P^4}{(P+1)^4} \sqrt{\frac{ (P-2)}{(3 P-2)}},$&$
   a_o^4\equiv\frac{4M^2}{(P+1)^4} \sqrt{\frac{ (2 P-1)}{(2 P-3)}},$\\
   $ a_o^5 \equiv{4 m^2 P^4}\sqrt{\frac{\left(P^2-2\right)}{3 P^2-2}},$&$ a_o^6 \equiv\frac{4m^2 P^4 }{\sqrt{3}},$\\
    $ a_o^7 \equiv\frac{4m^2}{\sqrt{3}}$&\\
\hline
\end{tabular}
 \end{table}
In Figures (\ref{Fig:TolorP1}), we also considered an extended region of the $\mathcal{P}$ parameters.
\begin{figure*}
\centering
  \includegraphics[width=4.75cm]{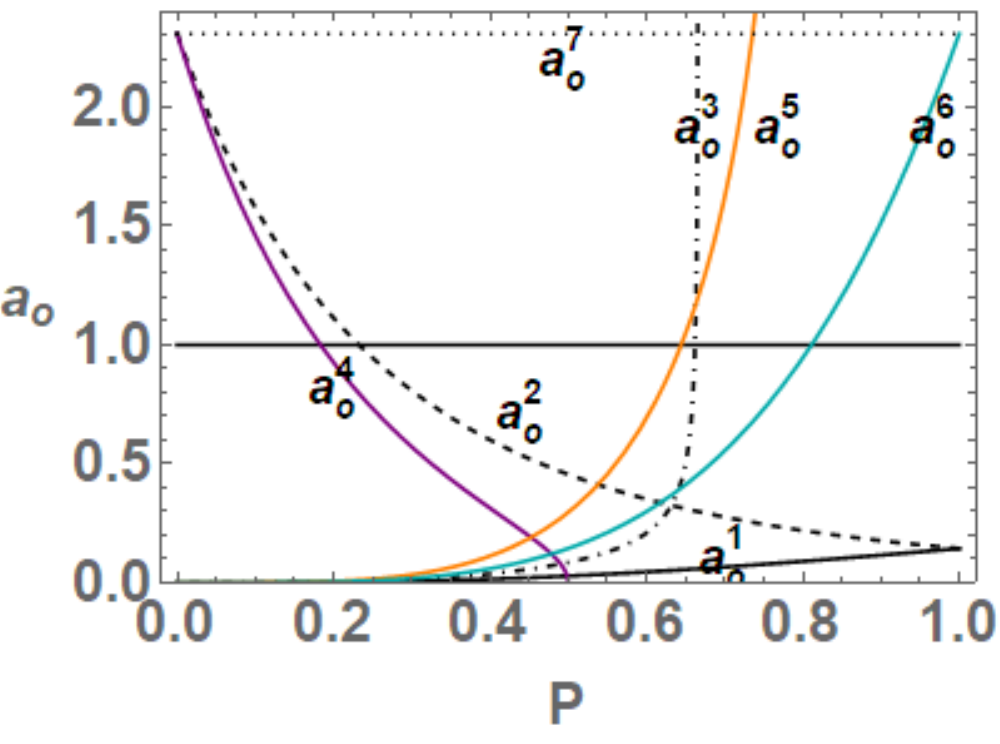}
    \includegraphics[width=5cm]{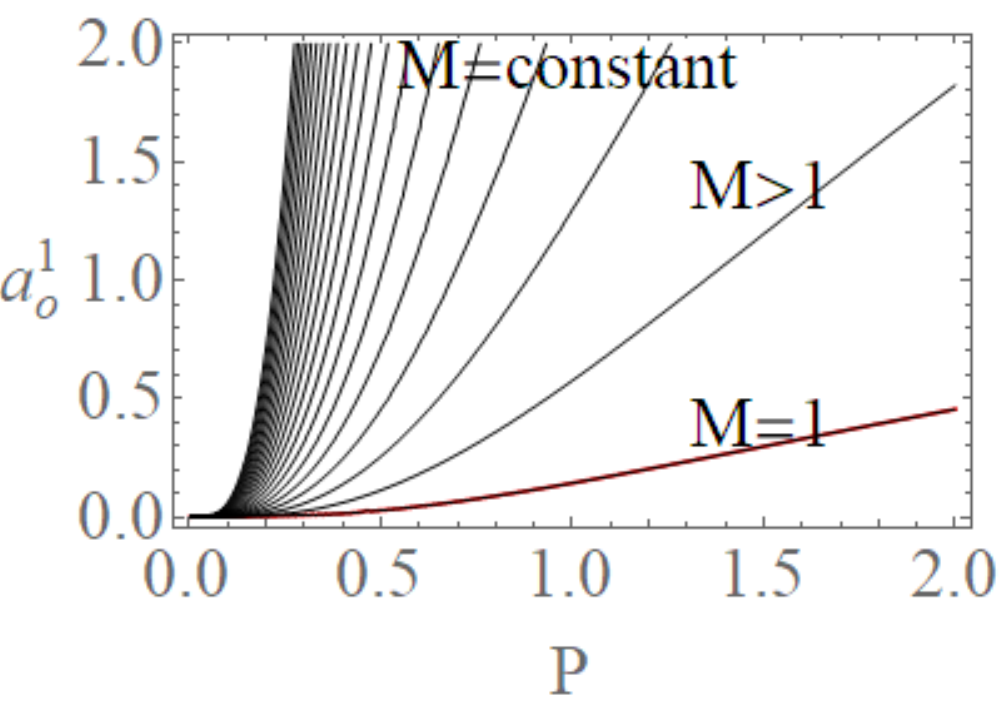}
      \includegraphics[width=5cm]{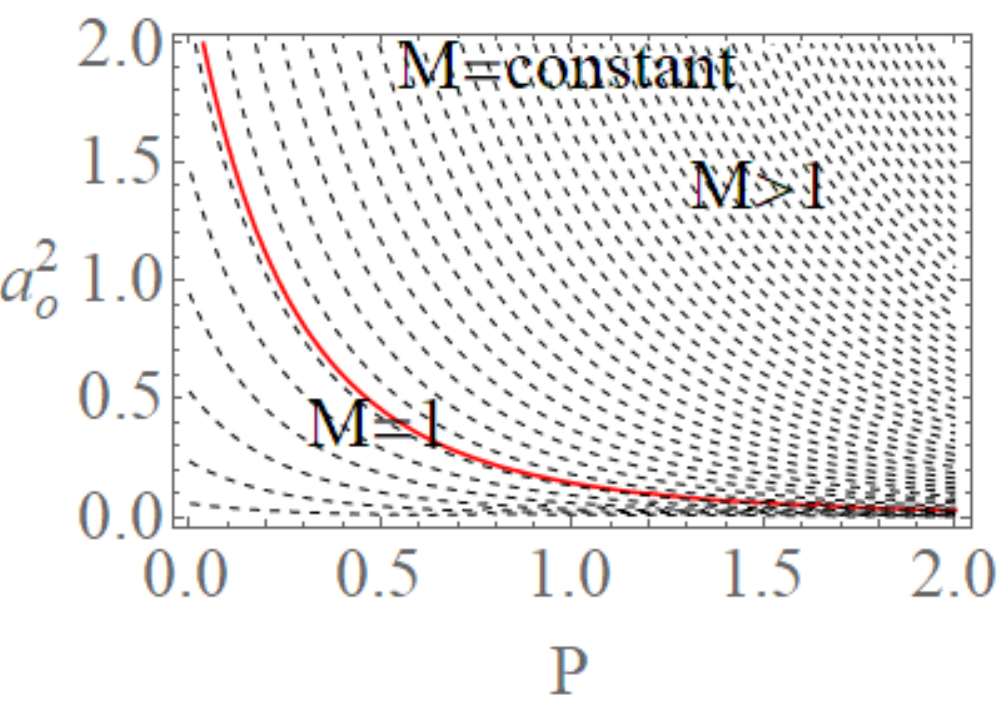}\\
    \includegraphics[width=5cm]{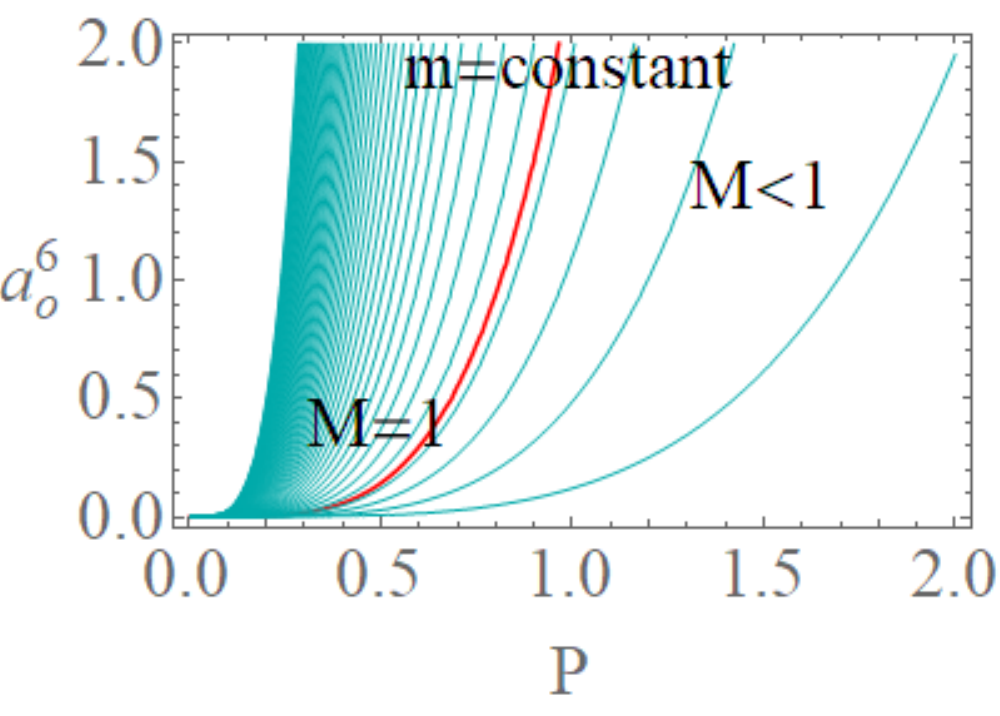}
          \includegraphics[width=4.75cm]{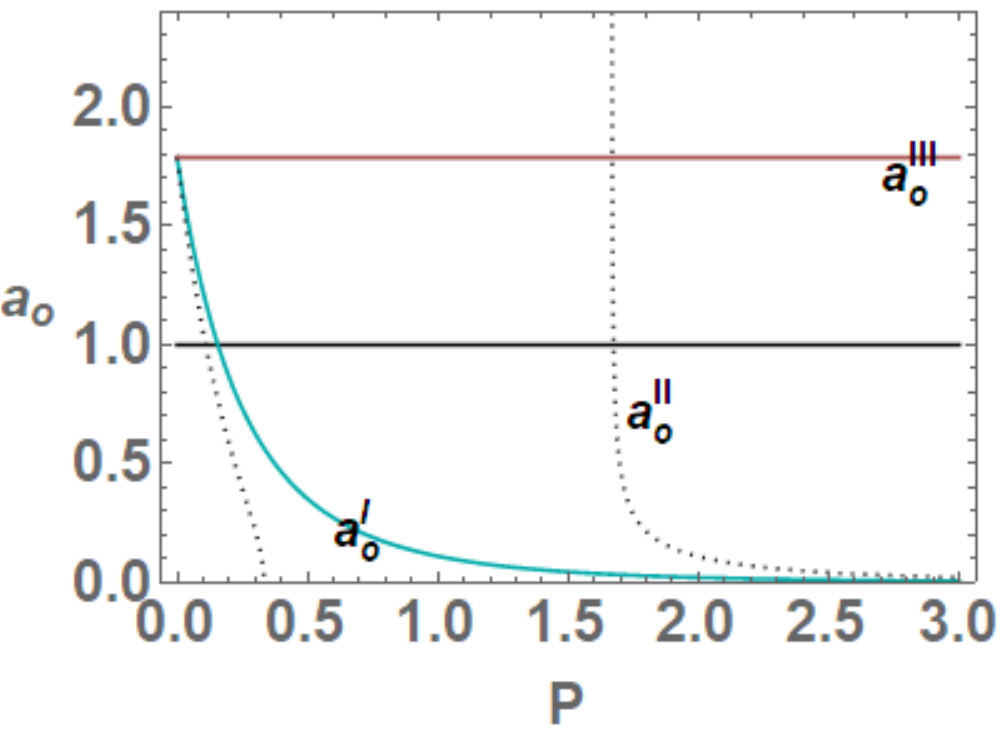}
    \includegraphics[width=5cm]{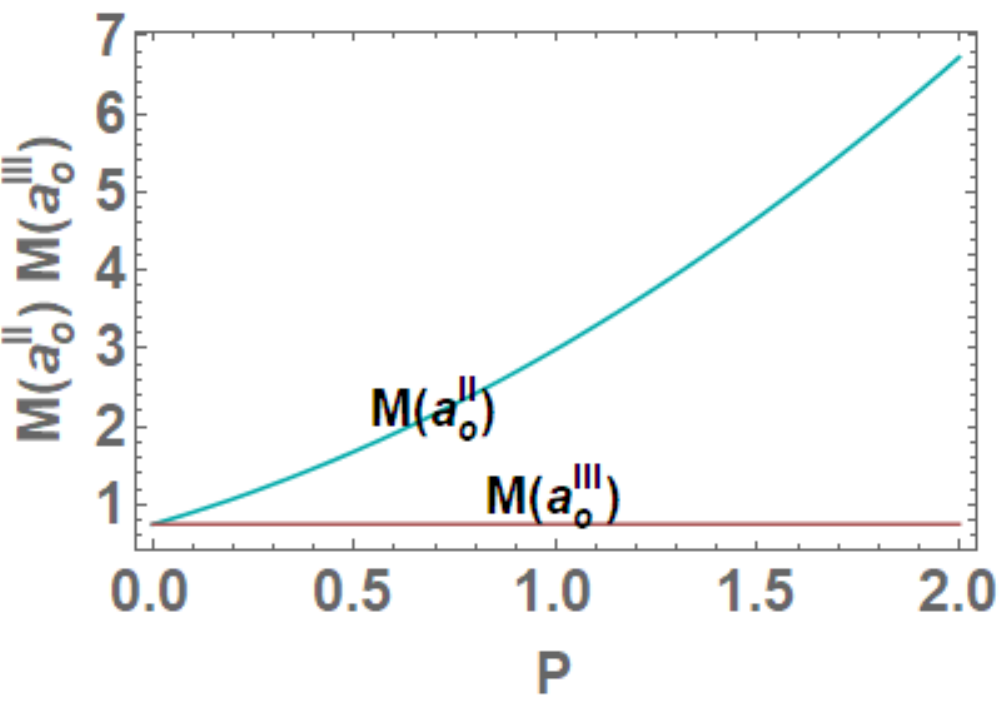}
  \caption{Left upper panel: quantities $a_o^{i}$ for $i\in\{1,...,7\}$ of  Table (\ref{Table:TolorP1}) as functions of the metric polymeric parameter $P\in[0,1]$ for $M=1$ or $m=1$ ($M$ is the  ADM mass in the Schwarzschild limit  and
 $m$ is  parameter depends  on  the polymeric function). Upper center and right panels and bottom-left panels show $a_o^{i}$ as functions of $P$ for $M=$constant and $m=$constant. The center bottom panel shows   $a_o^{\upsilon}$ for $\upsilon\in\{I,II,III\}$ of Equation~(\ref{Eq:det-rom}) solutions of $\partial_ {a_o}  L(M,P) = 0$, where  $a_o= A_{min}/8\pi$, is an area parameter where
 $A_{min}$ is  a minimum area appearing in LQG (minimum area gap of LQG). The bottom right panel represents $M(a_o^i)=$constant  in the plane $(P,a_o)$. See also Table (\ref{Table:pol-cy-multi}) for  further details on the notation.}\label{Fig:TolorP1}
\end{figure*}
\\
\item
\textbf{The temperatures: }
The evaluation of the temperature associated with the (regular) LBH  proceeds directly in terms of surface gravity $\kappa_+$:
\bea
T_{\mathbf{BH}}=\frac{\kappa_+}{2 \pi }\quad\mbox{or}\quad T_{\mathbf{BH}}(m)=\frac{(2 m)^3 \left(1-P^2\right)}{4 \pi  \left(a_o^2+(2 m)^4\right)}.
\eea
We actually evaluate the temperatures $T_{\mathbf{BH}}^{\pm}$ in terms of $\kappa_{\pm}$, respectively, for the outer and  inner horizons $r_{\pm}$. The interpretation and the evaluation of the temperature $T_{\mathbf{BH}}^-$ is debated in literature. In this analysis, while we intend clearly  $T_{\mathbf{BH}}^{+}\equiv T_{\mathbf{BH}} $ as the BH temperature, when we intend the BH in the extended plane, as in Figures (\ref{Fig:cononapo16}), then we need to consider $T_{\mathbf{BH}}^{\pm}$ .
On the other hand, considering the extended plane  $(P-r)$ or $(\epsilon-r)$, we expect the occurrence of an "extreme" case, where the temperature is null (similarly  to the case of extreme Kerr BH) as made evident from the study of the surface gravity $\kappa_{\pm}$.
Temperature is vanishing for $m\approx 0$, in the case $T_{\mathbf{BH}}^+(m,P)$:
\bea&&
\lim_{m\rightarrow\upsilon}T_{\mathbf{BH}}=0, \quad  \upsilon\equiv\{\infty,0\}. \\ &&\nonumber   T_{\mathbf{BH}}(m\rightarrow \infty)= \frac{1-P^2}{8 \pi  m}+\mathrm{O}\left(\left(\frac{1}{m}\right)^2\right),
\\
 &&T_{\mathbf{BH}}(P\approx0)=\frac{2 m^3}{\pi  (a_o^2+16 m^4)}-\frac{2 m^3 P^2}{\pi  a_o^2+16 \pi  m^4}+\mathrm{O}\left(P^4\right).
\eea
The analysis in Figures (\ref{Fig:cononapo16}) investigates {extended parameter regions}, where  negative surface gravity is possible.
(The negative $\kappa_{\pm}$ and therefore $T_{\mathbf{BH}}^{\pm}$  is a delicate and   intriguing aspect of classical and loop quantum BHs,  tightly connected to the white holes definition---\cite{Barrau:2018rts,Barrau:2019swg,Haggard}).
\\\item
\textbf{The luminosity}

In the analysis of luminosity, we consider \cite{Aleshi2012zz}. The luminosity can be estimated  by considering the Stefan--Boltzmann law as $L (m) =\alpha A_{\mathbf{BH}} (m) T_{\mathbf{BH}} ^4(m)$,  where $A_{\mathbf{BH}}$ is the $BH$ area  (on horizon $r_+$), and $\alpha$  is a factor depending on the evaluation model adapted for the luminosity. However, in this work, we mainly consider the quantity $L/\alpha$. By assuming $\alpha=$constant, we focus on the analysis of luminosity with the variation of the  {$\mathcal{P}$} parameters of the LQG graph and on the metric bundles.
 Studying  $L(m)/\alpha$  (or $L(M)/\alpha$), we investigate the regular BH mass  evaporation process (the energy flux particularly where BH evaporation occurs through
the Hawking emission in the proximity of the BH outer horizon $r_+$,  with a temperature evaluated according to the
Bekenstein--Hawking law and  connected therefore to the surface gravity $\kappa_+$). We perform our investigation  considering different values of $\mathcal{P}$ and on the geometries connected by the metric bundles.
The luminosity is, therefore, in terms of $m$:
\bea&&\label{Eq:luci}
L (m)=\alpha  A_{\mathbf{BH}}(m) T_{\mathbf{BH}}(m)^4=\frac{16 \alpha  m^{10} \left(1-P^2\right)^4}{\pi ^3 \left(a_o^2+16 m^4\right)^3}.
\eea
In the Schwarzschild limit, $P\rightarrow 0$ (correspondent to $m\rightarrow M$), there is
\bea&&
L=\frac{16 \alpha  m^{10}}{\pi ^3 \left(a_o^2+16 m^4\right)^3}-\frac{64 P^2 \left(\alpha  m^{10}\right)}{\pi ^3 \left(a_o^2+16 m^4\right)^3}+\mathrm{O}\left(P^4\right).
\eea
We should note that:
$L(m) = 0$ for $P = 1$ or $ m = 0$  {(in this special analysis, we consider $m$ and $P$ independent---for the limiting condition $m=0$ for the approximate geometry; see, for example, discussion in \cite{Aleshi2012zz})}.

It is, however, relevant to consider the extremes of luminosity function $L$  (related to  the BH evaporation process for mass loss).
Therefore, conveniently, we introduce here the following special values of the minimal length parameter $a_o$:
\bea&&\label{Eq:det-rom}
 a_o^I\equiv\frac{4 \sqrt{5} M^2}{5 (P+1)^4},\quad a_o^{II}\equiv4 \sqrt{\frac{M^4 (3 P-1)}{(P+1)^8 (3 P-5)}},\quad a_o^{III}\equiv\frac{4 \sqrt{5} m^2}{5},
\eea
represented in Figures (\ref{Fig:TolorP1}).
 Therefore, there is  $ \partial_m L(m) = 0$ for
$a_o=a_o^{III}$  and for some limiting cases on $\mathcal{P}$ (for example, vertices of the LBHs  triangle in Figures (\ref{Fig:vengplre}), i.e., limiting geometries for  parameters values as studied in Section (\ref{Sec:metric-bundles})).
Considering  explicitly dependence  on $P$, there is
 $\partial_P L(m) = 0 $ and $\partial_{a_o} L(m) = 0 $ for the limiting cases on the parameters $\mathcal{P}$.
 We now  focus on the situations when $m=m(M,P)$. In this case, there is  an extreme  for the minimal area.
(We include also the extremes  $\partial_M  L(M,P) = 0$  in the limiting cases and  for $ a = a_o^I$).
Then, there is
$\partial_{P} L(M,P) = 0$ in the limiting cases and $P\in[0, 1/3]$  for $a = a_o^{II}$, where
$\partial_ {a_o}  L(M,P) = 0$ only for the limiting cases.
\begin{figure*}
  \centering
  \includegraphics[width=5.3cm]{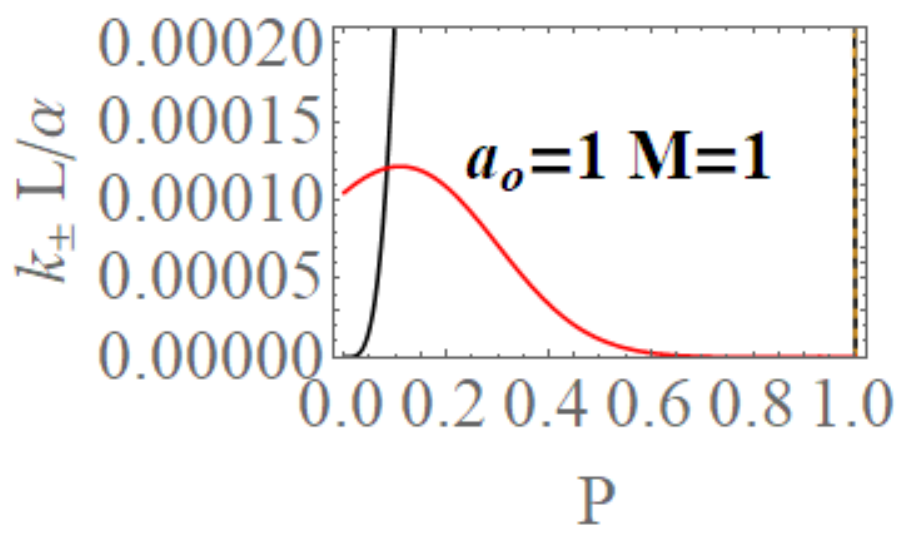}
    \includegraphics[width=5cm]{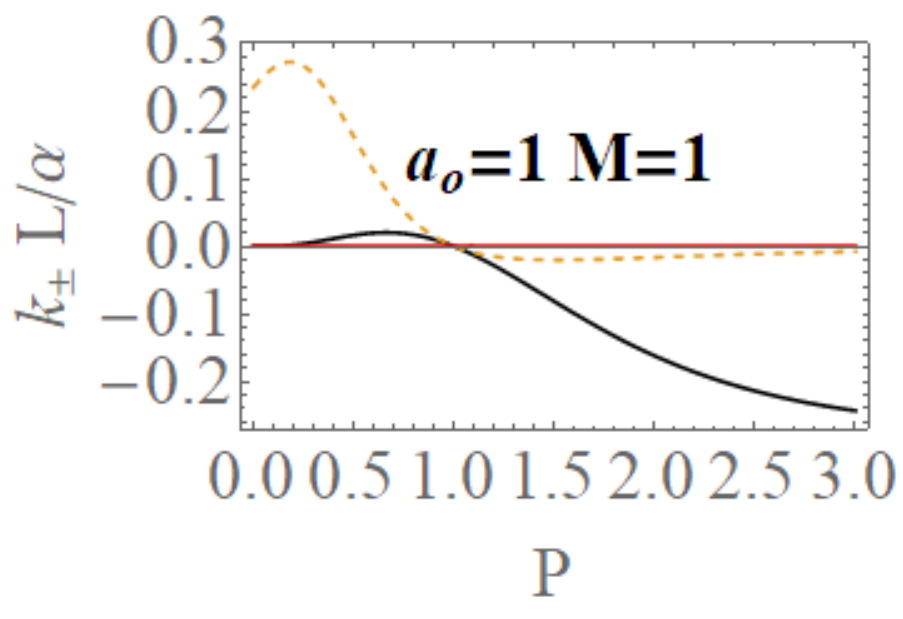}
      \includegraphics[width=5cm]{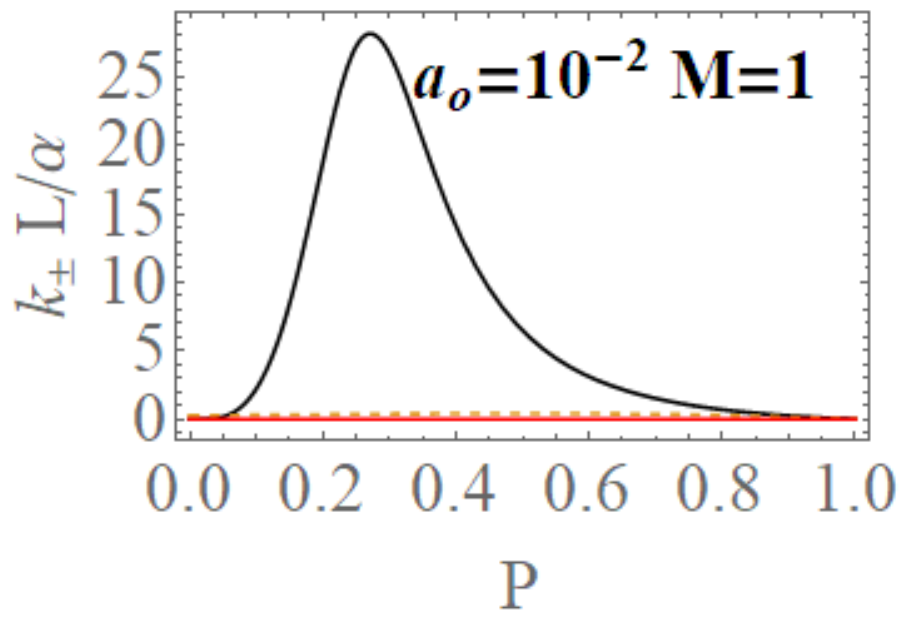}\\
        \includegraphics[width=5cm]{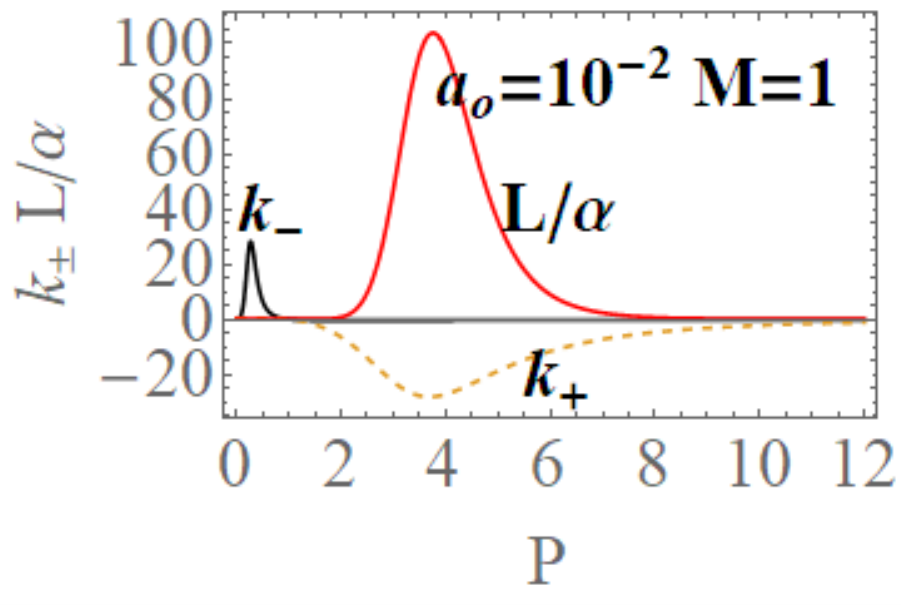}
          \includegraphics[width=5.3cm]{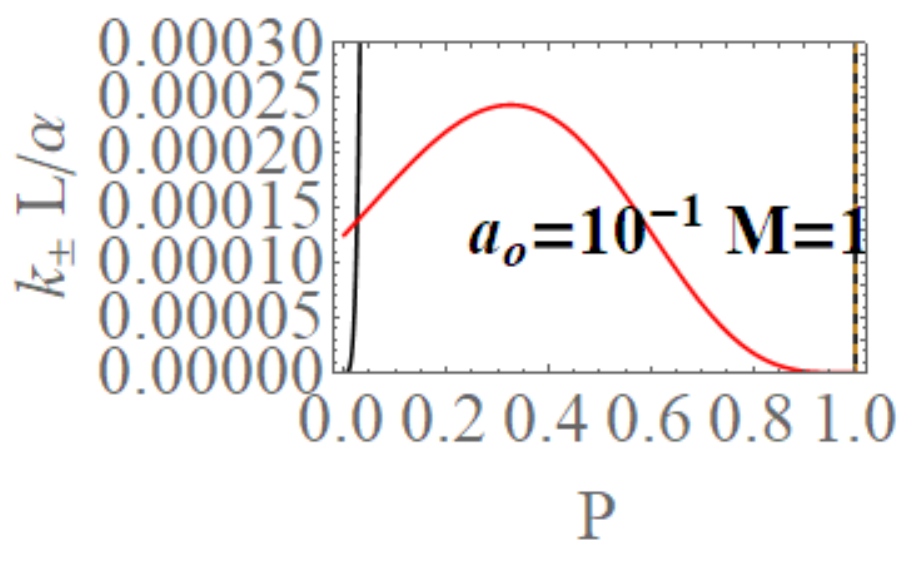}
            \includegraphics[width=5cm]{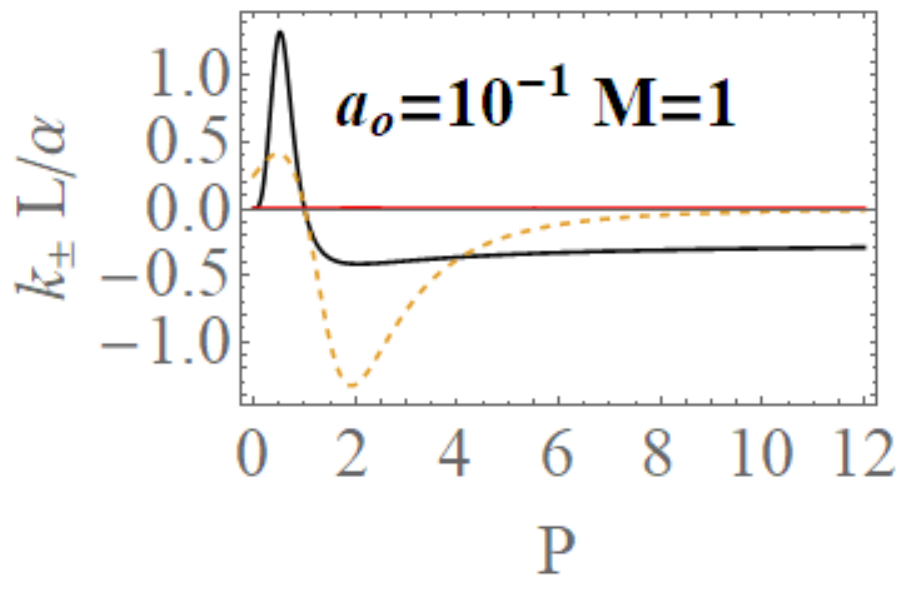}\\
              \includegraphics[width=5cm]{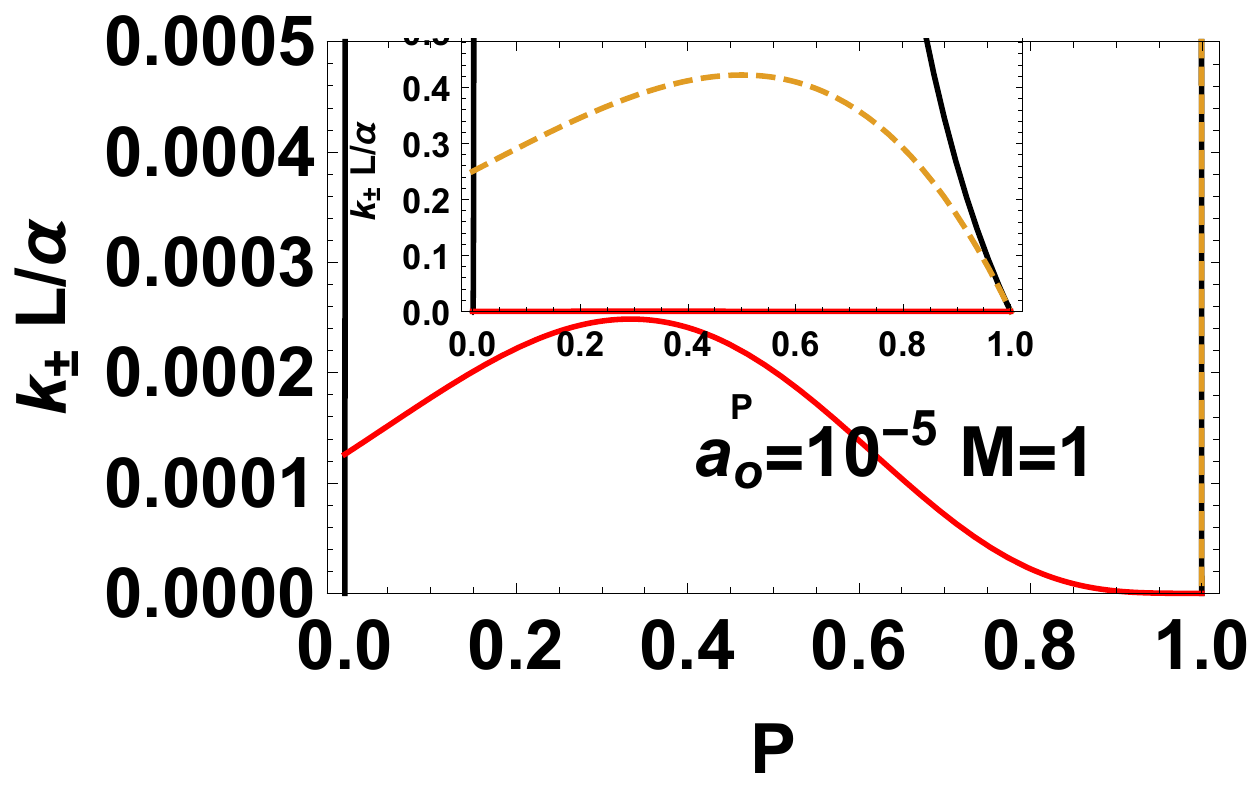}
                \includegraphics[width=5cm]{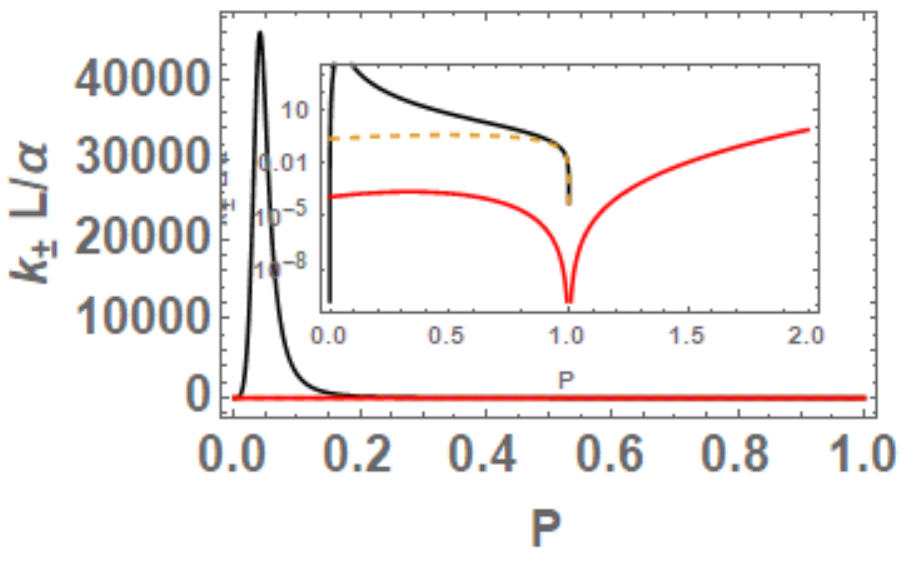}
                  \includegraphics[width=5cm]{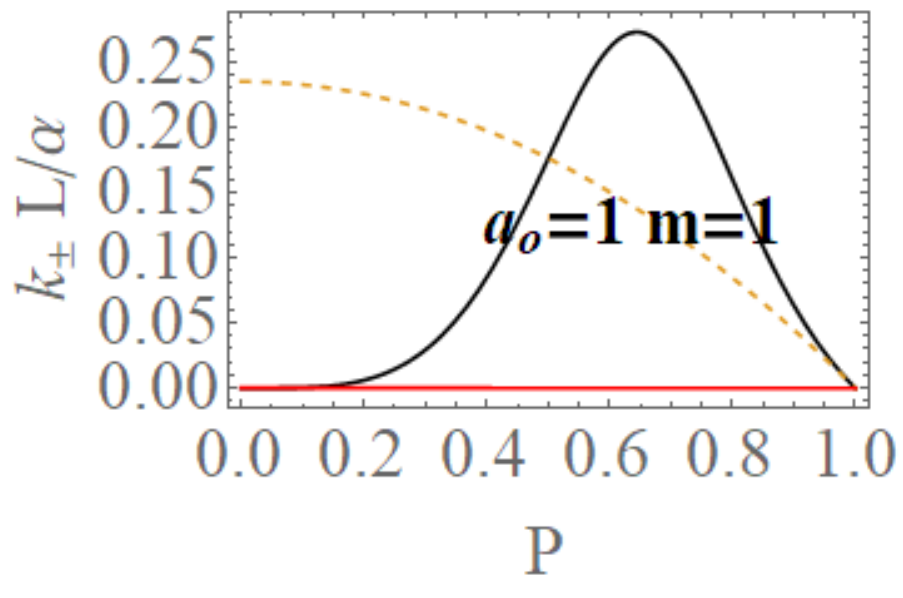}\\
                    \includegraphics[width=5cm]{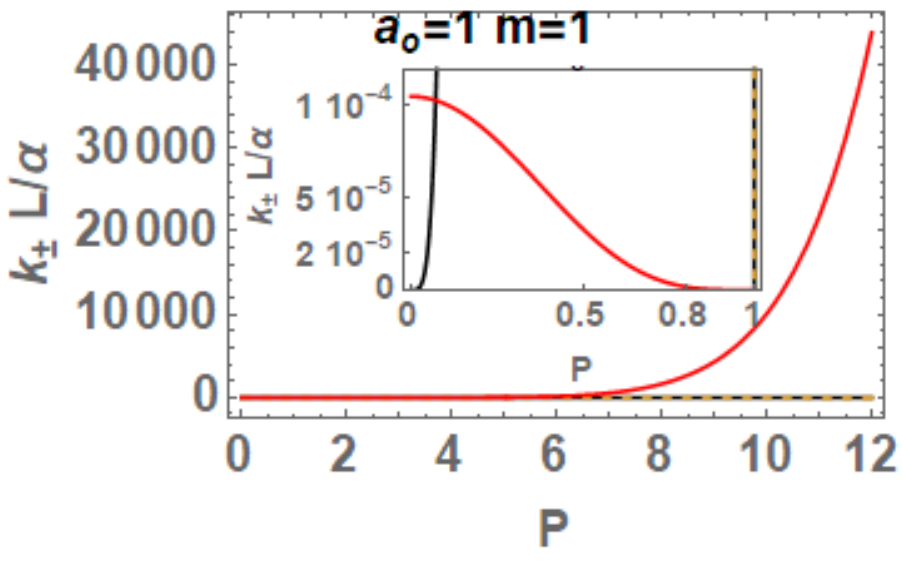}
                      \includegraphics[width=5cm]{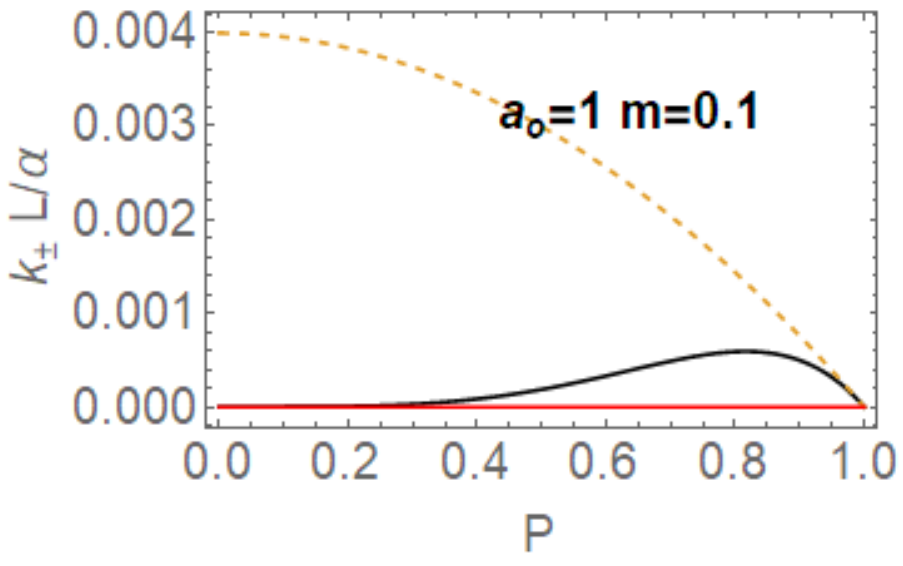}
                        \includegraphics[width=5cm]{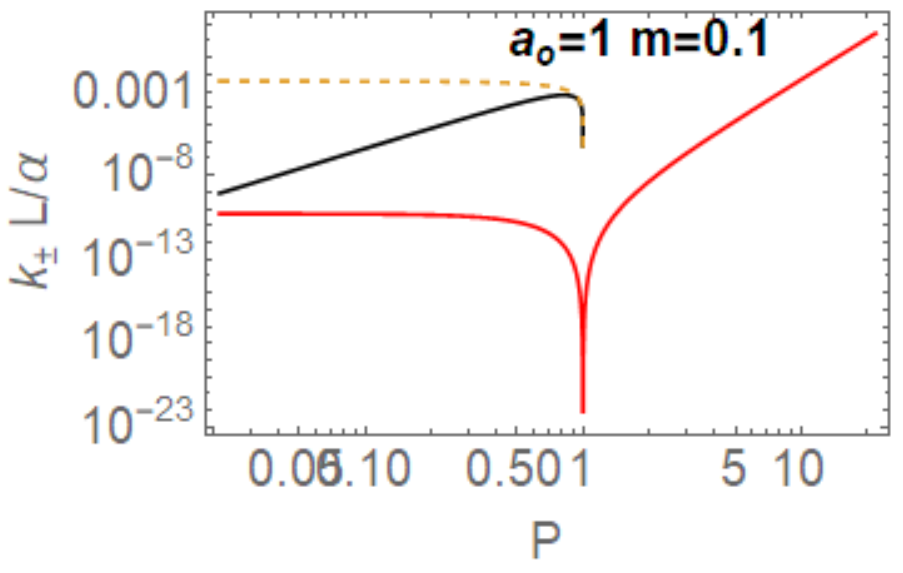}\\
      \includegraphics[width=5cm]{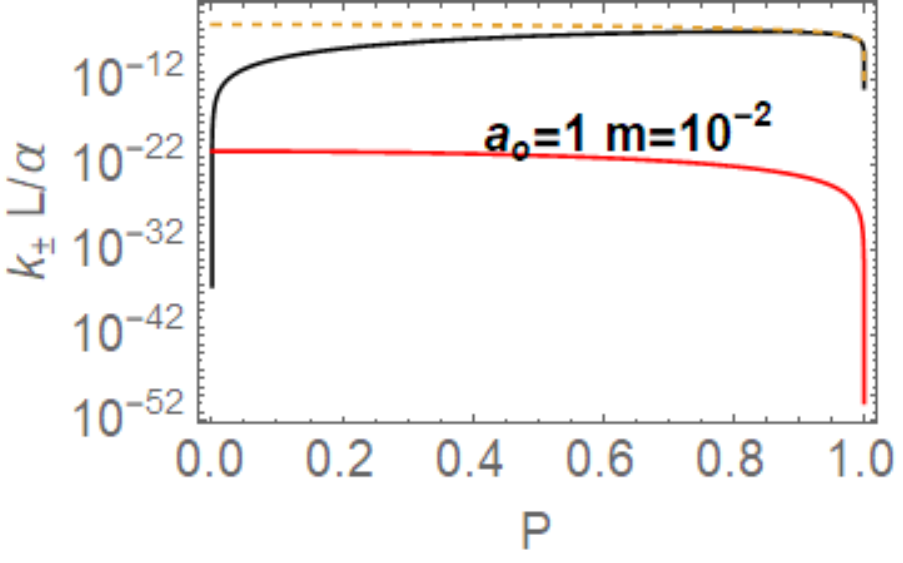}
        \includegraphics[width=5cm]{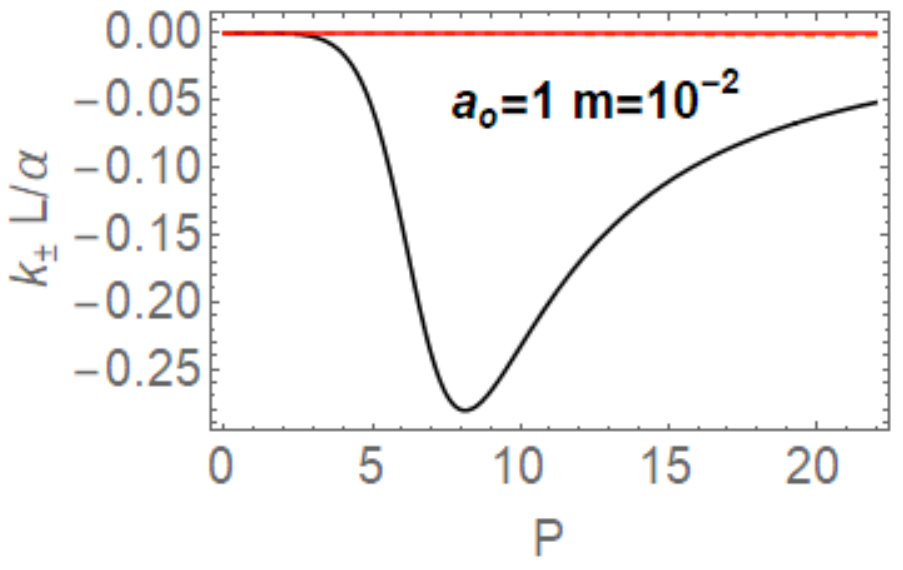}
          \includegraphics[width=5cm]{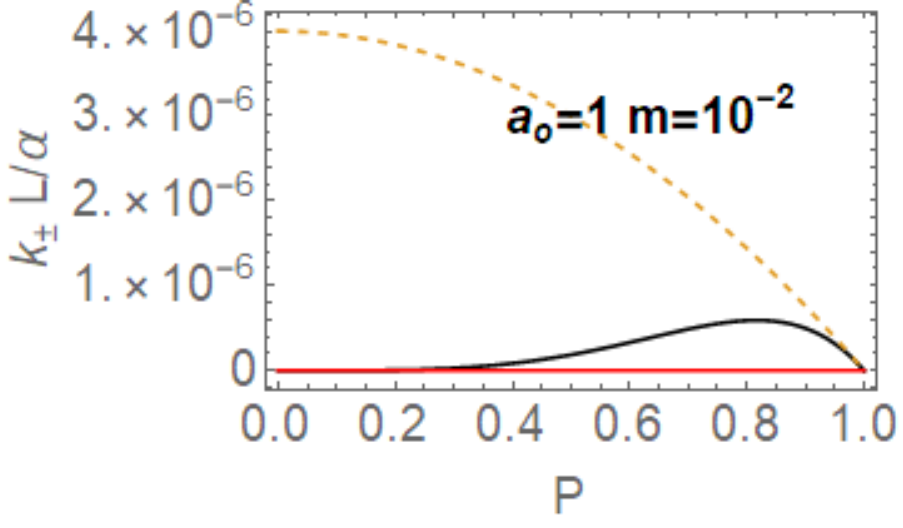}\\
            \includegraphics[width=5cm]{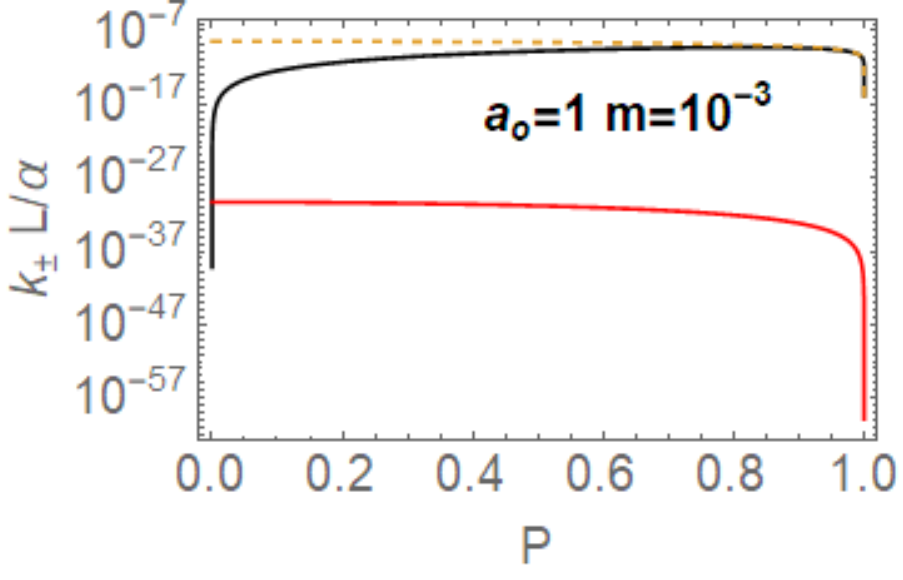}
            \includegraphics[width=5.4cm]{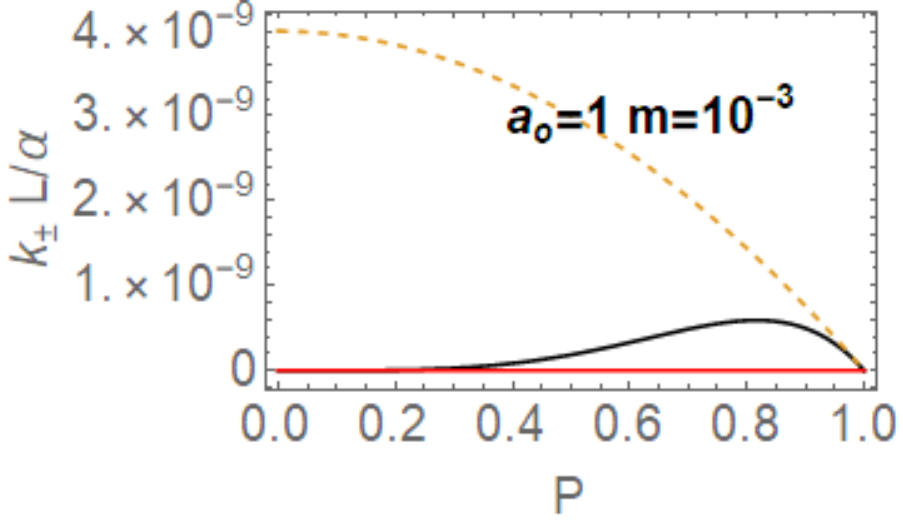}
            \includegraphics[width=5cm]{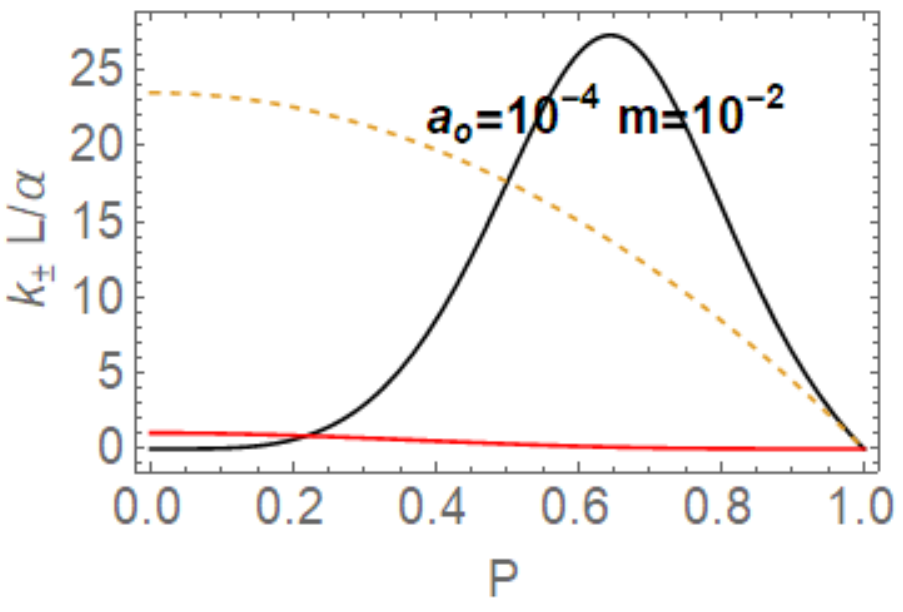}
  \caption{Plots of the surface gravity $\kappa_{\pm}$ and luminosity $L/\alpha$ as a function of  the polymeric parameter, ($\alpha$ is a constant)  evaluated  in the two different approaches and for selected values of  the parameters---see also Table (\ref{Table:pol-cy-multi}). $a_o= A_{min}/8\pi$ is an area parameter where
 $A_{min}$ is  a minimum area appearing in LQG (minimum area gap of LQG).
  $P$   is the  metric polymeric parameter,
  $M$ is the  ADM mass in the Schwarzschild limit, while
 $m$ is  a parameter that depends on the polymeric function.}\label{Fig:cononapo16}
\end{figure*}
In Figure(\ref{Fig:cononapo16}), we consider different  limiting cases on the LBH model parameters $\mathcal{P}=(P,m,a_o)$ on the BHs quantities $\kappa_{\pm}$, $L/\alpha$, and the temperature $T_{\mathbf{BH}}^{\pm}$, making  evident the presence of  extreme points and even negative values of temperature  in extended regions of parameters.
\\
\item
\textbf{LBHs thermodynamical properties and  MBs}
We now consider the  quantities  of the regular LBHs geometries, $\kappa_{\pm}$ (surface gravity) and  $ L/\alpha$ (luminosity)  evaluated on the  metric bundles of the geometry. This analysis will connect different geometries of the same metric  bundle through their thermodynamical properties. This  treatment of the thermodynamical properties and LQG-BH will also characterize the role of the graph parameters in shaping different solutions. Eventually, this analysis  connects the  extended plane  parameter variation  with the  transition from a LBH solution to another solution.
\begin{figure*}[h!]
  \centering
  \includegraphics[width=5cm]{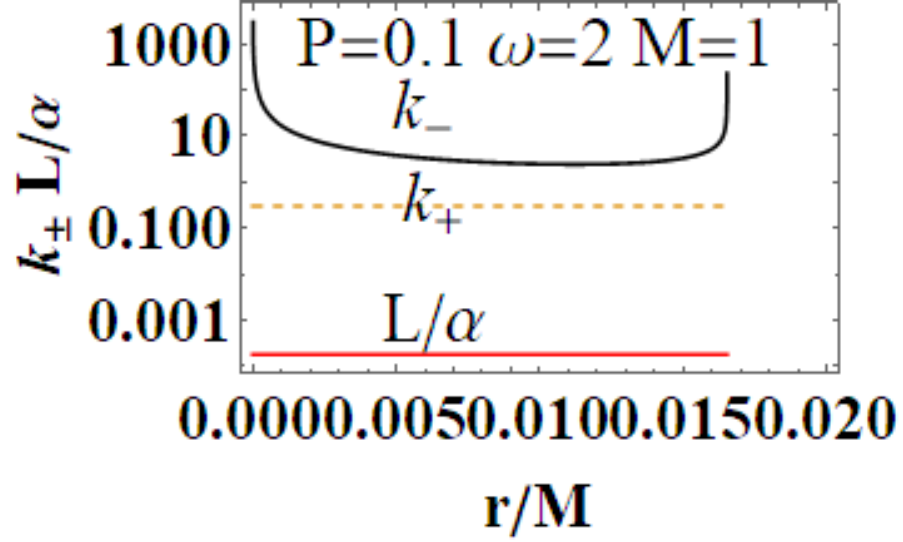}
    \includegraphics[width=5cm]{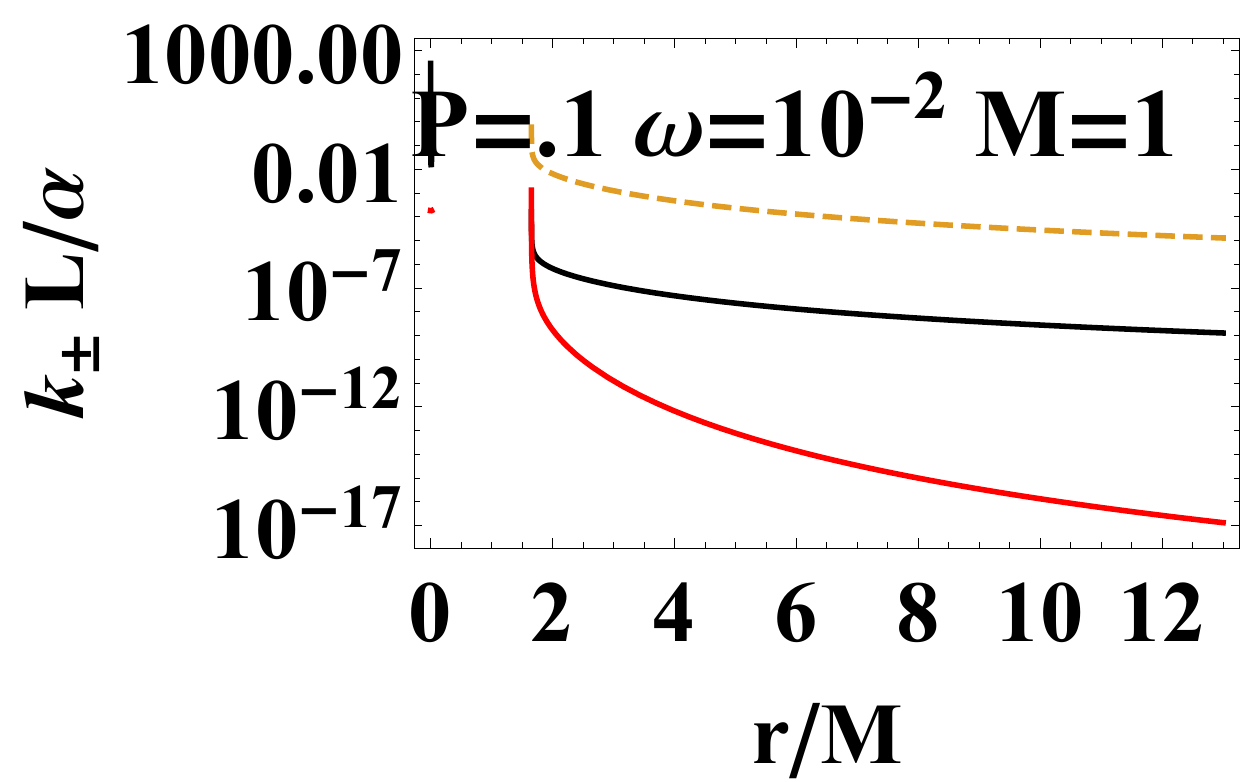}
      \includegraphics[width=5cm]{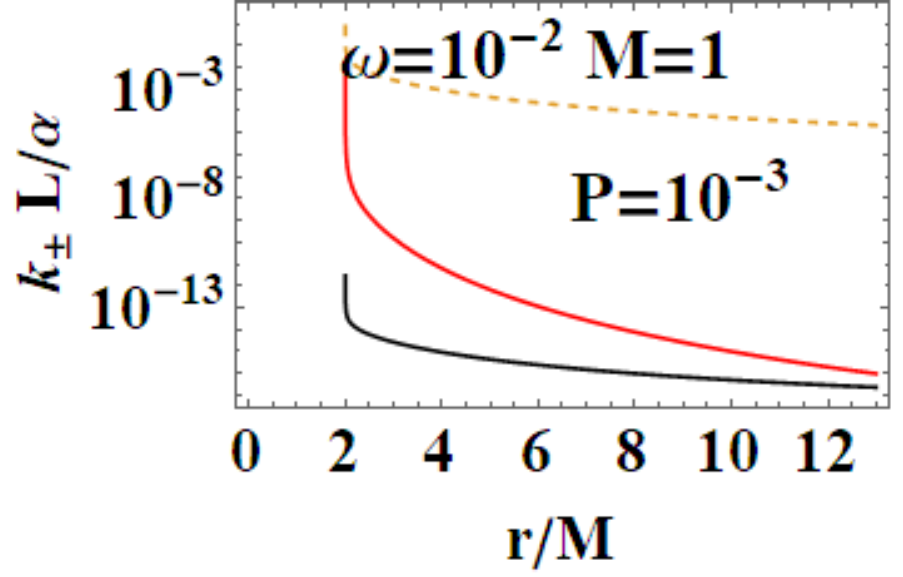}\\
        \includegraphics[width=5cm]{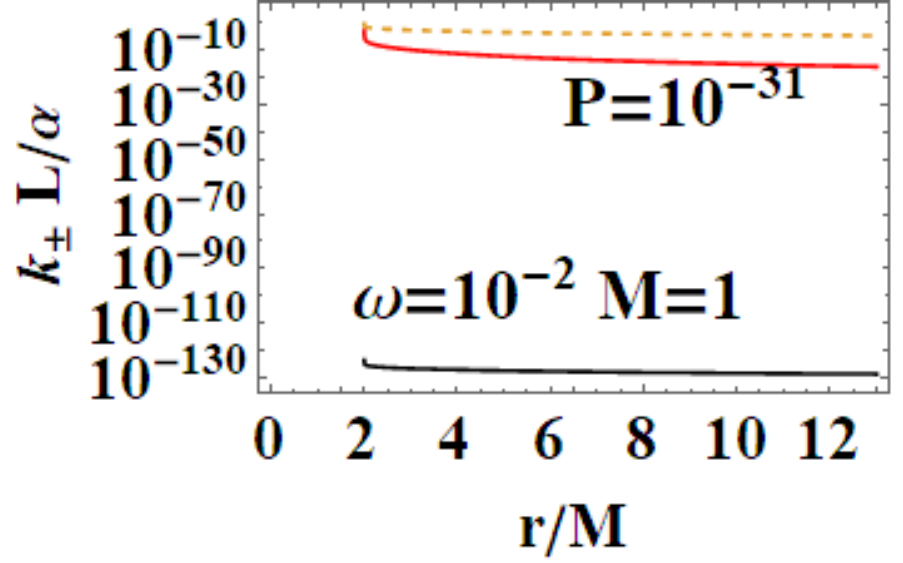}
          \includegraphics[width=5cm]{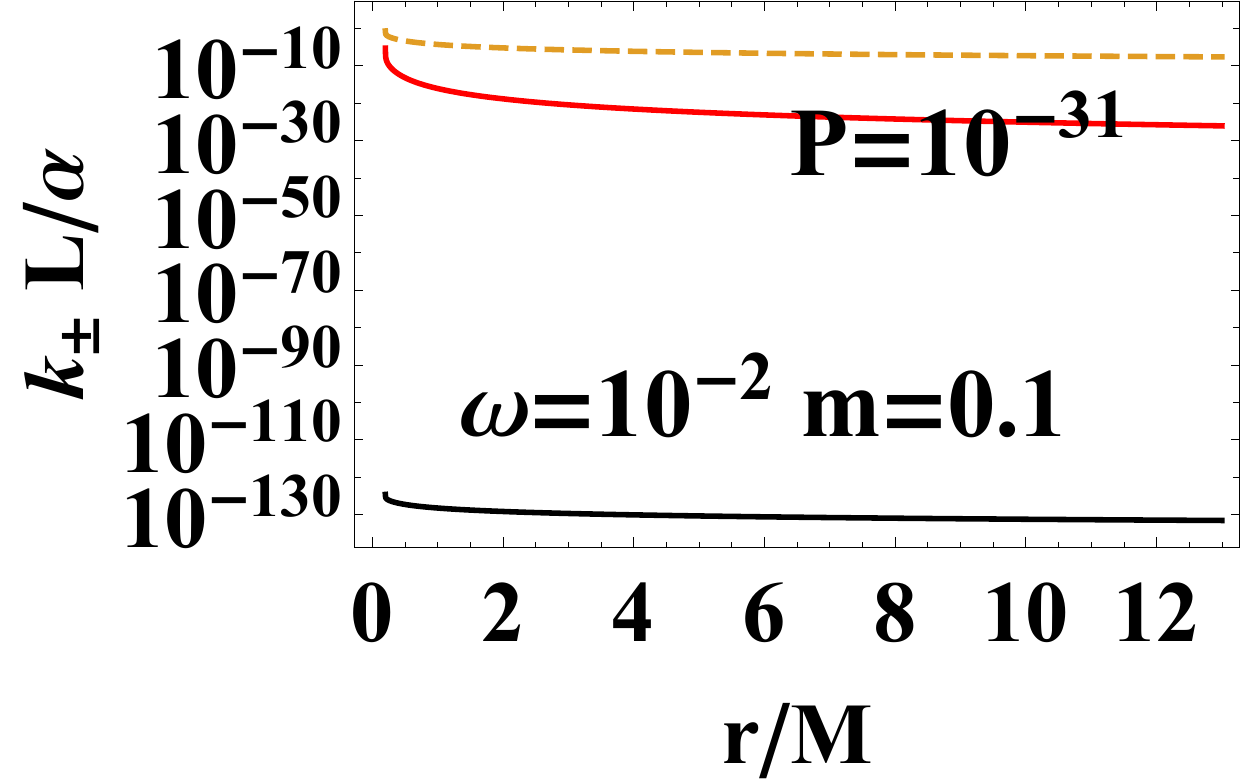}
            \includegraphics[width=5cm]{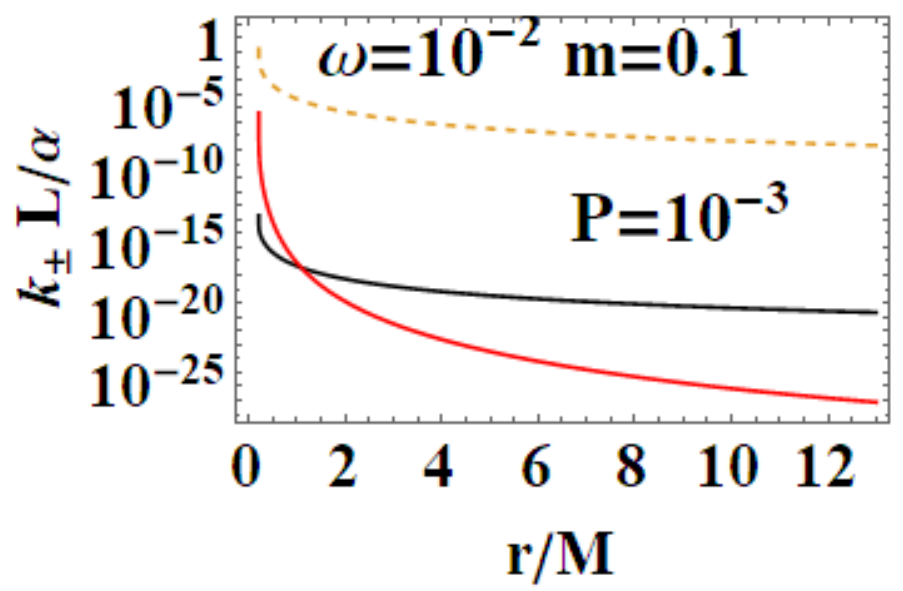}\\
              \includegraphics[width=5cm]{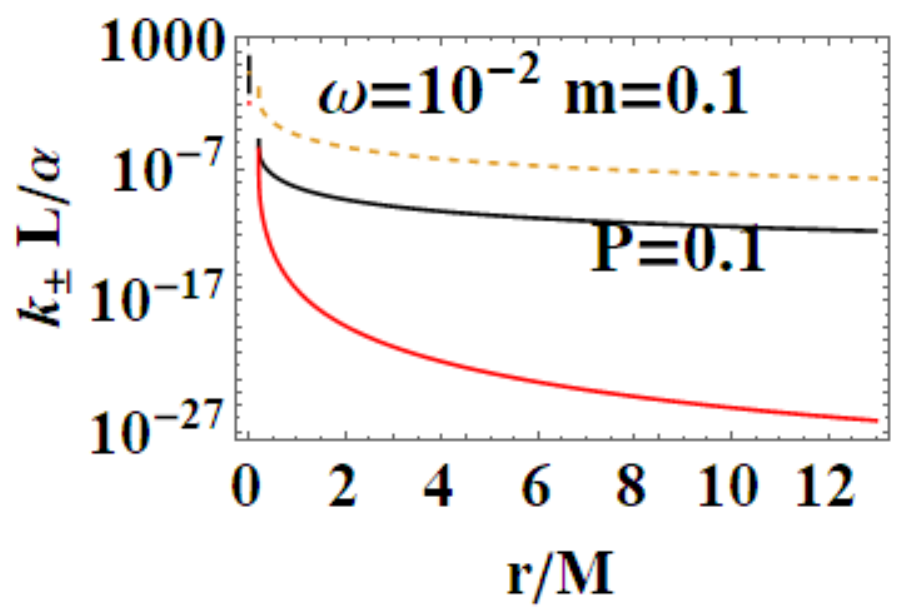}
                \includegraphics[width=5cm]{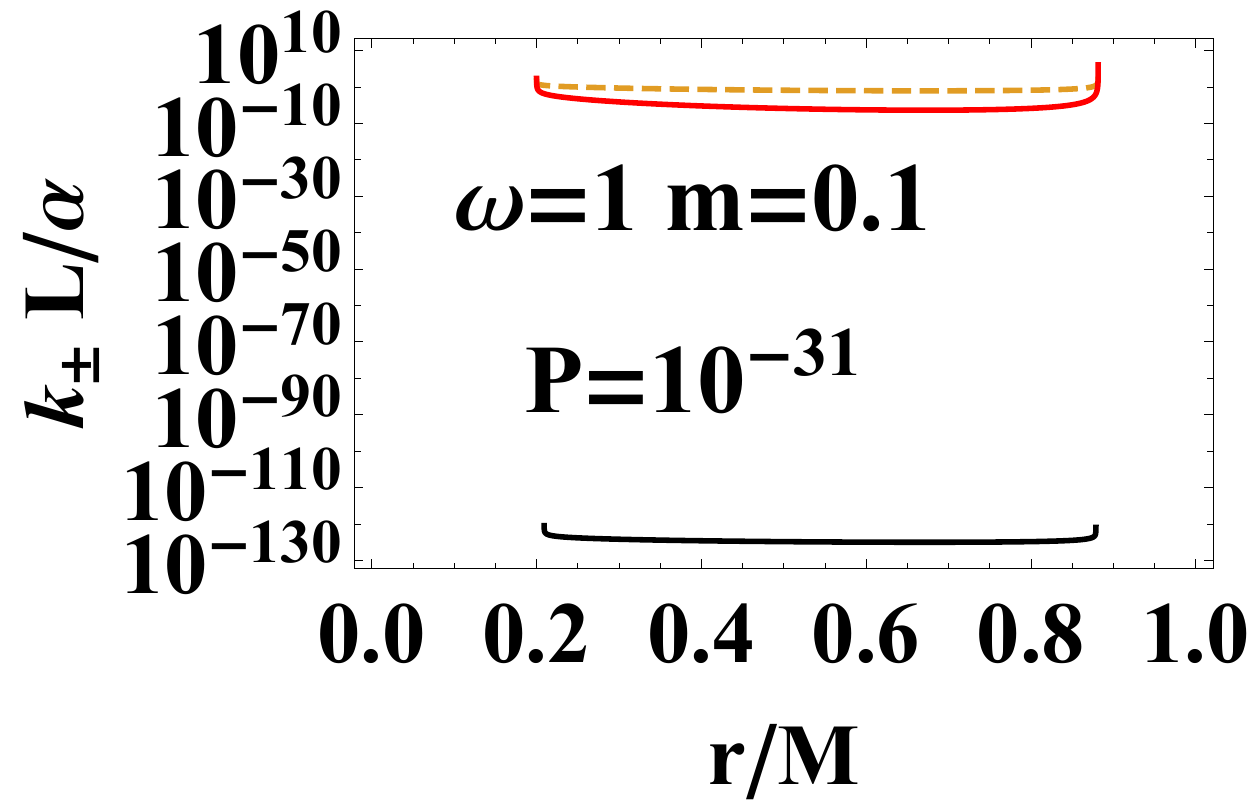}
\includegraphics[width=5cm]{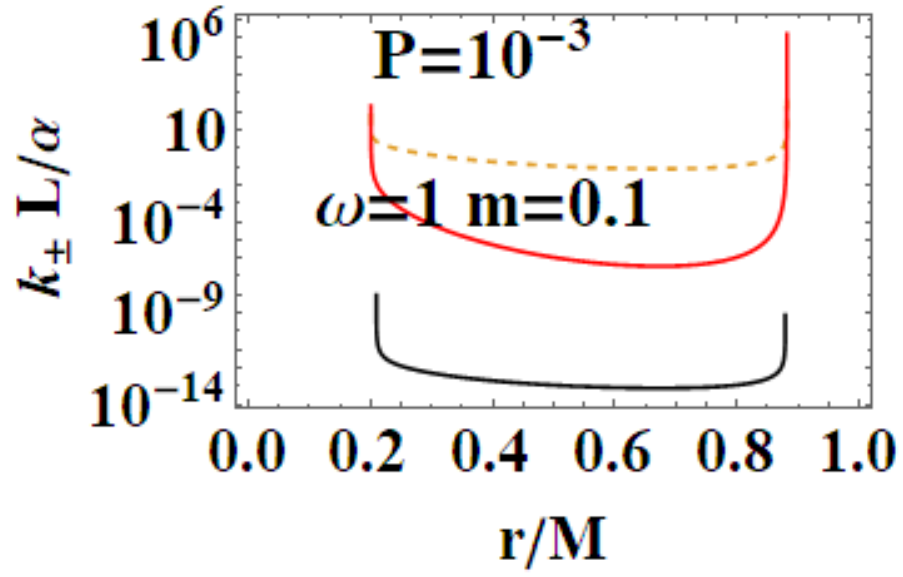}\\
      \includegraphics[width=5cm]{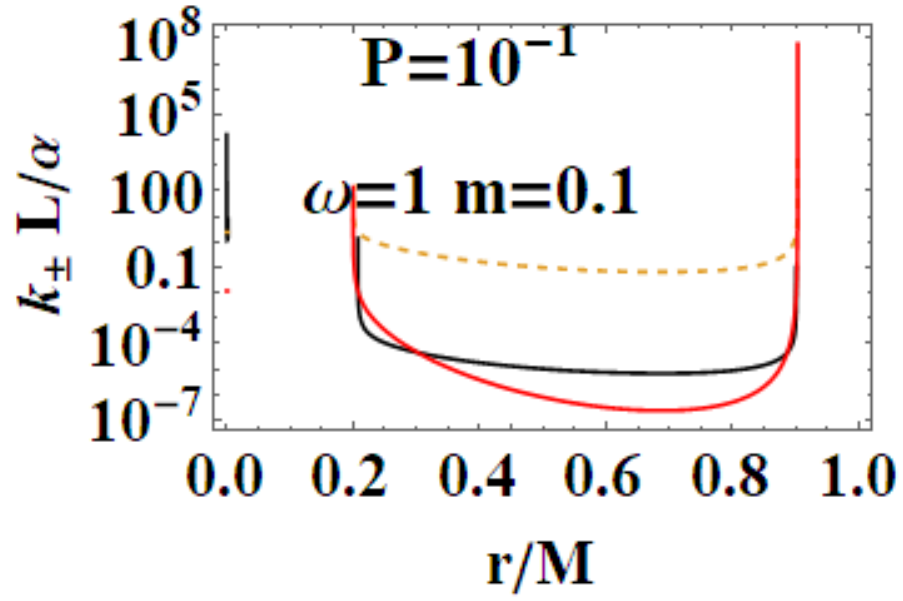}
 \includegraphics[width=5cm]{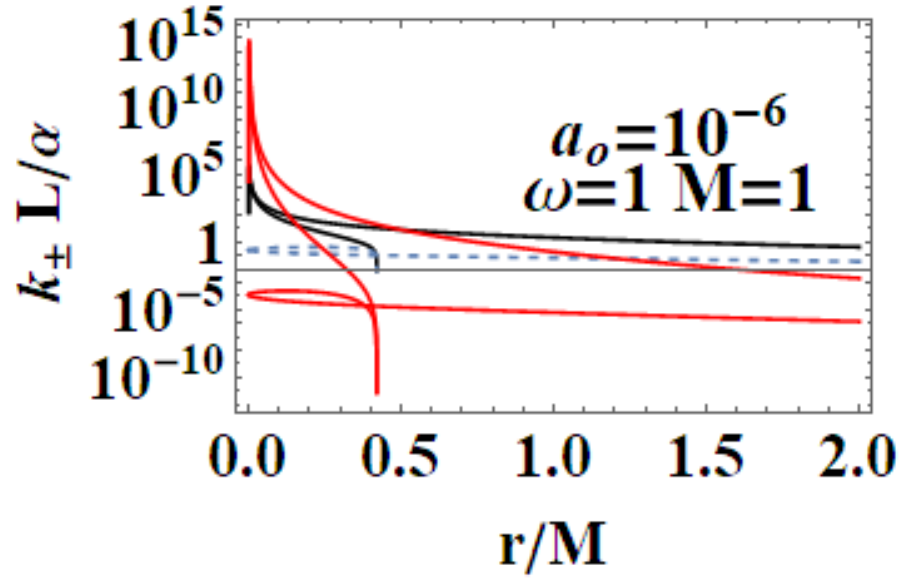}
\includegraphics[width=5cm]{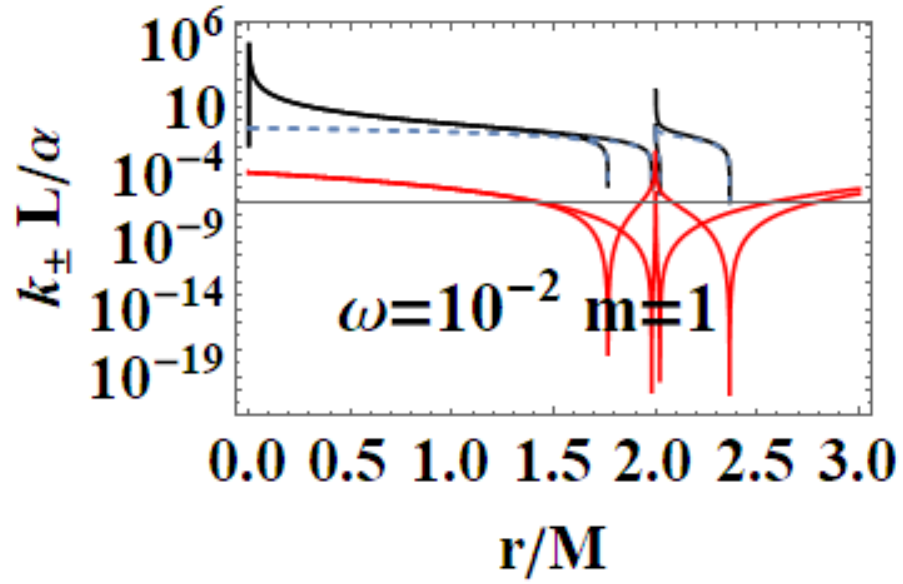}\\
 \includegraphics[width=5cm]{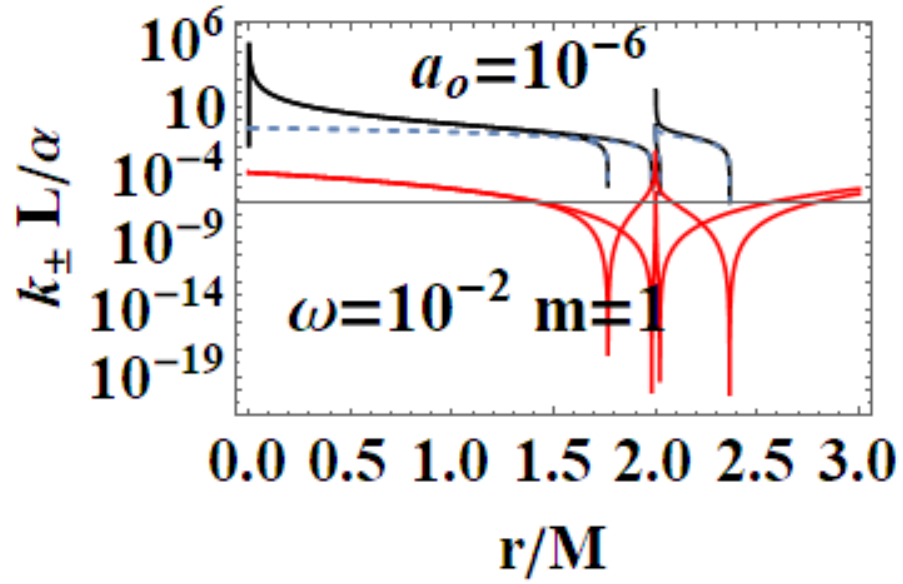}
 \includegraphics[width=5cm]{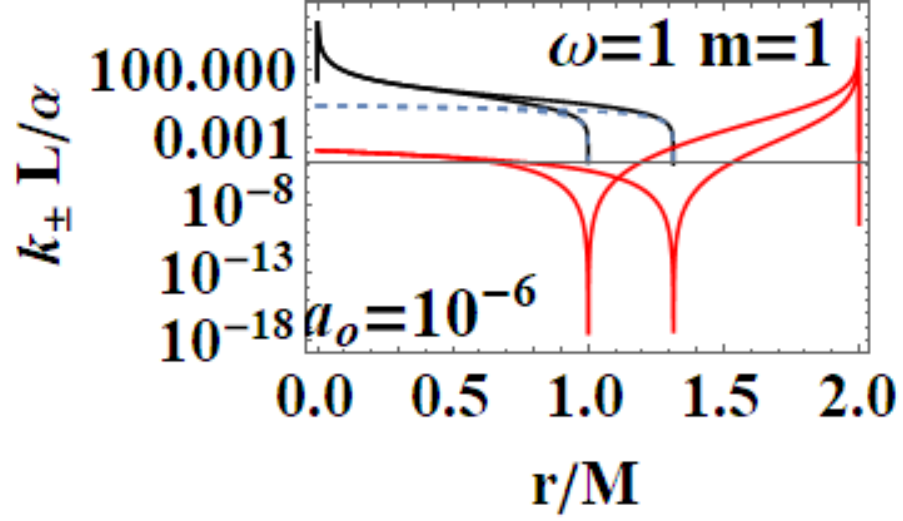}
 \includegraphics[width=5cm]{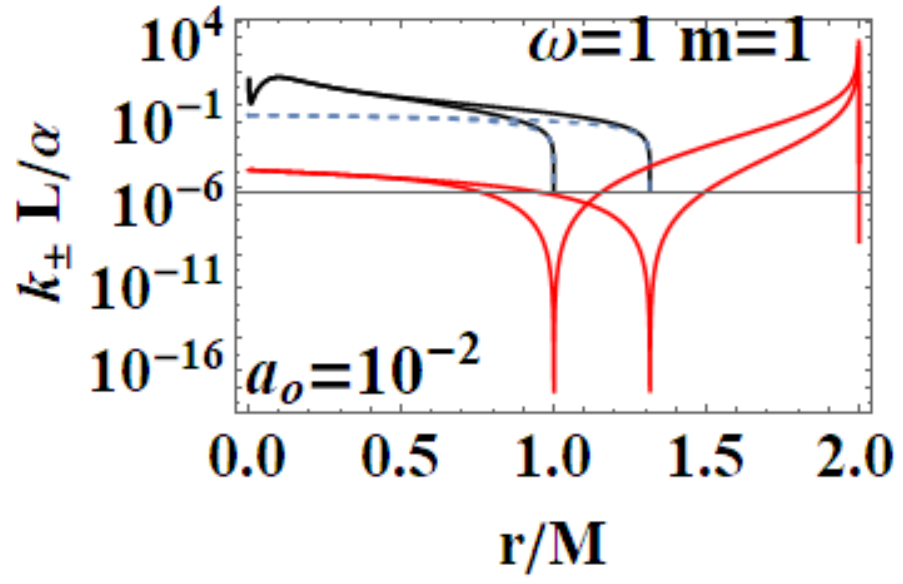}\\
 \includegraphics[width=5cm]{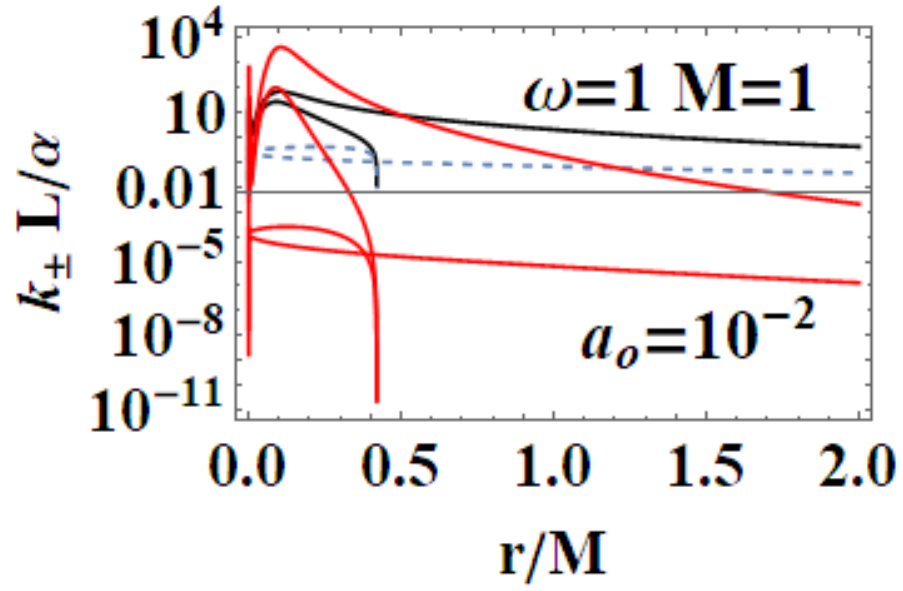}
  \includegraphics[width=5cm]{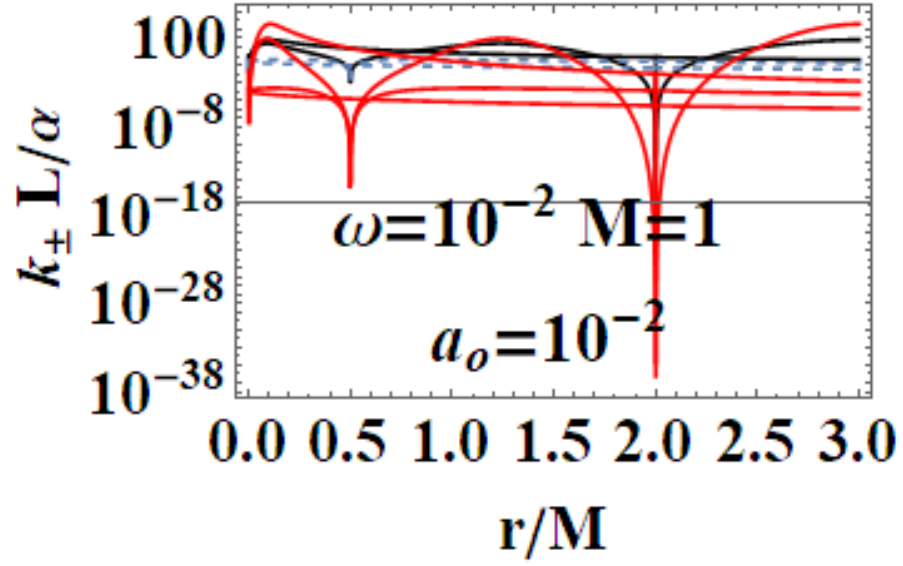}
 \includegraphics[width=5cm]{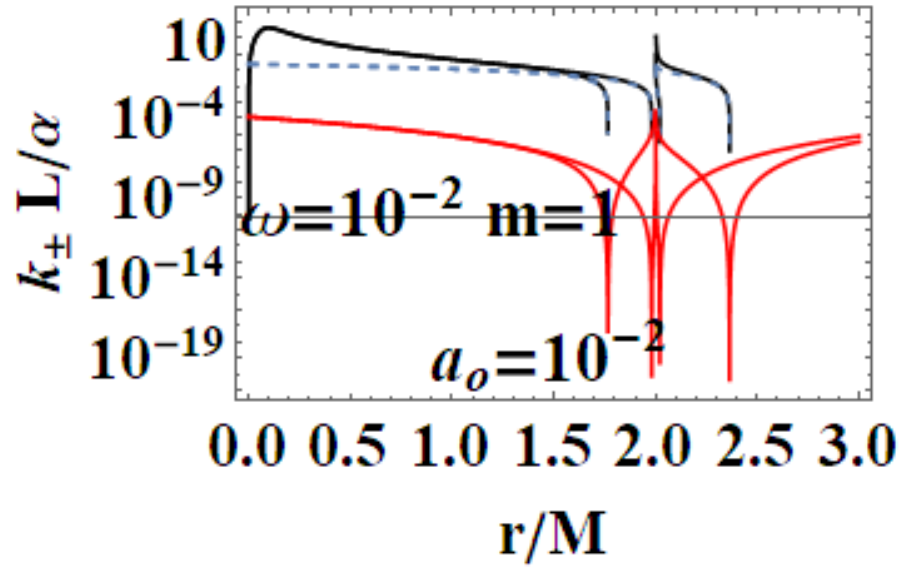}
  \caption{Plots of the surface gravity $\kappa_{\pm}$ and luminosity $L/\alpha$, $\alpha$ is a constant as functions of $r$ evaluated on the metric bundles of   $a_o(\omega)$  or $P_{\omega}$ solution for metric bundles, for different values of the parameters as signed on the panel. Table (\ref{Table:pol-cy-multi}) contains further details on the notation. $a_o= A_{min}/8\pi$ is an area parameter where
 $A_{min}$ is a minimum area appearing in LQG (minimum area gap of LQG).
  $P$   is the  metric polymeric parameter,
  $M$ is the  ADM mass in the Schwarzschild limit  while
 $m$ is a parameter that depends  on  the polymeric function, $\omega$ is the bundle (light-like orbital stationary frequency).}\label{Fig:comprPot1}
\end{figure*}
In Figures (\ref{Fig:comprPot1}), we note the  presence of singularities and the behaviors at increasing distance from the $r=0$ (the bundles' origins).
A transformation from one solution of the bundle to another follows transformations of $(\kappa_{\pm},L/\alpha)$ on the curves evaluated on the bundles.
{This analysis explores the possibility  of a transition  from  one solution of a  metric family to another geometry of the same family, which, for example, can occur after  interaction of the attractor with the surrounding matter environment in non-isolated  BH systems,  which is  the general  case in the most  common  astrophysical  environments. This process would lead a BH  from a point to another point of its extended plane representation. (This transition could also involve, of course,  for some other  diverse  processes, a transition of the   graph parameters). The relevant aspect of this  analysis is that this  transition must carry the system from one point  to another in the extended plane \emph{along} an MB curve. This means that the observer from the initial state will see a transition of the fixed  frequency from a point $r_1$ to  $r _2\neq r_1$ (in general, there are no fixed point along $r=$constant),  where  $r_1 $ and $ r_2$ are two  points along the bundle uniquely identified by the detection of the fixed photon frequency. Vice versa, the observer will be able to recognize at the fixed point  through the
photon orbital frequency variation in the external region any geometry transition in the extended plane (regulated by thermodynamic laws). At fixed frequency, there is always one and only one  bundle; furthermore, a bundle curve does not in general self-cross-- there is an absence of knots.  Therefore, we also test the hypothesis that the bundles, connecting the solutions uniquely through their characteristic frequencies and defining the associated light surfaces,  could have a role in such transitions.
Obviously, the thermodynamic onset provides in the new points of the plane a series of quantities as surface of gravity luminosity or temperature that have evolved on the bundles as shown in these analyses. Therefore, these results have to be compared  with the correspondent analysis of MB curves.  It should be also noted that, in Figures (\ref{Fig:comprPot1}), we have fixed, depending on the parametrization of the bundles,  different parameters  and the frequency. (The functions associated with these quantities are  generally well defined far from the horizons. In the analysis, we have taken advantage of this property  to evaluate in the extended plane these quantities also on  the horizon  curves as clear from the analysis in Figures (\ref{Fig:comprPot1}).  Whatever the parameterizations adopted and the fixed parameters set, the horizon points of the extended plane clearly highlighted by the  vertices of the correspondent  triangle in the representation of the Figures (\ref{Fig:vengplre}) indicate signs of singularity for these quantities).
}
\end{itemize}
\medskip

\section{Discussion and Final Remarks}\label{Sec:conlc}
We review the  steps of this analysis, discussing  the main results and further developments.

 In  Section (\ref{Sec:metric-bundles}), we found  the  metric bundles for the LQG metric approximation, for the geometries considered in  Section (\ref{Sec:metric}). Results are shown in Figures (\ref{Fig:govmentra}) and more extensively analyzed in Section (\ref{Eq:everi-chan}),  within different parameterizations, discussing through these structures  the main characteristics of the family of geometries and  the characteristics of the horizons $r_{\pm}$, through the light surfaces associated with the MBs. We have also clarified some aspects  of construction  of the extended plane, within the application discussed here, the  choice of metric bundles parametrization, and the first formulation of  BH thermodynamics within the MB scenario in  Section (\ref{Sec:termo}). We clarified aspects of MB definitions in the static and spherically symmetric spacetime; see discussion in Section (\ref{Sec:litig-colo}) and  (\ref{Sec:ext-plane}).  An extensive characterization is shown in  Figures (\ref{Fig:colorPP}) and (\ref{Fig:CausalP}), which leads to the construction of the extended plane for these solutions.
We explored   two  representations of the extended plane  for this LQG-BH solution in Figures (\ref{Fig:vengplre}) and (\ref{Fig:vengplrea}) and interpreted   by comparing with the  (stationary) Kerr geometry extended plane in
 Section (\ref{Sec:ext-plane}). In this section, we also  found  the horizons' replicas,   showed in  Figures (\ref{Fig:govmentra}) and Figures (\ref{Fig:govmentr}). These  steps lead to the comparison   of the LBH
with the case  of Reissner--Norstr\"om  (RN) geometry in Section (\ref{Sec:RN}). The LQG geometry under consideration shares different similarities  with RN metrics also from the MBs' stand point. Thus, the second part of this  investigation starts, where we analyzed the LBHs'  thermodynamical properties---Section (\ref{Sec:termo}).
We  characterized the thermodynamical properties of the LBH  solutions in the extended plane for different parameters---see Figures (\ref{Fig:pax12},\ref{Fig:pax12m}),
and on the metric bundles in Figures (\ref{Fig:cononapo16})--Figures (\ref{Fig:comprPot1}),
 showing the divergences from the reference GR  solution. These quantities are evaluated  on the horizon curves in the extended plane  and on the MBs, relating different geometries on the bundles curves with the different values of luminosity, temperatures, or BH areas.
Divergences  with respect to the expected results considering the reference (asymptotically) GR solution  are shown in
Figures (\ref{Fig:fut10}) with respect to the  limiting light-like frequencies $\omega$ used as characteristic frequencies of the bundles. (The analysis of bundles  in the extended plane compares intrinsically with the asymptotical solution which in the  plane is contained  as  points in line $\mathcal{P}=0$).
In Figures (\ref{Fig:pcolorP1}), we show results of the analysis on the horizontal line of the bundles' structures, correspondent to the light surfaces on a specific geometry, and this analysis points out very clearly the presence of non-monotone behaviors of the frequencies $\omega$ with dependence on the metric parameters $m$, depending also on the ADM mass (therefore eventually after a mass  shift following the BH  interaction with the matter environment, or, possibly due  to a  "transition", the "graph state"  may undertake from one value of its characteristic parameters to another).
Similar behavior is shown with the presence of maxima and minima in the LBHs areas and temperatures (surfaces gravity) as in  Figures (\ref{Fig:pax12},\ref{Fig:pax12m}),
and evaluated on the metric bundles in  Figures (\ref{Fig:cononapo16}) and Figures (\ref{Fig:comprPot1}).
A  relevant aspect  of this analysis is that  the replicas  relate a region  close to the horizons virtually in the sense of Figures (\ref{Fig:govmentr})  and (\ref{Fig:govmentra}) and a  region far from the "central"  BH where there is  a copy of the frequency $\omega$. We can measure the discrepancies on the light frequencies (and consequently the timelike frequencies) as measured in these regions expecting a GR solution.
Notably, the analysis may be interpreted as a deformation of  aspects of causal structure (in the sense of causal ball, for example) within the extended plane representation of Figures (\ref{Fig:govmentra}) or Figures (\ref{Fig:govmentr}), relating graph properties to BHs thermodynamics with MBs.

We summarize  below results of the investigation with some comments and contextualization.

\textbf{\emph{1. }Constraining LQG solutions.}
One purpose of this analysis is to provide constraints to the  LQG  mini-super-space   polymeric regular BHs    of Equation~(\ref{Eq:dom-gionr}) within the framework provided by  metric Killing  bundles. The special framework is therefore  here firstly applied to {LQG--BH }and BH thermodynamics to  discern possible  LQG imprints in the  characteristics  of regions close to   BHs horizons and particularly within the idea to constrain the underlining graph features. Constraints are provided as   restrictions of  the graph features    with respect to ADM and polymeric mass $(M,m)$, the $(\epsilon, P)$ polymeric metric parameters, and the minimal LQG area parameter $a_o$.
(Eventually, we enlarged the parameters' value ranges to test the model, bracing  the hypothesis of interacting attractor in an  astrophysical BH scenario, where the  attractor, and consequently the  graph, eventually may evolve---inducing a transition from a point to another on a bundle  in  the extended plane).
 Discrepancies are  highlighted by the comparative  analysis with the reference solution of the  general relativistic onset.
As metric bundles are particular  sets of BH  solutions which are defined  by  properties of  special associated  light surfaces,
 this issue has been addressed  with the investigations of the  properties  of the orbital null like frequencies $\omega$ (characteristic bundle frequencies---all the geometries of the bundle \emph{per}-definition have  equal values of orbital  light frequency).  Particularly, we are interested in the properties of light-like limiting frequencies  of  stationary  observers tracing some characteristics of the regions close to the  horizons  in the sense of the extended plane and  the replicas.
 A  goal  of this analysis was   to provide constraints to the  graph construction, which, in various ways, underlines the space(-time) structure of the model bridging the classical limit and quantum theory.  In this respect, this analysis  actually  was  inspired by the idea to relate the graph to light propagation, within Killing horizons and thermodynamics through the MB formalism.
Besides the problem of the  observation  of the possible quantum effects on large (macroscopical) scale structure, there are on the theoretical grounds  several  aspects related to these theories to be clarified.
The graph model quite naturally encodes  the  geometry discretization  with the
 loop quantum gravity  states. %
 The key point  of these quantum gravity approaches is actually the  geometrical interpretation---in other words, to assign
an opportune and adapted geometry to (the set of LQG) states.
 LQG states (quanta of space) were  provided as  spin network
states, which are associated with a graph for the  3D (quantum) geometry.
(The graph  granulates the geometry, constituents and structure; therefore, there is  large interest in the analysis of this very fundamental idea beyond a novel gravity geometrization/universalism).
A second key question is of course to reconcile the quantum approach to a classical or (semi-classical) continuous geometry.
The different regimes of the theory are provided generally by the graphs.
LQG is generally based  on a fixed (lattice) graph  replacing  and "fine-graining" the texture of GR geometry and establishing part of the relational structure.  The Hilbert space, formed by the states
 on the graph, provides a benchmarking between these.
 It should be also noted that indeed these special light-surfaces related  aspects of GR causal structure (delimiting  existence of static and stationary observers)  to the graph geometry, in different parts of the same spacetime (through replicas) and different geometries (bundles).
We also provided constraints considering possible thermodynamic transformations intended as a shift from one solution to another (a transition of horizontal lines of the extended plane). We focused in this  investigation particularly on  the analysis of the     BH thermodynamical properties considering  luminosity and surface gravity. In the MB frame, we  found variations of these quantities on the  bundle curves, and thus relate  the different geometries, according to the metric parameters,  with their thermodynamic characteristics.

\textbf{\emph{2.} MBs for LQG-BH solutions.} This analysis ultimately also clarifies aspects of MBs introduced in \cite{renmants,ergon,observers,Pugliese:2019rfz,
Pugliese:2019efv} when applied to the spherically symmetric case generalizing the tangency conditions of MBs with the horizons' curves in the extended plane with a notion of approximations in the sense of Figure~(\ref{Fig:govmentr}).    As a sideline of this analysis, the  MBs approach provided a novel frame of analysis and representation of the  families of geometries with Killing horizons.
This reinterpretation started from  the construction of  the  extended plane for  LQG polymeric BHs solutions in Section (\ref{Sec:ext-plane})   comparing  with  the case of the  Reissner--Norstr\"om geometry in  Section (\ref{Sec:RN}), whereas, in  Section (\ref{Sec:wordsee-h}), we introduced the
{metric bundles of the LBHs}. One goal was to explore the MBs for the  spherically symmetric static solutions as limiting solutions in the extended plane,  hence the comparative analysis  with the extended plane in Kerr geometries and the analysis with RN geometries.  For one side, the metric (\ref{Eq:dom-gionr}) has similarities in RN spacetimes. On the other side, we used the electrically charged and spherically symmetric  spacetime to enlighten properties of the extended plane. In the extended plane, the RN geometries were interpreted in \cite{renmants} as limiting geometries of the (stationary  electro-vacuum) Kerr-Newman solution occurring when the total charge is  $\Qa_T=Q/M$. In this respect, the RN geometry  was seen as a point in the  Kerr-Newman extended plane of $\mathcal{P}-r/M$, where $\mathcal{P}$ is  pair of two parameters.

\textbf{\emph{3.} MBs and  BHs thermodynamics.} The analysis has been completed with   the re-formulation in Section (\ref{Sec:termo}) of
 several aspects of the BH thermodynamics, where particularly
in Section (\ref{Sec:termo-1})  the BHs  surfaces gravity $\kappa_{\pm}$, the luminosity $L$, and the temperature $T_{\mathbf{BH}}^{\pm}$ have been investigated  in terms of the loop model parameters $\mathcal{P}$, thus these quantities are  considered  on metric bundles.
Relevant for emission analysis is the region $r\in[2M,3M]$, which is in the limiting Schwarzschild geometry  ($M$ is the ADM mass) the region between the outer horizon and the last photon circular orbits, which here has a special role in the bundles analysis as we showed in Figures (\ref{Fig:govmentra}).
In conclusion, we constrain the graph-metric bundles and thermodynamics.  
The notion of horizons' replicas  provided by the collections of the MBs allows for reinterpreting the  (classical) thermodynamical BH physics  in terms of transitions from one point  to  another (on the vertical line) of the extended plane.   From this standpoint, we focused on the explorations of the main quantities of  LBHs physics entering into the analysis of BH thermodynamical transformations.

\textbf{\emph{4. }Astrophysical relevance and phenomenological impact.}  A further  goal of our work was therefore also  to test the MBs  in the context of modified gravity. The goal is ultimately to detect (and interpret) the hints of  quantum modifications outside the horizons, constraining eventually the graph properties---here the parameters $\mathcal{P}$ we used to construct the extended plane or the masses (the loop and  ADM mass), and to evaluate a possible shift in the model parameters  evolving  from one solution in the extended plane to another.  Our analysis presents  an observational frame  provided by the MBs onset  grounded on the analysis of certain light-surfaces of the geometries.  These light surfaces are the basis of several constraints of aspects of accretion disks. The characteristic bundles frequencies and connects regions close to the horizons to far regions.  A further advantage  of  this method  with respect to others (for example, the analysis of particle motions or  spectra emission) is for its astrophysical interests, opening a wide window of different applications centered on the concept of MBs.
The definition of these particular surfaces and their associated orbital frequencies are the basis of constraints  to different  results of the  High Energy Astrophysics of BHs,   providing therefore  a  powerful  and ample investigation scenario.
Up to now, no observational
evidence has   outlined a clear distinctive  signature of quantum gravity  scenario but  on
constraints  on the existing proposals.
There is therefore a great deal of attention on noticing
any discrepancies in the current observations with respect to the predictions of the
standard theoretical setup enclosed in GR theory that could be possibly explained in a quantum model. In this sense, the astrophysical setting offers
the most natural arena for investigating the phenomenology of quantum gravity; in particular, one can search
for new phenomena, which are unpredicted by the current GR model, but explained in a  quantum gravity framework.
 Such observation could provide  a strong constraint  on the  validity  of many models.
In this analysis, by  comparing the predictions of  the model  with
the features of the ordinary  general relativistic astrophysics, within the analysis of these  light-surfaces and the derived concept of metric bundles, we highlight
some small but finite discrepancies, expectably detectible from the observations.
In this perspective,
the construction of the extended  plane and metric bundles, with the replica definition,   has consistently proved the possibility to detect the  existence of  divergences from expected prediction of the GR model of reference.
Several observational channels are  opened  today  in this context applicable to the examination   of the  light-surfaces; for example, we mention
the recent window of Gravitational Wave analysis  and especially the   Event Horizon Telescope to explore from  different (independent) angles the physics of BHs and their horizons.
 We tested   the viability of this method to constrain the  theory and possible observational evidence on the Astrophysical  phenomena related to the BH physics  focusing on the BH events' horizon analysis.
For these reasons, this investigation obviously could not exclude considerations on BH thermodynamics, directly governed by the  BH horizon. The discrepancies
highlighted between the predictions of these models, and the general relativistic ones
are small but may be detectable.

In conclusion, the  light-surfaces are at the base of MB definitions, as  their characteristic frequencies have a wide field of application in many different aspects of BH astrophysics considering magnetic fields and different features of accretion physics. Therefore, we believe this approach  could be a  fruitful environment in which we can highlight the details attributable to transition
   from classical to quantum scales, characterized by non-trivial modification  from the
corresponding GR counterpart. It can be expected that this departure from
the results  can   be evident also  in   the extended matter configurations, as  accretion disks, their   dynamics,  and morphology.
We expect therefore  to apply this method in different exact and approximated solutions.

\authorcontributions{All authors have contributed to the overall elaboration of the article, its different stages of development and realization.}

\funding{{This research received} no external funding.}

\conflictsofinterest{The authors declare no conflict of interest.}

\reftitle{References}

\end{document}